\documentclass{book}

\usepackage[paperwidth=17cm, paperheight=24cm, includeheadfoot,
            left=2cm, right=2cm, top=2cm, bottom=1.3cm, twoside]{geometry}

\usepackage%
 {
  amsmath, amssymb, appendix, caption, cite, enumitem, fancyhdr, fancyvrb,
  graphicx, import, lineno, longtable, remreset, shorttoc, sidecap, url
 }

\newcommand{\tmpstring}{}
\newcommand{\settmpstring}[1]{\renewcommand{\tmpstring}{#1}}
\newcommand{\SourceCodeLines}[1]
 {%
  \settmpstring{{\ttfamily\bfseries\tiny\theFancyVerbLine}}
  \ifnum#1>9
    \settmpstring
     {\parbox[b]{7.5pt}{\ttfamily\bfseries\tiny\rightline\theFancyVerbLine}}
  \fi
  \ifnum#1>99
    \settmpstring
     {\parbox[b]{11.2pt}{\ttfamily\bfseries\tiny\rightline\theFancyVerbLine}}
  \fi
 }

\def\thepart{\Alph{part}}

\makeatletter
\@removefromreset{figure}{chapter}
\makeatother

\makeatletter
\renewcommand{\thefigure}{\@arabic\c@figure}
\makeatother

\makeatletter
\@removefromreset{table}{chapter}
\makeatother

\makeatletter
\renewcommand{\thetable}{\@arabic\c@table}
\makeatother

\makeatletter
\@removefromreset{equation}{chapter}
\makeatother

\makeatletter
\renewcommand{\theequation}{\@arabic\c@equation}
\makeatother

\usepackage{tocloft} 
\usepackage{minitoc} 

\cftsetpnumwidth{1.73em}

\makeatletter
\renewcommand{\@tocrmarg}{4em}
\makeatother

\mtcsetformat{minitoc}{tocrightmargin}{2.55em plus 1fil}

\newcommand{\Author}{}
\newcommand{\AuthorLastName}[1]{\renewcommand{\Author}{#1}}

\def\Title#1{\chapter[\thepart\thelecture\ \protect\mbox{\protect%
\parbox[t]{110mm}{#1 \textit{(\Author)}}}\smallskip]{\Large #1}}

\newsavebox{\shortTitleBox}
\def\shortTitle#1{\savebox{\shortTitleBox}{#1 \textit{(\Author)}}}

\newcounter{lecture}[part]

\newcommand{\SwitchToFancy}
 {%
  \pagestyle{fancy}
  \fancyhf{}
  \fancyhead[OR]{\rightmark}
  \renewcommand{\headrulewidth}{0.4pt}
  \fancyfoot[OR]{\thepage}
  \fancyfoot[EL]{\thepage}
  \fancyhead[EL]{\usebox{\shortTitleBox}}
 }

\newenvironment{example}[1]
 {
  \noindent\rule{\textwidth}{0.5mm}\vspace{-0.1cm}
  \begin{quote}
  \underline{Example}: #1
  \vspace{0.1cm}

 }
 {
  \vspace{\baselineskip}

  \end{quote}
  \vspace*{-0.7cm}\rule{\textwidth}{0.5mm}

 }

\makeatletter
\def\thebibliography#1
{
 \section*{References\@mkboth{References}{References}}
 \list{[\arabic{enumi}]}
  {
   \settowidth\labelwidth{[#1]}\leftmargin\labelwidth\advance\leftmargin
   \labelsep\usecounter{enumi}
  }
 \def\newblock{\hskip .11em plus .33em minus .07em}
 \sloppy\clubpenalty4000\widowpenalty4000\sfcode`\.=1000\relax
}
\makeatother

\makeatletter
\renewcommand{\@makefntext}[1]{\setlength{\parindent}{0pt}%
\begin{list}{}{\setlength{\labelwidth}{1.5em}%
\setlength{\leftmargin}{\labelwidth}%
\setlength{\labelsep}{3pt}\setlength{\itemsep}{0pt }%
\setlength{\parsep}{0pt}\setlength{\topsep}{0pt}%
\footnotesize}\item[\hfill\@makefnmark]#1%
\end{list}}
\makeatother

\input{syntaxHighlighting_melchert.tex}

\begin{document}

\dominitoc

\faketableofcontents

\renewcommand{\cftsecfont}{\bfseries}
\renewcommand{\cftsecleader}{\bfseries\cftdotfill{\cftdotsep}}
\renewcommand{\cftsecpagefont}{\bfseries}

\fancypagestyle{plain}
 {
  \renewcommand{\headrulewidth}{0pt}
  \fancyhead[OL]
   {
    \vspace{-8.5pt}
    Published in:\\
    \textit{Reinhard Leidl and Alexander K.\ Hartmann (eds.)}\\
    \textit{\emph{Modern Computational Science 2012 -- Optimization}}\\
    \textit{Lecture Notes from the 4$^\text{th}$ International Summer School}\\
    \textit{BIS-Verlag, Oldenburg, 2012. ISBN 978-3-8142-2269-1}
   }
 }

\stepcounter{lecture}
\setcounter{figure}{0}
\setcounter{equation}{0}
\setcounter{table}{0}

\AuthorLastName{Melchert}
\Title{Basic Data Analysis and More -- A Guided Tour Using {\tt python}}
\shortTitle{Basic data analysis}
\SwitchToFancy
\bigskip
\bigskip
\begin{raggedright}
  \itshape Oliver Melchert\\
  Institute of Physics\\
  Faculty of Mathematics and Science\\
  Carl von Ossietzky Universit\"at Oldenburg\\
  D-26111 Oldenburg\\
  Germany
  \bigskip
  \bigskip
\end{raggedright}

\paragraph{Abstract.}
In these lecture notes, a selection of frequently required statistical tools
will be introduced and illustrated. 
They allow to post-process data that stem from, e.g.,  large-scale numerical
simulations (aka sequence of random experiments).
From a point of view of data analysis, the concepts and techniques introduced 
here are of general interest and are, at best, employed  by computational aid.
Consequently, an exemplary implementation of the presented techniques
using the {\tt python} programming language is  provided.
The contents of these lecture notes is rather selective and represents
a computational experimentalist's view on the subject of basic data analysis,
ranging from the simple computation of moments for distributions of random 
variables to more involved topics such as hierarchical cluster analysis
and the parallelization of {\tt python} code.
\medskip

\begin{center}
  \framebox
   {%
    \parbox{0.9\textwidth}
     {%
      Note that in order to save space, all {\tt{python}} snippets presented
      in the following are undocumented. In general, this has to be considered
      as bad programming style. However, the supplementary material,
      i.e., the example programs you can download from the MCS homepage,
      is well documented (see Ref.\ \cite{supplMat}).%
     }%
   }
\end{center}

\newpage

\minitoc

\section{Basic {\tt python}: selected features}
\label{sect:basicPython}

Although {\tt python} syntax is almost as readable as pseudocode
(allowing for an intuitive understanding of the code-snippets listed in
the present section),
it might be useful to discuss a minimal number of its features, needed 
to fully grasp the examples and scripts in the supplementary material (see Ref.\ \cite{supplMat}).
In this regard, the intention of the present notes is not to demonstrate
every nut, bolt and screw of the {\tt python} programming language,
it rather illustrates some basic data structures and shows how to manipulate them. 
To get a more comprehensive introduction to the {\tt python} programming 
language, Ref.\ \cite{pythonTutorial} is a good starting point. 

There are two elementary data structures which one should be aware of when
using {\tt python}: \emph{lists} and \emph{dictionaries}. Subsequently, the use of these
is illustrated by means of the interactive {\tt python} mode. One can enter this 
mode by simply typing \verb!python! on the command line.
\paragraph{Lists.} A list is denoted by a pair of square braces. 
The list elements are indexed by integer numbers, where the smallest index has value $0$. 
Generally speaking, lists can contain any kind of data. 
Some basic options to manipulate lists are shown below:
\SourceCodeLines{99}
\begin{Verbatim}
 >>> ## set up an empty list
 ... a=[]
 >>> ## set up a list containing integers
 ... a=[4,1]
 >>> ## append a further element to the list
 ... a.append(2)
 >>> ## print the list and the length of the list
 ... print "list=",a,"length=",len(a)
 list= [4, 1, 2] length= 3
 >>> ## lists are circular, list has indices 0...len(a)-1
 ... print a[0], a[len(a)-1], a[-1]
 4 2 2
 >>> ## print a slice of the list (upper bound is exclusive)
 ... print a[0:2]
 [4, 1]
 >>> ## loop over list elements
 ... for element in a:
 ...   print element,
 ... 
 4 1 2
 >>> ## the command range() generates a special list
 ... print range(3)
 [0, 1, 2]
 >>> ## loop over list elements (alternative)
 ... for i in range(len(a)):
 ...   print i, a[i]
 ... 
 0 4
 1 1
 2 2
\end{Verbatim}

\paragraph{Dictionaries.} 
A dictionary is an unordered set of \verb!key:value! pairs, surrounded by 
curled braces.
Therein, the keys serve as indexes to access the associated values 
of the dictionary.
Some basic options to manipulate dictionaries are shown below
\SourceCodeLines{99}
\begin{Verbatim}
 >>> ## set up an empty dictionary
 ... m = {}
 >>> ## set up a small dictionary
 ... m = {'a':[1,2]}
 >>> ## add another key:value-pair
 ... m['b']=[8,0]
 >>> ## print the full dictionary
 ... print m
 {'a': [1, 2], 'b': [8, 0]}
 >>> ## print only the keys/values
 ... print m.keys(), m.values()
 ['a', 'b'] [[1, 2], [8, 0]]
 >>> ## check if key is contained in map
 ... print 'b' in m
 True
 >>> print 'c' in m
 False
 >>> ## loop over key:value pairs
 ... for key,list in m.iteritems():
 ...   print key, list, m[key]
 ... 
 a [1, 2] [1, 2]
 b [8, 0] [8, 0]
\end{Verbatim}

\paragraph{Handling files.} 
Using {\tt python} it takes only a few lines to fetch and 
decompose data contained in a file. Say you want to 
pass through the file \verb!myResults.dat!, which contains
the results of your latest numerical simulations:
\SourceCodeLines{99}
\begin{Verbatim}
 0 17.48733
 1 8.02792
 2 7.04104
\end{Verbatim}
Now, disassembling the data can be done as shown below:
\SourceCodeLines{99}
\begin{Verbatim}
 >>> ## open file in reade-mode
 ... file=open("myResults.dat","r")
 >>> ## pass through file, line by line
 ... for line in file:
 ...   ## separate line at blank spaces
 ...   ## and return a list of data-type char
 ...   list = line.split()
 ...   ## eventually cast elements
 ...   print list, int(list[0]), float(list[1])
 ... 
 ['0', '17.48733'] 0 17.48733
 ['1', '8.02792'] 1 8.02792
 ['2', '7.04104'] 2 7.04104
 >>> ## finally, close file
 ... file.close()
\end{Verbatim}

\paragraph{Modules and functions.}
In {\tt python} the basic syntax for defining a function reads
\verb!def funcName(args): <indented block>!. 
{\tt python} allows to emphasize on code modularization. In doing so, it offers the 
possibility to gather function definitions within some file, i.e.\ a module,  
and to import the respective module to an interactive session or to some 
other script file. E.g., consider the following module (\verb!myModule.py!)
that contains the function \verb!myMean()!:
\SourceCodeLines{99}
\begin{Verbatim}
 def myMean(array):
         return sum(array)/float(len(array))
\end{Verbatim}
Within an interactive session this module might be imported
as shown below
\SourceCodeLines{99}
\begin{Verbatim}
 >>> import myModule
 >>> myModule.myMean(range(10))
 4.5
 >>> sqFct = lambda x : x*x
 >>> myModule.myMean(map(sqFct,range(10)))
 28.5
\end{Verbatim}
This example also illustrates the use of namespaces in {\tt python}:
the function\linebreak \verb!myModule()! is contained in an external
module and must be imported in order to be used. As shown above,
the function can be used if the respective module name is prepended.
An alternative would have been to import the function directly 
via the statement \verb!from myModule import myMean!. Then 
the simpler statement \verb!myMean(range(10))! would have
sufficed to obtain the above result. Furthermore, simple functions 
can be defined ``on the fly'' by means of the \verb!lambda! 
statement. Line 4 above shows the definition of the function \verb!sqFct!,
a lambda function that returns the square of a supplied number. In order to 
apply \verb!sqFct! to each element in a given list, the \verb!map(sqFct,range(10))!
statement, equivalent to the list comprehension \verb![sqFct(el) for el in range(10)]!
(i.e.\ an inline looping construct), might be used.

\paragraph{Basic sorting.}
Note that the function \verb!sorted(a)! (used in the script \verb!pmf.py!, see 
Section \ref{sect:pf} and supplementary material), which returns a new sorted list using the 
elements in \verb!a!, is not available for {\tt python} versions prior to version 
$2.4$. As a remedy, you might define your own sorting function.
In this regard, one possibility to accomplish the task of sorting the elements contained
in a list reads 
\SourceCodeLines{99}
\begin{Verbatim}
 >>> ## set up unsorted list
 ... a = [4,1,3]
 >>> ## def func that sorts list in place
 ... def sortedList(a):
 ...   a.sort(); return a
 ... 
 >>> for element in sortedList(a):
 ...   print element,
 ... 
 1 3 4
\end{Verbatim}
Considering dictionaries, its possible to recycle the function \verb!sortedList()!
to obtain a sorted list of the keys by defining a function similar to
\SourceCodeLines{99}
\begin{Verbatim}
 def sortedKeys(myDict): return sortedList(myDict.keys())
\end{Verbatim}
As remark, note that {\tt python} uses \emph{Timsort} \cite{timsort}, a hybrid sorting 
algorithm based on \emph{merge sort} and \emph{insertion sort} \cite{clrs2001}.

To sum up, {\tt python} is an interpreted (no need for compiling) high-level programming 
language with a quite simple syntax.
Using {\tt python}, it is easy to write modules that can serve as small libraries. Further,
a regular {\tt python} installation comes with an elaborate suite of general purpose libraries that
extend the functionality of {\tt python} and can be used out of the box. These are e.g., 
the \verb!random! module containing functions for random number generation,
the \verb!gzip! module for reading and writing compressed files,
the \verb!math! module containing useful mathematical functions,
the \verb!os! module providing a cross-platform interface to the functionality
of most operating systems, and 
the \verb!sys! module containing functions that interact with the {\tt python} interpreter.
%

\section{Basic data analysis}
\label{sect:basicDataAnalysis}
\subsection{Distribution of random variables}
\label{sect:pf}

Numerical simulations, performed by computational means, can be 
considered as being \emph{random experiments}, i.e., experiments 
with outcome that is not predictable.
For such a random experiment, the \emph{sample space} $\Omega$ specifies 
the set of all possible outcomes (also referred to as \emph{elementary events}) of the experiment. 
By means of the sample space, a \emph{random variable} $X$ 
(accessible during the random experiment), can be 
understood as a function
\begin{eqnarray}
X:\Omega \rightarrow \mathbb{R} \qquad \Big(\text{\parbox{0.25\textwidth}{$X=$ random variable  $\Omega=$ sample space}}\Big)
\end{eqnarray}
that relates some numerical value to each possible outcome of the 
random experiment thus considered. To be more precise, 
for a possible outcome $\omega\in\Omega$ of a random experiment, $X$ 
yields some numerical value $x=X(\omega)$. To facilitate intuition, 
an exemplary random variable for a $1D$ random walk is considered
in the example below. 
Note that it is also possible to combine several random variables $\{X^{(i)}\}_{i=0}^{k}$
to define a new random variable as a function of those, i.e.\ \begin{eqnarray}
Y=f(X^{(0)},\ldots,X^{(k)}) \qquad \Big(\text{\parbox{0.3\textwidth}{combine several random variables to a new one}}\Big).
\end{eqnarray}
To compute 
the outcome $y$ related to $Y$, one needs to perform random experiments for the $X^{(i)}$, 
resulting in the outcomes $x^{(i)}$. Finally, the numerical value of $y$
is obtained by equating $y=f(x^{(0)},\ldots,x^{(k)})$, as illustrated 
in the example below.

\begin{figure}
\centering
\includegraphics[width=0.95\textwidth]{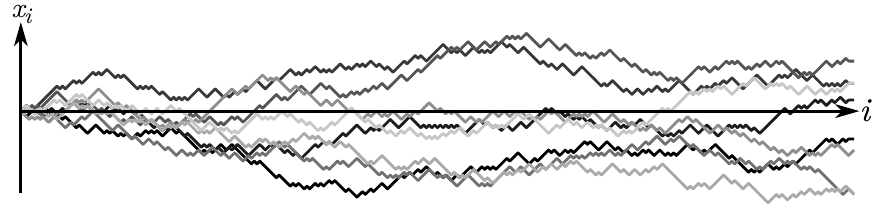}
\caption{A number of $8$ independent symmetric $1D$ random walks, where
each walk takes a number of $N=300$ steps. The position after the 
$i$th step (horizontal axis) is referred to as $x_i$ (vertical axis).
The individual line segments connect subsequent positions along the $1D$ 
random walk. \label{fig:1DrandWalk_samples}}
\end{figure}
\begin{example}{The symmetric $1D$ random walk}
The $1D$ random walk, see Fig.\ \ref{fig:1DrandWalk_samples}, is a very basic example of a trajectory that
evolves along the line of integer numbers $\mathbb{Z}$. Let us agree
that each step along the walk has length $1$, leading to the left or right with 
equal probability (such a walk is referred to as symmetric).

Now, consider a random experiment that reads: \emph{Take a step!}
Then, the sample space which specifies the set of all 
elementary events for the random experiment is just
$\Omega=\{\text{left},\text{right}\}$. To signify the effect of taking 
a step we might use a random variable $X$ that results in $X(\text{left})=-1$ and
$X(\text{right})=1$. Note that this is an example of a \emph{discrete}
random variable. 

Further, let us consider a symmetric $1D$ random walk that starts at the distinguished 
point $x_0=0$ and takes $N$ successive steps, where the position after the $i$th step
is $x_i$. 

As random experiment one might ask for the end position of the walk after $N$ successive steps.
Thus, the random experiment reads: 
\emph{Determine the end position $x_N$ for a symmetric $1D$ random walk, attained after 
$N$ independent steps!} 
In order to accomplish this, we might refer to the same random variable as above, where
$\Omega=\{\text{left},\text{right}\}$, $X(\text{left})=-1$, and
$X(\text{right})=1$.
A proper random variable that tells the end position of the walk after an overall
number of $N$ steps is simply $Y=\sum_{i=0}^{N-1} X^{(i)}$.
The numerical value of the end position is given by $x_N=\sum_{i=0}^{N-1} X(\omega_i)$,
wherein $\omega_i$ signifies the outcome of the $i$th random experiment (i.e.\ step).
\end{example}

The behavior of such a random variable is fully captured by the probabilities 
of observing outcomes smaller or equal to a given value $x$. 
To put the following arguments on solid ground, we need the concept of a \emph{probability
function} 
\begin{eqnarray}
P: 2^{\Omega} \rightarrow [0,1] \qquad \Big(\text{\parbox{0.3\textwidth}{$P=$ probability function  $2^\Omega=$ power set}}\Big),
\end{eqnarray}
wherein $2^{\Omega}$ specifies the set of all 
subsets of the sample space $\Omega$ (also referred to as \emph{power set}). In general,
the probability function satisfies $P(\Omega)=1$ and for two disjoint \emph{events} $A^{(1)}$ 
and $A^{(2)}$ one has $P(A^{(1)}\cup A^{(2)})=P(A^{(1)}) + P(A^{(2)})$. 
Further, if one performs a random experiment twice, and if the experiments are performed
\emph{independently}, the total probability of a particular event for the combined experiment
is the product of the single-experiment probabilities. I.e., for two events $A^{(1),(2)}$ 
one has $P(A^{(1)},A^{(2)})=P(A^{(1)})P(A^{(2)})$.

\begin{example}{Probability function}
For the sample space $\Omega=\{\text{left},\text{right}\}$, associated to the random variable $X$ 
considered in the above example on the symmetric $1D$ random walk, one has
the power set $2^\Omega=\{\emptyset,\{\text{left}\},\{\text{right}\},\{\text{left},\text{right}\}\}$.
Consequently, the probability function reads: $P(\emptyset)=0$,
$P(\{\text{left}\})=P(\{\text{right}\})=0.5$, and $P(\{\text{left},\text{right}\})=1$.
Further, if two successive steps of the symmetric $1D$ random walk are considered, the probability
for the (exemplary) combined outcome $(\text{left},\text{left})$ reads: $P(\{(\text{left},\text{left})\})=P(\{\text{left}\})P(\{\text{left}\})=0.25$.
\end{example}

By means of the probability function $P$, the (\emph{cumulative}) \emph{distribution function} $F_X$ of a random 
variable $X$ signifies a function
\begin{eqnarray}
 F_{X}: \mathbb{R} \rightarrow [0,1],~\text{where}~F_{X}=P(X\leq x)
\qquad \Big(\text{\parbox{0.35\textwidth}{$F_X=$ distribution function $P=$ probability function}}\Big).
\end{eqnarray}
The distribution is non-decreasing, implying that $F_X(x_1) \leq F_X(x_2)$ for $x_1<x_2$, and 
normalized, i.e., $\lim_{x\to -\infty}F_X(x)\to 0$ and $\lim_{x\to \infty}F_X(x)\to 1$. 
Further, it holds that $P(x_0 < X \leq x_1) = F_X(x_1)-F_X(x_0)$.

Considering the result of a sequence of random experiments, it is useful to draw 
a distinction between different types of data, to be able to choose a proper set 
of methods and tools for post-processing.
Subsequently, we will distinguish between discrete random variables (as, e.g., the $1D$ random 
walk used in the above examples) and continuous random variables.

\subsection*{Discrete probability distributions}
\label{subsect:discreteVars}

Besides the concept of the distribution function, an alternative description
of a \emph{discrete} random variable $X$ is possible by means of its 
associated \emph{probability mass function} (pmf),
\begin{eqnarray}
 p_{X}: \mathbb{R} \rightarrow [0,1],~\text{where}~p_{X}(x)=P(X= x)
\qquad \Big(\text{\parbox{0.35\textwidth}{$p_X=$ prob. mass function $P=$ probability function}}\Big).\label{pmf}
\end{eqnarray}
Related to this, note that a discrete random variable can only 
yield a countable number of outcomes (for an elementary step with unit step length 
in the $1D$ random walk problem these where just $\pm 1$) and hence, the pmf is zero almost
everywhere.
E.g., the nonzero entries of the pmf related to the random variable $X$ considered in the context of the 
symmetric $1D$ random walk are $p_X(-1)=p_X(1)=0.5$. 
Finally, the distribution function is related to the pmf via $F_X(x)=\sum_{\tilde{x}_i\leq x} p_X(\tilde{x}_i)$
(where $\tilde{x}_i$ refers to those outcomes $u$ for which $p_X(u)>0$).
\medskip

\begin{figure}
\centering
\includegraphics[width=0.95\textwidth]{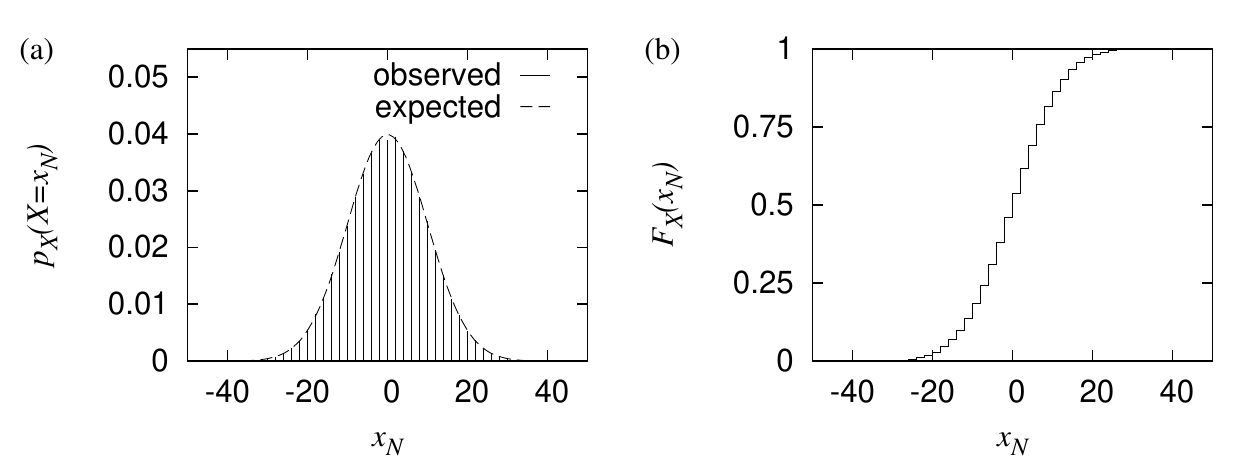}
\caption{Results for the symmetric $1D$ random walk. 
(a) approximate pmf associated to the end positions 
after $N=100$ steps (solid vertical lines) for the 
symmetric $1D$ random walk as compared to a Gaussian distribution
with zero mean and width $\sqrt{N}$ (dashed curve), 
and (b) related distribution function.
For the figure, a number of $n=10^5$ independent random 
walks where considered.
\label{fig:1DrandWalk_pmf}}
\end{figure}

\begin{example}{Monte Carlo simulation for the symmetric $1D$ random walk}
Let us fix the number of steps in an individual symmetric $1D$ random walk to 
$N=100$ and consider an ensemble of $n=10^5$ independent walks. 
Now, what does the distribution of end points $x_N$ of the 
ensemble of walks thus considered look like?

If you want to tackle that question by means of computer simulations, you 
may follow the subsequent three steps:

(i) Implement the symmetric $1D$ random walk model using your favorite programming
language. Using $\tt{python}$ \cite{pyRef}, a minimal program to simulate the above 
model (\verb!1D_randWalk.py!, see supplementary material) reads:
\SourceCodeLines{9}
\begin{Verbatim}[fontfamily=txtt,commandchars=\\\{\}]
 \PY{k+kn}{from} \PY{n+nn}{random} \PY{k+kn}{import} \PY{n}{seed}\PY{p}{,} \PY{n}{choice}
 \PY{n}{N} \PY{o}{=} \PY{l+m+mi}{100}
 \PY{n}{n} \PY{o}{=} \PY{l+m+mi}{100000}
 \PY{k}{for} \PY{n}{s} \PY{o+ow}{in} \PY{n+nb}{range}\PY{p}{(}\PY{n}{n}\PY{p}{)}\PY{p}{:}
   \PY{n}{seed}\PY{p}{(}\PY{n}{s}\PY{p}{)}	
   \PY{n}{endPos} \PY{o}{=} \PY{l+m+mi}{0}			 	
   \PY{k}{for} \PY{n}{i} \PY{o+ow}{in} \PY{n+nb}{range}\PY{p}{(}\PY{n}{N}\PY{p}{)}\PY{p}{:}
      \PY{n}{endPos} \PY{o}{+}\PY{o}{=} \PY{n}{choice}\PY{p}{(}\PY{p}{[}\PY{o}{-}\PY{l+m+mi}{1}\PY{p}{,}\PY{l+m+mi}{1}\PY{p}{]}\PY{p}{)} 
   \PY{k}{print} \PY{n}{s}\PY{p}{,}\PY{n}{endPos}
\end{Verbatim}

In line 1, the convenient \verb!random! module that implements pseudo-random number generators (PRNGs)
for different distributions is imported. The function \verb!seed()! is used to set an 
initial state for the PRNG, and \verb!choice()! returns an element (chosen uniformly at random) 
from a specified list. In lines 2 and 3, the number of steps in a single walk and the 
overall number of independent walks are specified, respectively. In lines 6--7, a 
single path is constructed and the seed as well as the resulting final position are sent to 
the standard outstream in line 9. 
It is a bit tempting to include data post-processing directly within the 
simulation code above.
However, from a point of view of data analysis you are more
flexible if you store the output of the simulation in an external file.
Then it is more easy to ``revisit'' the data, in case you want to (or are asked to) 
perform some further analyses. Hence, we might invoke the 
{\tt python} script on the command line and redirect its output as follows  
\SourceCodeLines{9}
\begin{Verbatim}
 > python 1D_randWalk.py > N100_n100000.dat
\end{Verbatim}
After the script has terminated successfully, the file \verb!N100_n100000.dat! keeps the 
simulated data available for post-processing at any time. 

(ii) An approximate pmf associated to the distribution of end points can be constructed
from the raw data contained in file \verb!N100_n100000.dat!.
We might write a small, stand-alone {\tt python} script that handles that issue. 
However, from a point of view of code-recycling and modularization it is 
more rewarding to collect all ``useful'' functions, i.e., functions that might be used
again in a different context, in a particular file that serves as some kind of 
tiny library. If we need a data post-processing script, we can then easily include
the library file and use the functions defined therein. As a further benefit, the 
final data analysis scripts will consist of a few lines only. Here, let us adopt the 
name \verb!MCS2012_lib.py! (see supplementary material) for the tiny library file 
and define two functions as listed below:
\SourceCodeLines{99}
\begin{Verbatim}[fontfamily=txtt,commandchars=\\\{\}]
 \PY{k}{def} \PY{n+nf}{fetchData}\PY{p}{(}\PY{n}{fName}\PY{p}{,}\PY{n}{col}\PY{o}{=}\PY{l+m+mi}{0}\PY{p}{,}\PY{n}{dType}\PY{o}{=}\PY{n+nb}{int}\PY{p}{)}\PY{p}{:}
 	\PY{n}{myList} \PY{o}{=} \PY{p}{[}\PY{p}{]}
 	\PY{n+nb}{file} \PY{o}{=} \PY{n+nb}{open}\PY{p}{(}\PY{n}{fName}\PY{p}{,}\PY{l+s}{"}\PY{l+s}{r}\PY{l+s}{"}\PY{p}{)}
 	\PY{k}{for} \PY{n}{line} \PY{o+ow}{in} \PY{n+nb}{file}\PY{p}{:} 
 		\PY{n}{myList}\PY{o}{.}\PY{n}{append}\PY{p}{(}\PY{n}{dType}\PY{p}{(}\PY{n}{line}\PY{o}{.}\PY{n}{split}\PY{p}{(}\PY{p}{)}\PY{p}{[}\PY{n}{col}\PY{p}{]}\PY{p}{)}\PY{p}{)}
 	\PY{n+nb}{file}\PY{o}{.}\PY{n}{close}\PY{p}{(}\PY{p}{)}
 	\PY{k}{return} \PY{n}{myList}
 
 \PY{k}{def} \PY{n+nf}{getPmf}\PY{p}{(}\PY{n}{myList}\PY{p}{)}\PY{p}{:}
 	\PY{n}{pMap} \PY{o}{=} \PY{p}{\PYZob{}}\PY{p}{\PYZcb{}}
 	\PY{n}{nInv} \PY{o}{=} \PY{l+m+mf}{1.}\PY{o}{/}\PY{n+nb}{len}\PY{p}{(}\PY{n}{myList}\PY{p}{)}
 	\PY{k}{for} \PY{n}{element} \PY{o+ow}{in} \PY{n}{myList}\PY{p}{:}
 	   \PY{k}{if} \PY{n}{element} \PY{o+ow}{not} \PY{o+ow}{in} \PY{n}{pMap}\PY{p}{:}
 	      \PY{n}{pMap}\PY{p}{[}\PY{n}{element}\PY{p}{]} \PY{o}{=} \PY{l+m+mf}{0.}
	   \PY{n}{pMap}\PY{p}{[}\PY{n}{element}\PY{p}{]} \PY{o}{+}\PY{o}{=} \PY{n}{nInv}
	\PY{k}{return} \PY{n}{pMap}
\end{Verbatim}

Lines 1--7 implement the function {\tt fetchData(fName,col=0,dType=int)}, used 
to collect data of type {\tt dType} from column number {\tt col} (default column is
0 and default data type is {\tt int}) from the file named {\tt fName}.
In function {\tt getPmf(myList)}, defined in lines 9--16, 
the integer numbers (stored in the list {\tt myList}) are used to approximate 
the underlying pmf.  

So far, we started to build the tiny library. A small data post-processing 
script (\verb!pmf.py!, see supplementary material) that uses the library in 
order to construct the pmf from the simulated data reads:
\SourceCodeLines{99}
\begin{Verbatim}[fontfamily=txtt,commandchars=\\\{\}]
 \PY{k+kn}{import} \PY{n+nn}{sys}
 \PY{k+kn}{from} \PY{n+nn}{MCS2012\PYZus{}lib} \PY{k+kn}{import} \PY{o}{*}
 
 \PY{c}{\PYZsh{}\PYZsh{} parse command line arguments}
 \PY{n}{fileName} \PY{o}{=} \PY{n}{sys}\PY{o}{.}\PY{n}{argv}\PY{p}{[}\PY{l+m+mi}{1}\PY{p}{]}
 \PY{n}{col}      \PY{o}{=} \PY{n+nb}{int}\PY{p}{(}\PY{n}{sys}\PY{o}{.}\PY{n}{argv}\PY{p}{[}\PY{l+m+mi}{2}\PY{p}{]}\PY{p}{)}
 
 \PY{c}{\PYZsh{}\PYZsh{} construct approximate pmf from data}
 \PY{n}{rawData}  \PY{o}{=} \PY{n}{fetchData}\PY{p}{(}\PY{n}{fileName}\PY{p}{,}\PY{n}{col}\PY{p}{)}
 \PY{n}{pmf}      \PY{o}{=} \PY{n}{getPmf}\PY{p}{(}\PY{n}{rawData}\PY{p}{)}
 
 \PY{c}{\PYZsh{}\PYZsh{} dump pmf/distrib func. to standard outstream}
 \PY{n}{FX} \PY{o}{=} \PY{l+m+mf}{0.}
 \PY{k}{for} \PY{n}{endpos} \PY{o+ow}{in} \PY{n+nb}{sorted}\PY{p}{(}\PY{n}{pmf}\PY{p}{)}\PY{p}{:}
	\PY{n}{FX} \PY{o}{+}\PY{o}{=} \PY{n}{pmf}\PY{p}{[}\PY{n}{endpos}\PY{p}{]}
	\PY{k}{print} \PY{n}{endpos}\PY{p}{,}\PY{n}{pmf}\PY{p}{[}\PY{n}{endpos}\PY{p}{]}\PY{p}{,}\PY{n}{FX}
\end{Verbatim}

This already illustrates lots of the {\tt python} functionality
that is needed for a decent data post-processing. In line 1, a basic 
{\tt python} module, called {\tt sys}, is imported. 
Among other things, it allows to access command line parameters stored as a list with the 
default name {\tt sys.argv}. Note that the first entry of the list is
reserved for the file name. All ``real'' command line parameters start at
the list index 1.
In line 2, all functions contained in the tiny library \verb!MCS2012_lib.py! are 
imported and are available for data post-processing by means of their genuine 
name (no \verb!MCS2012_lib.! statement has to precede a functions name).
In lines 14--16, the approximate pmf as well as the related distribution function
are sent to the standard outstream (for a comment on the built-in function \verb!sorted()!,
see paragraph ``Basic sorting'' in Section \ref{sect:basicPython}).

To cut a long story short, the approximate pmf 
for the end positions of the symmetric $1D$ random walks stored in the file 
\verb!N100_n100000.dat! can be obtained by calling the script on the command line via  
\SourceCodeLines{9}
\begin{Verbatim}
 > python pmf.py N100_n100000.dat 1 > N100_n100000.pmf
\end{Verbatim}
Therein, the digit $1$ indicates the column of the supplied file where the end position 
of the walks is stored, and the approximate pmf and the related distribution function 
are redirected to the file \verb!N100_n100000.pmf!. 

(iii) On the basis of analytical theory, one can expect that the enclosing curve
of the pmf is well represented by a Gaussian distribution with 
mean $0$ and width $\sqrt{N}$. 
However, we need to rescale the approximate pmf, i.e., the observed probabilities, 
by a factor of $2$ if we want to compare it to the \emph{expected} 
probabilities given by the Gaussian distribution. 
Therein, the factor $2$ reflects the fact that if we consider walks with an even (or odd)
number of steps only, the \emph{effective} length-scale that characterizes the distance 
between two neighboring positions is $2$.
A less handwaving way to arrive at that conclusion is to start with the proper
end point distribution for a symmetric $1D$ random walk, given by a (discrete) symmetric 
binomial distribution (see Section \ref{sect:chisquare}), and to approximate it, using Stirlings expansion, by a 
(continuous) distribution. The factor $2$ is then immediate.
Using the convenient {\tt gnuplot} plotting program \cite{gnuplotRef}, you can visually inspect the 
difference between the observed and expected probabilities by 
creating a file, e.g.\ called \verb!endpointDistrib.gp! (see supplementary material), with content:
\SourceCodeLines{99}
\begin{Verbatim}[fontfamily=txtt,commandchars=\\\{\}]
 \PY{k}{set} \PY{n+nb}{multiplot}

 \PY{c}{\PYZsh{}}\PY{c}{\PYZsh{}}\PY{c}{ }\PY{c}{p}\PY{c}{r}\PY{c}{o}\PY{c}{b}\PY{c}{a}\PY{c}{b}\PY{c}{i}\PY{c}{l}\PY{c}{i}\PY{c}{t}\PY{c}{y}\PY{c}{ }\PY{c}{m}\PY{c}{a}\PY{c}{s}\PY{c}{s}\PY{c}{ }\PY{c}{f}\PY{c}{u}\PY{c}{n}\PY{c}{c}\PY{c}{t}\PY{c}{i}\PY{c}{o}\PY{c}{n}
  \PY{k}{set} \PY{n+nb}{origin} \PY{l+m+mf}{0.}\PY{o}{,}\PY{l+m+mf}{0.} 
  \PY{k}{set} \PY{n+nb}{size} \PY{l+m+mf}{0.5}\PY{o}{,}\PY{l+m+mf}{0.65}
  \PY{k}{set} \PY{n+nb}{key} \PY{n}{samplen} \PY{l+m+mf}{1.}
  \PY{k}{set} \PY{n+nb}{yr} \PY{p}{[}\PY{o}{:}\PY{l+m+mf}{0.05}\PY{p}{]}\PY{p}{;} \PY{k}{set} \PY{n+nb}{ytics} \PY{p}{(}\PY{l+m+mf}{0.00}\PY{o}{,}\PY{l+m+mf}{0.02}\PY{o}{,}\PY{l+m+mf}{0.04}\PY{p}{)}
  \PY{k}{set} \PY{n+nb}{xr} \PY{p}{[}\PY{l+m+mi}{-50}\PY{o}{:}\PY{l+m+mi}{50}\PY{p}{]}
  \PY{k}{set} \PY{n+nb}{xlabel} \PY{l+s}{"}\PY{l+s}{x\PYZus{}N}\PY{l+s}{"} \PY{n}{font} \PY{l+s}{"}\PY{l+s}{Times-Italic}\PY{l+s}{"}
  \PY{k}{set} \PY{n+nb}{ylabel} \PY{l+s}{"}\PY{l+s}{p\PYZus{}X(X=x\PYZus{}N)}\PY{l+s}{"} \PY{n}{font} \PY{l+s}{"}\PY{l+s}{Times-Italic}\PY{l+s}{"}

  \PY{c}{\PYZsh{}}\PY{c}{\PYZsh{}}\PY{c}{ }\PY{c}{e}\PY{c}{x}\PY{c}{p}\PY{c}{e}\PY{c}{c}\PY{c}{t}\PY{c}{e}\PY{c}{d}\PY{c}{ }\PY{c}{p}\PY{c}{r}\PY{c}{o}\PY{c}{b}\PY{c}{a}\PY{c}{b}\PY{c}{i}\PY{c}{l}\PY{c}{i}\PY{c}{t}\PY{c}{y}
  \PY{k}{f}\PY{p}{(}\PY{n}{x}\PY{p}{)}\PY{o}{=}\PY{n+nf}{exp}\PY{p}{(}\PY{o}{-}\PY{p}{(}\PY{n}{x}\PY{o}{-}\PY{n}{mu}\PY{p}{)}\PY{o}{*}\PY{o}{*}\PY{l+m+mi}{2}\PY{o}{/}\PY{p}{(}\PY{l+m+mi}{2}\PY{o}{*}\PY{n+nb}{s}\PY{o}{*}\PY{n+nb}{s}\PY{p}{)}\PY{p}{)}\PY{o}{/}\PY{p}{(}\PY{n+nb}{s}\PY{o}{*}\PY{n+nf}{sqrt}\PY{p}{(}\PY{l+m+mi}{2}\PY{o}{*}\PY{n}{pi}\PY{p}{)}\PY{p}{)}
  \PY{n+nv}{mu}\PY{o}{=}\PY{l+m+mi}{0}\PY{p}{;} \PY{n+nv}{s}\PY{o}{=}\PY{l+m+mi}{10}

  \PY{k}{p} \PY{l+s}{"}\PY{l+s}{N100\PYZus{}n100000.pmf}\PY{l+s}{"} \PY{n+nb}{u} \PY{l+m+mi}{1}\PY{o}{:}\PY{p}{(}\PY{err}{\PYZdl{}}\PY{l+m+mi}{2}\PY{o}{/}\PY{l+m+mi}{2}\PY{p}{)} \PY{n+nb}{w} \PY{n}{impulses} \PY{n+nb}{t} \PY{l+s}{"}\PY{l+s}{observed}\PY{l+s}{"}\PYZbs{}
  \PY{o}{,} \PY{n+nf}{f}\PY{p}{(}\PY{n}{x}\PY{p}{)} \PY{n+nb}{t} \PY{l+s}{"}\PY{l+s}{expected}\PY{l+s}{"}

 \PY{c}{\PYZsh{}}\PY{c}{\PYZsh{}}\PY{c}{ }\PY{c}{d}\PY{c}{i}\PY{c}{s}\PY{c}{t}\PY{c}{r}\PY{c}{i}\PY{c}{b}\PY{c}{u}\PY{c}{t}\PY{c}{i}\PY{c}{o}\PY{c}{n}\PY{c}{ }\PY{c}{f}\PY{c}{u}\PY{c}{n}\PY{c}{c}\PY{c}{t}\PY{c}{i}\PY{c}{o}\PY{c}{n}
  \PY{k}{set} \PY{n+nb}{origin} \PY{l+m+mf}{0.5}\PY{o}{,}\PY{l+m+mf}{0.}
  \PY{k}{set} \PY{n+nb}{size} \PY{l+m+mf}{0.5}\PY{o}{,}\PY{l+m+mf}{0.65}
  \PY{k}{set} \PY{n+nb}{yr} \PY{p}{[}\PY{l+m+mi}{0}\PY{o}{:}\PY{l+m+mi}{1}\PY{p}{]}\PY{p}{;} \PY{k}{set} \PY{n+nb}{ytics} \PY{p}{(}\PY{l+m+mf}{0.}\PY{o}{,}\PY{l+m+mf}{0.25}\PY{o}{,}\PY{l+m+mf}{0.5}\PY{o}{,}\PY{l+m+mf}{0.75}\PY{o}{,}\PY{l+m+mf}{1.}\PY{p}{)}
  \PY{k}{set} \PY{n+nb}{xr} \PY{p}{[}\PY{l+m+mi}{-50}\PY{o}{:}\PY{l+m+mi}{50}\PY{p}{]}
  \PY{k}{set} \PY{n+nb}{ylabel} \PY{l+s}{"}\PY{l+s}{F\PYZus{}X(x\PYZus{}N)}\PY{l+s}{"} \PY{n}{font} \PY{l+s}{"}\PY{l+s}{Times-Italic}\PY{l+s}{"}

  \PY{k}{p} \PY{l+s}{"}\PY{l+s}{N100\PYZus{}n100000.pmf}\PY{l+s}{"} \PY{n+nb}{u} \PY{l+m+mi}{1}\PY{o}{:}\PY{p}{(}\PY{err}{\PYZdl{}}\PY{l+m+mi}{3}\PY{p}{)} \PY{n+nb}{w} \PY{n}{steps} \PY{n+nb}{notitle}

 \PY{k}{unset} \PY{n+nb}{multiplot}
\end{Verbatim}

Calling the file via \verb!gnuplot -persist endpointDistrib.gp!,
the output should look similar to Fig.\ \ref{fig:1DrandWalk_pmf}.
\end{example}


\subsection*{Continuous probability distributions}
\label{subsect:continuousVars}

Given a continuous distribution function $F_X$, the density of a \emph{continuous} 
random variable $X$, referred to as \emph{probability density function} 
(pdf), reads
\begin{eqnarray}
 p_{X}: \mathbb{R} \rightarrow [0,1],~\text{where}~p_{X}(x)=\frac{dF_X(x)}{dx}
\qquad \Big(\text{\parbox{0.3\textwidth}{$p_X=$ prob. density function $F_X=$ distribution functin}}\Big). \label{eq:pdf}
\end{eqnarray}
The pdf is strictly nonnegative, and, as should be clear from the 
definition, the probability that $X$ falls within a certain interval, 
say $x \to x+\Delta x$, is given by the integral of the pdf $p_X(x)$ 
over that interval, i.e.\
$P(x < X \leq x+\Delta x) = \int_{x}^{x+\Delta x}~p_X(u)~du$.
Since the pdf is normalized, one further has $1=\int_{-\infty}^{\infty} p_X(u)~du$.

\begin{figure}
\centering
\includegraphics[width=0.95\textwidth]{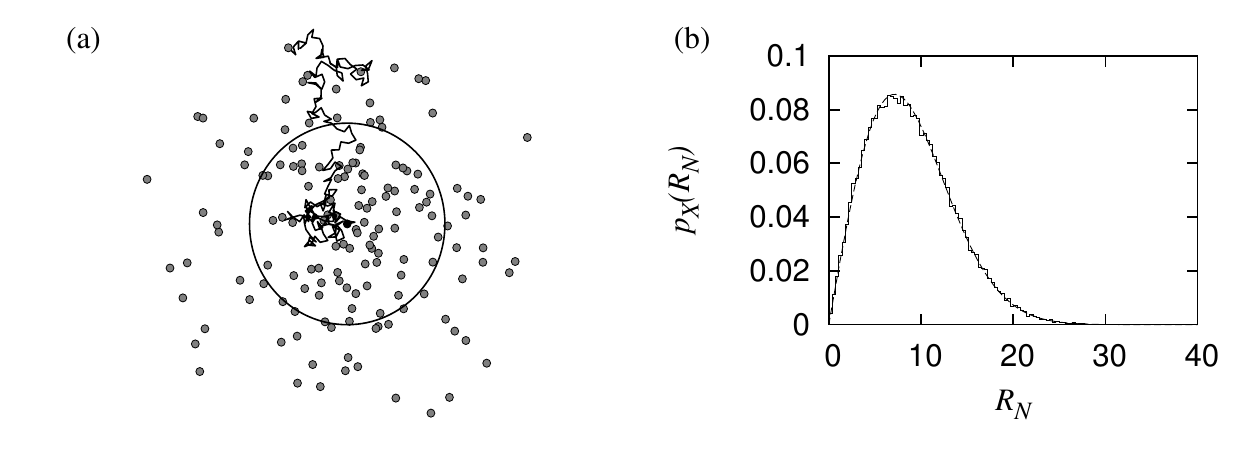}
\caption{
Results for the $2D$ random walk. 
(a) Snapshot of an ensemble of $n=150$ independent walkers 
(grey dots) after $N=200$ steps. The trail of a single walker 
is indicated by the zigzag line. The origin of that line also 
indicates the common starting point of all walkers. 
Further, the circle denotes the average distance 
$\langle R_N \rangle$ of the walks to the starting point after 
$N=200$ steps. 
(b) Approximate pdf for the distance to the starting point (using $100$ 
bins indicated by the solid step-curve, see Section \ref{sect:hist} 
on histograms below), estimated from $n=10^5$ walks after $N=100$ steps,
compared to the functional form (dashed line) explained in the text.
\label{fig:2DrandWalk_pdf}}
\end{figure}

\begin{example}{The continuous $2D$ random walk}

The continuous $2D$ random walk (see Fig.\ \ref{fig:2DrandWalk_pdf}) is an example for a random 
walk with fixed step-length (for simplicity assume unit
step length), where the direction of the consecutive steps 
is drawn uniformly at random.
As continuous random variable $X$ we may choose the distance
of the walk to its starting point after a number of $N$ 
steps, referred to as $R_N$.
To implement this random walk model, we can recycle
the {\tt python} script for the $1D$ random walk, written
earlier. A proper code (\verb!2d_randWalk.py!, see supplementary material) to simulate 
the model is listed below:
\SourceCodeLines{99}
\begin{Verbatim}[fontfamily=txtt,commandchars=\\\{\}]
 \PY{k+kn}{from} \PY{n+nn}{random} \PY{k+kn}{import} \PY{n}{seed}\PY{p}{,} \PY{n}{random}
 \PY{k+kn}{from} \PY{n+nn}{math} \PY{k+kn}{import} \PY{n}{sqrt}\PY{p}{,} \PY{n}{cos}\PY{p}{,} \PY{n}{sin}\PY{p}{,} \PY{n}{pi}
 
 \PY{n}{N} \PY{o}{=} \PY{l+m+mi}{100}
 \PY{n}{n} \PY{o}{=} \PY{l+m+mi}{100000}
 \PY{k}{for} \PY{n}{s} \PY{o+ow}{in} \PY{n+nb}{range}\PY{p}{(}\PY{n}{n}\PY{p}{)}\PY{p}{:}	
   \PY{n}{seed}\PY{p}{(}\PY{n}{s}\PY{p}{)}
   \PY{n}{x} \PY{o}{=} \PY{n}{y} \PY{o}{=} \PY{l+m+mf}{0.}
   \PY{k}{for} \PY{n}{i} \PY{o+ow}{in} \PY{n+nb}{range}\PY{p}{(}\PY{n}{N}\PY{p}{)}\PY{p}{:}
      \PY{n}{phi} \PY{o}{=} \PY{n}{random}\PY{p}{(}\PY{p}{)}\PY{o}{*}\PY{l+m+mi}{2}\PY{o}{*}\PY{n}{pi}
      \PY{n}{x}  \PY{o}{+}\PY{o}{=} \PY{n}{cos}\PY{p}{(}\PY{n}{phi}\PY{p}{)}
      \PY{n}{y}  \PY{o}{+}\PY{o}{=} \PY{n}{sin}\PY{p}{(}\PY{n}{phi}\PY{p}{)}
   \PY{k}{print} \PY{n}{s}\PY{p}{,}\PY{n}{sqrt}\PY{p}{(}\PY{n}{x}\PY{o}{*}\PY{n}{x}\PY{o}{+}\PY{n}{y}\PY{o}{*}\PY{n}{y}\PY{p}{)}
\end{Verbatim}

Therein, in line 10, the direction of the upcoming step is drawn 
uniformly at random, and, in lines 11--12, the two independent 
walk coordinates are updated. Note that in line 2, the basic 
mathematical functions and constants are imported. Finally, 
in line 13, the distance $R_N$ for a single walk is sent to the 
standard output.
Invoking the script on the command line and redirecting its output 
to the file \verb!2dRW_N100_n100000.dat!, we can further obtain 
an approximation to the underlying pdf by constructing a normalized 
histogram of the distances $R_N$ by means of the script \verb!hist.py!
(see supplementary material) as follows:
\SourceCodeLines{9}
\begin{Verbatim}
 > python hist.py 2dRW_N100_n100000.dat 1 100 \
 > 				> 2dRW_N100_n100000.pdf
\end{Verbatim}
The two latter numbers signify the column of the file, where the relevant
data is stored, and the desired number of bins, respectively.
At this point, let us just use the script \verb!hist.py! as a ``black-box'' and
postpone the discussion of histograms until Section \ref{sect:hist}.
After the script has terminated successfully, the file 
\verb!2dRW_N100_n100000.pdf! contains the normalized histogram. 
For values of $N$ large enough, one can expect that 
$R_N$ is properly characterized by the Rayleigh distribution 
\begin{eqnarray}
p_N(R_N) = \frac{R_N}{\sigma^2} \exp\{-R_N^2 / (2 \sigma^2)\}~,\label{eq:Rayleigh}
\end{eqnarray}
where $\sigma^2=(2n)^{-1}\sum_{i=0}^{n-1}R_{N,i}^2$. Therein, the 
values $R_{N,i}$, $i=0\ldots n-1$, comprise the sample of observed
distances. 
In order to compute $\sigma$ for the sample of observed distances, 
one could use a cryptic {\tt python} one-liner. 
However, it is more readable to accomplish that task in the following way 
(\verb!sigmaRay.py!, see supplementary material):
\SourceCodeLines{9}
\begin{Verbatim}[fontfamily=txtt,commandchars=\\\{\}]
 \PY{k+kn}{import} \PY{n+nn}{sys}\PY{o}{,}\PY{n+nn}{math}
 \PY{k+kn}{from} \PY{n+nn}{MCS2012\PYZus{}lib} \PY{k+kn}{import} \PY{n}{fetchData} 
 
 \PY{n}{rawData} \PY{o}{=} \PY{n}{fetchData}\PY{p}{(}\PY{n}{sys}\PY{o}{.}\PY{n}{argv}\PY{p}{[}\PY{l+m+mi}{1}\PY{p}{]}\PY{p}{,}\PY{l+m+mi}{1}\PY{p}{,}\PY{n+nb}{float}\PY{p}{)}
 
 \PY{n}{sum2}\PY{o}{=}\PY{l+m+mf}{0.}
 \PY{k}{for} \PY{n}{val} \PY{o+ow}{in} \PY{n}{rawData}\PY{p}{:} \PY{n}{sum2}\PY{o}{+}\PY{o}{=}\PY{n}{val}\PY{o}{*}\PY{n}{val}
 \PY{k}{print} \PY{l+s}{"}\PY{l+s}{sigma=}\PY{l+s}{"}\PY{p}{,}\PY{n}{math}\PY{o}{.}\PY{n}{sqrt}\PY{p}{(}\PY{n}{sum2}\PY{o}{/}\PY{p}{(}\PY{l+m+mf}{2.}\PY{o}{*}\PY{n+nb}{len}\PY{p}{(}\PY{n}{rawData}\PY{p}{)}\PY{p}{)}\PY{p}{)}
\end{Verbatim}

Invoking the script on the command line yields:
\SourceCodeLines{9}
\begin{Verbatim}
 > python sigmaRay.py 2dRW_N100_n100000.dat
 sigma= 7.09139394259
\end{Verbatim}
Finally, an approximate pdf of the distance to the starting point of the walkers,
i.e., a histogram using 100 bins, as well as the Rayleigh probability distribution function 
with $\sigma=7.091$ is shown in Fig.\ \ref{fig:2DrandWalk_pdf}(b).
\end{example}

\subsection*{Basic parameters related to random variables}
\label{subsect:moments}

To characterize the gross features of a distribution function, it is useful 
to consider statistical measures that are related to its \emph{moments}.
Again, drawing a distinction between discrete and continuous random variables,
the $k$th moment of the distribution function, related to some random variable $X$,
reads
\begin{eqnarray}
 E[X^k] = \begin{cases}
\sum_i x_i^k~ p_X(x_i),& \text{for $X$ \emph{discrete}, and $p_X=$ pmf,}\\
\int_{-\infty}^{\infty} x^k~ p_X(x)~dx, & \text{for $X$ \emph{continuous}, and $p_X=$ pdf.}
\end{cases}
\end{eqnarray}
Therein, $E[\cdot]$ denotes the \emph{expectation operator}.

Usually, numerical simulations result in a large but finite amount of data and 
the random variables considered therein yield a sequence of effectively discrete
numerical values. 
As a consequence, from a practical point of view, the pdf underlying the data is 
not known precisely.
Bearing this in mind, we proceed by estimating the gross
statistical features of a \emph{sample} of $N$ statistically independent numerical values 
$x=\{x_0,\ldots,x_{N-1}\}$, i.e., the associated random variables are 
assumed to be independent and identically distributed. 
For simplicity, we may assume the uniform pmf $p_X(x_i)=1/N$ for $i=0\ldots N-1$.
Commonly, $N$ is referred to as \emph{sample size}.
The \emph{average} or \emph{mean} value of the sample is then properly estimated by the finite sum
\begin{eqnarray}
{\rm av}(x) = \frac{1}{N} \sum_{i=0}^{N-1} x_i ~, \qquad \text{(average or mean value)}, \label{eq:mean}
\end{eqnarray}
which simply equates the finite sequence to the first moment of 
the (effectively unknown) underlying distribution.
Note that for values that stem from a probability distribution that decreases slowly as
$x\to\infty$ (aka having a broad tail), the convergence properties of the sum 
might turn out to be very poor for increasing $N$.
Further, one might be interested in the spread of the values
within the sample. A convenient measure related to that is the 
\emph{variance}, defined by
\begin{eqnarray}
{\rm Var}(x) = \frac{1}{N-1} \sum_{i=0}^{N-1} [x_i -{\rm av}(x)]^2 ~. \qquad \text{(variance)}. \label{eq:Var}
\end{eqnarray}
Essentially, the variance measures the mean squared deviation of the values contained in the 
sequence, relative to the mean value. Usually, the mean is not known \emph{a priori} and has to 
be estimated from the data beforehand, e.g.\ by using Eq.\ (\ref{eq:mean}). 
This reduces the number of independent terms of the sum by one 
and leads to the prefactor of $N-1$. To be more precise, Eq.\ (\ref{eq:Var}) defines 
the \emph{corrected} variance, as opposed to the \emph{uncorrected} variance 
${\rm uVar}(x)=(N-1)/N \times {\rm Var}(x)$. The latter one can also be written 
as ${\rm uVar}(x)={\rm av}( [x - {\rm av}(x)]^2 )$.
While the corrected variance is an unbiased estimator for the spread of the values within 
the sequence, the uncorrected variance is biased (see discussion below). 
A note on implementation: in order to improve on a naive implementation, and so to
reduce the round-off error in Eq.\ (\ref{eq:Var}), the so-called ``corrected two-pass 
algorithm'' (see Ref.\ \cite{nrc1992}) might be used.
The square root of the variance yields the \emph{standard deviation} 
\begin{eqnarray}
{\rm sDev}(x) = \sqrt{{\rm Var}(x)} ~. \qquad \text{(standard deviation)} \label{eq:sDev}
\end{eqnarray}
At this point, note that the variance and standard deviation depend on the second
moment of the underlying distribution.
Again, the subtleties of an underlying distribution with a broad tail might lead to 
a non-converging variance or standard deviation as $N\to\infty$ (see example below).
For a finite sequence of values, the standard error of the mean, referred to as {\rm sErr}, is 
also a measure of great treasure. Assuming that the values in the sample are statistically 
independent, it is related to the standard deviation by means of
\begin{eqnarray}
{\rm sErr}(x)=\frac{{\rm sDev}(x)}{\sqrt{N}} ~. \qquad \text{(standard error).} \label{eq:sErr}
\end{eqnarray}
For a finite sample size $N$, the standard error is of interest since it gives an idea of 
how accurate the sample mean approximates the true mean (attained in the limit $N\to\infty$).

\begin{example}{Basic statistics}

We may amend the tiny library \verb!MCS2012_lib.py! by implementing the function 
\verb!basicStatistics()! as listed below:
\SourceCodeLines{99}
\begin{Verbatim}[fontfamily=txtt,commandchars=\\\{\}]
 \PY{k}{def} \PY{n+nf}{basicStatistics}\PY{p}{(}\PY{n}{myList}\PY{p}{)}\PY{p}{:}
	\PY{n}{av} \PY{o}{=} \PY{n}{var} \PY{o}{=} \PY{n}{tiny} \PY{o}{=} \PY{l+m+mf}{0.}
	\PY{n}{N} \PY{o}{=} \PY{n+nb}{len}\PY{p}{(}\PY{n}{myList}\PY{p}{)} 
	\PY{k}{for} \PY{n}{el} \PY{o+ow}{in} \PY{n}{myList}\PY{p}{:} 
		\PY{n}{av} \PY{o}{+}\PY{o}{=} \PY{n}{el}
	\PY{n}{av} \PY{o}{/}\PY{o}{=} \PY{n}{N}
	\PY{k}{for} \PY{n}{el} \PY{o+ow}{in} \PY{n}{myList}\PY{p}{:}
		\PY{n}{dum}   \PY{o}{=} \PY{n}{el} \PY{o}{-} \PY{n}{av}
		\PY{n}{tiny} \PY{o}{+}\PY{o}{=} \PY{n}{dum}
		\PY{n}{var}  \PY{o}{+}\PY{o}{=} \PY{n}{dum}\PY{o}{*}\PY{n}{dum}
	\PY{n}{var}  \PY{o}{=} \PY{p}{(}\PY{n}{var} \PY{o}{-} \PY{n}{tiny}\PY{o}{*}\PY{n}{tiny}\PY{o}{/}\PY{n}{N}\PY{p}{)}\PY{o}{/}\PY{p}{(}\PY{n}{N}\PY{o}{-}\PY{l+m+mi}{1}\PY{p}{)}
	\PY{n}{sDev} \PY{o}{=} \PY{n}{sqrt}\PY{p}{(}\PY{n}{var}\PY{p}{)}
	\PY{n}{sErr} \PY{o}{=} \PY{n}{sDev}\PY{o}{/}\PY{n}{sqrt}\PY{p}{(}\PY{n}{N}\PY{p}{)}
	\PY{k}{return} \PY{n}{av}\PY{p}{,} \PY{n}{sDev}\PY{p}{,} \PY{n}{sErr}
\end{Verbatim}

Therein, the variance of the numerical values contained in the list \verb!myList! is computed 
by means of the corrected two-pass algorithm. To facilitate the computation of the 
standard deviation, we further need to import the square root function, available in the \verb!math!
module, by adding the line \verb!from math import sqrt! at the very beginning of the file. 

\paragraph{Good convergence (Gaussian distributed data):}
In a preceding example we gained the intuition that the pmf of the end points
for a symmetric $1D$ random walk, involving a number of $N$ steps, can be expected 
to compare well to a Gaussian distribution with mean $\langle x_N \rangle=0$ ($={\tt av}(x)$ as $n\to\infty$; $n=$sample size) and width
$\sigma=\sqrt{N}$ ($={\tt sDev}(x)$ as $n\to\infty$).
In order to compute averages of data stored in some file, we may write 
a small data analysis script (\verb!basicStats.py!, see supplementary material) 
as listed below:
\SourceCodeLines{99}
\begin{Verbatim}[fontfamily=txtt,commandchars=\\\{\}]
 \PY{k+kn}{import} \PY{n+nn}{sys}
 \PY{k+kn}{from} \PY{n+nn}{MCS2012\PYZus{}lib} \PY{k+kn}{import} \PY{o}{*}

 \PY{c}{\PYZsh{}\PYZsh{} parse command line arguments}
 \PY{n}{fileName}     \PY{o}{=} \PY{n}{sys}\PY{o}{.}\PY{n}{argv}\PY{p}{[}\PY{l+m+mi}{1}\PY{p}{]}
 \PY{n}{col}          \PY{o}{=} \PY{n+nb}{int}\PY{p}{(}\PY{n}{sys}\PY{o}{.}\PY{n}{argv}\PY{p}{[}\PY{l+m+mi}{2}\PY{p}{]}\PY{p}{)}

 \PY{c}{\PYZsh{}\PYZsh{} construct approximate pmf from data}
 \PY{n}{rawData}      \PY{o}{=} \PY{n}{fetchData}\PY{p}{(}\PY{n}{fileName}\PY{p}{,}\PY{n}{col}\PY{p}{)}
 \PY{n}{av}\PY{p}{,}\PY{n}{sDev}\PY{p}{,}\PY{n}{sErr} \PY{o}{=} \PY{n}{basicStatistics}\PY{p}{(}\PY{n}{rawData}\PY{p}{)}

 \PY{k}{print} \PY{l+s}{"}\PY{l+s}{av   = }\PY{l+s+si}{\PYZpc{}4.3lf}\PY{l+s}{"}\PY{o}{\PYZpc{}}\PY{n}{av}
 \PY{k}{print} \PY{l+s}{"}\PY{l+s}{sErr = }\PY{l+s+si}{\PYZpc{}4.3lf}\PY{l+s}{"}\PY{o}{\PYZpc{}}\PY{n}{sErr}
 \PY{k}{print} \PY{l+s}{"}\PY{l+s}{sDev = }\PY{l+s+si}{\PYZpc{}4.3lf}\PY{l+s}{"}\PY{o}{\PYZpc{}}\PY{n}{sDev}
\end{Verbatim}

Basically, the script imports all the functions defined in the 
tiny library \verb!MCS2012_lib.py! (line 2), reads the data 
stored in a particular column of a supplied file (line 9) and 
computes the mean, standard deviation and error associated to 
the sequence of numerical values (line 10).
If we invoke the script for the data accumulated for the symmetric $1D$ random walk 
(bear in mind to indicate the column number where the data can be found), it yields:
\SourceCodeLines{9}
\begin{Verbatim}
 > python basicStats.py N100_n100000.dat 1
 av   = 0.008
 sErr = 0.032
 sDev = 10.022
\end{Verbatim}
Apparently, the results of the Monte Carlo simulation are in agreement with the
expectation.

\paragraph{Poor convergence (power-law distributed data):}
The probability density function $p_X(x)$ for power-law distributed 
continuous real variables $x$ can be written as
\begin{eqnarray}
p_X(x)=\frac{\alpha-1}{x_0}\Big(\frac{x}{x_0}\Big)^{-\alpha}\sim x^{-\alpha},\qquad \text{(power-law pdf),} \label{eq:powerLaw}
\end{eqnarray}
where, in order for the pdf to be normalizable, we might assume $\alpha>1$, 
and where $x_0$ shall denote the smallest 
$x$-value for which the power-law behavior holds.
Pseudo random numbers, drawn according to this pdf, can be obtained via the 
\emph{inversion method}. For that, one draws a random real number $r$ 
uniformly in $[0,1)$ and generates a power-law distributed random number $x$ as
$x=x_0(1-r)^{-1/(\alpha-1)}$, where $x_0\leq x <\infty$.
A small, stand-alone {\tt python} script (\verb!poorConvergence.py!; 
see supplementary material) that implements the inversion method to 
obtain a sequence of $N=10^5$ random real numbers drawn from the power-law
pdf with $\alpha=2.2$ and $x_0=1$ is listed
below:

\SourceCodeLines{99}
\begin{Verbatim}[fontfamily=txtt,commandchars=\\\{\}]
 \PY{k+kn}{from} \PY{n+nn}{random} \PY{k+kn}{import} \PY{n}{seed}\PY{p}{,} \PY{n}{random}
 \PY{k+kn}{from} \PY{n+nn}{MCS2012\PYZus{}lib} \PY{k+kn}{import} \PY{n}{basicStatistics}
 
 \PY{c}{\PYZsh{}\PYZsh{} inversion method to obtain power law}
 \PY{c}{\PYZsh{}\PYZsh{} distributed pseudo random numbers}
 \PY{n}{N} \PY{o}{=} \PY{l+m+mi}{100000}
 \PY{n}{seed}\PY{p}{(}\PY{l+m+mi}{0}\PY{p}{)}
 \PY{n}{alpha} \PY{o}{=} \PY{l+m+mf}{2.2}
 \PY{n}{myList} \PY{o}{=} \PY{p}{[}\PY{p}{]}
 \PY{k}{for} \PY{n}{i} \PY{o+ow}{in} \PY{n+nb}{range}\PY{p}{(}\PY{n}{N}\PY{p}{)}\PY{p}{:}
	\PY{n}{r} \PY{o}{=} \PY{n+nb}{pow}\PY{p}{(}\PY{l+m+mf}{1.}\PY{o}{-}\PY{n}{random}\PY{p}{(}\PY{p}{)} \PY{p}{,} \PY{o}{-}\PY{l+m+mf}{1.}\PY{o}{/}\PY{p}{(}\PY{n}{alpha}\PY{o}{-}\PY{l+m+mf}{1.}\PY{p}{)}  \PY{p}{)}
	\PY{n}{myList}\PY{o}{.}\PY{n}{append}\PY{p}{(}\PY{n}{r}\PY{p}{)}
 
 \PY{c}{\PYZsh{}\PYZsh{} basic statistics for different sample sizes to}
 \PY{c}{\PYZsh{}\PYZsh{} assess convergence properties for av and sDev}
 \PY{n}{M} \PY{o}{=} \PY{l+m+mi}{100}\PY{p}{;} \PY{n}{dN} \PY{o}{=} \PY{n}{N}\PY{o}{/}\PY{n}{M}
 \PY{k}{for} \PY{n}{Nmax} \PY{o+ow}{in} \PY{p}{[}\PY{n}{dN}\PY{o}{+}\PY{n}{i}\PY{o}{*}\PY{n}{dN} \PY{k}{for} \PY{n}{i} \PY{o+ow}{in} \PY{n+nb}{range}\PY{p}{(}\PY{n}{M}\PY{p}{)}\PY{p}{]}\PY{p}{:}
	\PY{n}{av}\PY{p}{,}\PY{n}{sDev}\PY{p}{,}\PY{n}{sErr} \PY{o}{=} \PY{n}{basicStatistics}\PY{p}{(}\PY{n}{myList}\PY{p}{[}\PY{p}{:}\PY{n}{Nmax}\PY{p}{]}\PY{p}{)}
	\PY{k}{print} \PY{n}{Nmax}\PY{p}{,} \PY{n}{av}\PY{p}{,} \PY{n}{sErr}\PY{p}{,} \PY{n}{sDev}
\end{Verbatim}

\begin{figure}
\includegraphics[width=0.95\textwidth]{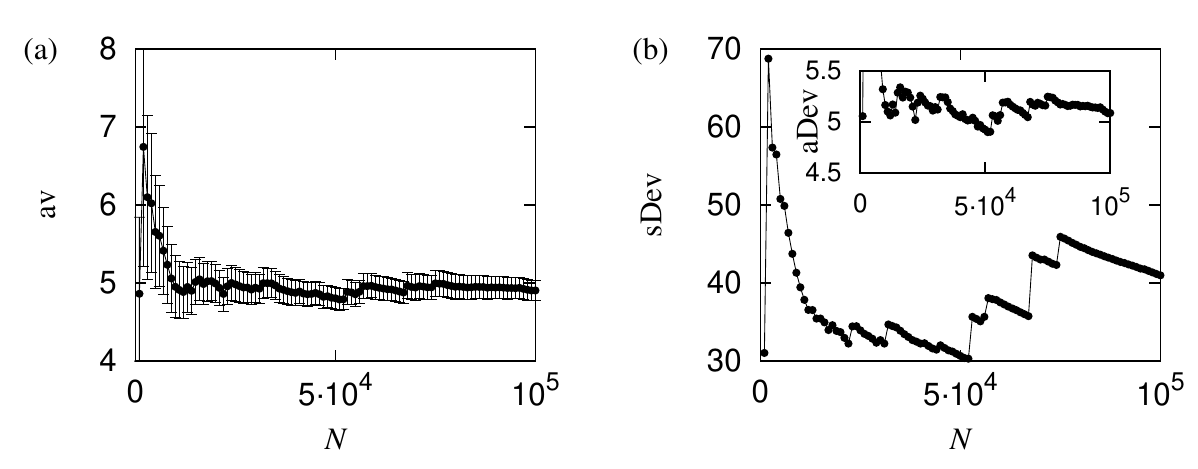}
\caption{Results for estimating (a) the mean (error bars 
denote the standard error), and (b) the standard deviation, 
for a sequence of $N=10^3\ldots 10^5$ random numbers (in 
steps of $\Delta N = 1000$) drawn from a power-law pdf with 
$\alpha=2.2$ and $x_0=1$. Apparently, the mean converges well and 
the standard deviation exhibits rather poor convergence 
properties.
However, the absolute deviation shown in the inset of (b) represents
a more robust estimation of the deviation within the data set.
\label{fig:poorConvergence}}
\end{figure}

In line 2, the function \verb!basicStatistics()!, as
defined in \verb!MCS2012_lib.py!, is imported and
available for data post-processing. The
inversion method to obtain random numbers according to 
Eq.\ (\ref{eq:powerLaw}) is implemented in lines 10 and 11.
The resulting numbers are stored in a list which is 
subsequently used to estimate the mean, standard error 
of the mean and standard deviation for a number of 
$N=10^3\ldots10^5$ (in steps of $\Delta N=1000$) samples (lines 16--19).
As a result, the convergence properties of the mean and 
standard error are shown in Fig.\ \ref{fig:poorConvergence}.
Apparently, the estimate for the mean converges quite well 
(see Fig.\ \ref{fig:poorConvergence}(a)), while
the standard deviation exhibits poor convergence properties
(see main plot of Fig.\ \ref{fig:poorConvergence}(b)).
In such a case, the standard deviation is said to be \emph{not robust}.
On second thought, this is intuitive, since for the 
moments of a power-law distribution it holds that
\begin{eqnarray}
\langle x^k \rangle = \frac{\alpha-1}{x_0^{-\alpha+1}} \int_{x_0}^{\infty} x^{-\alpha+k}~dx=\frac{\alpha-1}{x_0^{-\alpha+1}} \Big[ \frac{x^{-\alpha+k+1}}{(-\alpha+k+1)} \Big]_{x_0}^{\infty}~,
\end{eqnarray}
and thus one arrives at the conclusion that $\langle x^k\rangle$ is finite only if $k+1<\alpha$ while all higher moments diverge.
In particular, for $\alpha=2.2$, the second moment of the distribution, needed to compute the variance, does not converge as the sample size increases.
Note that a more \emph{robust} estimate of the deviations within the 
data set is provided by the \emph{absolute deviation}
\begin{eqnarray}
{\rm aDev}(x) = \frac{1}{N} \sum_{i=0}^{N-1} | x_i - {\rm av}(x)| ~, \qquad \text{(absolute deviation),} \label{eq:aDev}
\end{eqnarray}
shown in the inset of Fig.\ \ref{fig:poorConvergence}(b).
\end{example}

\subsection*{Estimators with(out) bias}
\label{subsect:bias}

In principle, there is no strict definition of how to ultimately 
estimate a specific parameter $\phi$ for a given data set $x$. 
To obtain an estimate for $\phi$, an \emph{estimator}
(for convenience let us refer to it as $\hat{\phi}(x)$) is used.
The hope is that the estimator maps a given data set 
to a particular value that is close to $\phi$.
However, 
regarding a specific parameter, it appears that some estimators are 
better suited than others.   
In this regard, a general distinction is drawn between
\emph{biased} and \emph{unbiased} estimators.
To cut a long story short, an estimator is said to be unbiased if 
it holds that $E[\hat{\phi}(x)]=\phi$. Therein, the computation of 
the expected value is with respect to all possible data sets $x$.
In words: an estimator is unbiased if its expectation value is equal 
to the true value of the parameter. Otherwise, $\hat{\phi}$ is biased.

To be more specific, consider a finite set of $N$ numerical values 
$x=\{x_0,\ldots,x_{N-1}\}$, where the associated random variables are 
assumed to be independent and identically distributed, following a pdf 
with (true) mean value $\mu$ and (true) variance $\sigma^2$.
As an example for an unbiased estimator, consider the computation 
of the mean value, i.e.\ $\phi\equiv\mu$. As an estimator
we might use the definition of the sample mean according to Eq.\ (\ref{eq:mean}),
i.e.\ $\hat{\phi}(x)\equiv{\rm av}(x)$.
We thus have
\begin{eqnarray}
E[{\rm av}(x)]=\frac{1}{N}  \sum_{i=0}^{N-1} E[x_i ] = \frac{1}{N}  (N \mu)=\mu~.
\end{eqnarray}
Further, the \emph{mean square error} (MSE) $E[(\hat{\phi}(x) -\phi)^2]$ of the estimator ${\rm av}(x)$ reads
\begin{eqnarray}
E[({\rm av}(x)-\mu)^2]=\frac{1}{N^2}\sum_{i=0}^{N-1} E[(x_i -\mu)^2 ] = \frac{1}{N^2} (N \sigma^2)=\frac{\sigma^2}{N}~.
\end{eqnarray}
In principle, the MSE measures both, the variance of an estimator and its 
bias. Consequently, for an unbiased estimator as, e.g., the definition of the 
sample mean used above, the MSE is equal to its variance.
Further, since $\lim_{N\to \infty}E[({\rm  av}(x)-\mu)^2]=0$, 
the estimator is said to be \emph{consistent}.

As an example for a biased estimator, consider the computation of the 
variance, i.e.\ $\phi\equiv\sigma^2$. As an estimator we might use
the uncorrected variance, as defined above, to find
\begin{eqnarray}
E[{\rm uVar}(x)] =  \frac{1}{N} \sum_{i=0}^{N-1} E[(x_i - \mu)^2] - E[({\rm av}(x)-\mu)^2]
= \frac{N-1}{N} \sigma^2~.
\end{eqnarray}
Since here it holds that $E[{\rm uVar}(x)]\neq \sigma^2$, the uncorrected 
variance is biased.
Note that the (corrected) variance, defined in Eq.\ (\ref{eq:Var}),
provides an unbiased estimate of $\sigma^2$.
Finally, note that the question whether bias arises is solely related
to the estimator, not the estimate (obtained from a particular data
set).


\subsection[Histograms]{Histograms: binning and graphical representation of data}
\label{sect:hist}

In the preceding section, we have already used histograms as a tool 
to construct approximate distribution functions from a finite 
set of data. Now, to provide a more precise definition 
of a \emph{histogram}, 
consider a data set $x\equiv\{x_i\}_{i=0}^{N-1}$ that stems from repeated 
measurements of a continuous random variable $X$ during a random experiment. 
To get a gross idea of the properties of the underlying 
continuous distribution function,
and to allow for a graphical representation of the respective
data, e.g., for the purpose of communicating results to the 
scientific community, a histogram of the observed data is of great use.

The idea is simply to accumulate the elements of the data set $x$ in a
finite number of, say, $n$ distinct intervals (or classes) 
$C_i = [c_i,c_{i+1})$, $i=0\ldots n-1$, called \emph{bins}, 
where the $c_i$ specify the interval boundaries. 
The frequency density $h_i$ (i.e., the relative frequency per unit-interval) 
associated with the $i$th bin can easily be obtained as $h_i=n_i/(N \Delta c_i)$, 
where $n_i$ specifies the number of elements that fall into bin $C_i$, 
and $\Delta c_i= c_{i+1}-c_i$ is the respective bin width.
The resulting set $H$ of tuples $(C_i,h_i)$, i.e.\
\begin{eqnarray}
H = \{(C_i,h_i)\}_{i=0}^{n-1}~,
\qquad \Big(\text{\parbox{0.38\textwidth}{$C_i=[c_i,c_{i+1})$, i.e.\ the $i$th bin $h_i=$ frequency density}}\Big),
\end{eqnarray}
specifies the histogram and give a discrete approximation 
to the pdf underlying the random variable.
Note that if one considers a finite sample size $N$,
it is only possible to construct an approximate pdf. However,
as the sample size increases one can be confident to approximate
the true pdf quite well.
All the values that fall within a certain bin $C_i$ are further
represented by a particular value $x_i^\prime\in C_i$. 
Often, $x_i^\prime$ is chosen as the center of the bin. 
Also note that, in order to properly represent the observed data, it might 
be useful to choose different widths $\Delta c_i=c_{i+1}-c_i$ for the different bins $b_i$.
As an example, one may decide to go for \emph{linear} or \emph{logarithmic binning}, detailed below.

\subsection*{Linear binning}
Considering \emph{linear binning}, the whole range of data $[x_{\rm min},x_{\rm max})$ is collected
using $n$ bins of equal width $\Delta c = (x_{\rm max}-x_{\rm min})/n$. 
Therein,
a particular bin $C_i$ accumulates all elements in the interval $[c_i,c_{i+1})$,
where the interval bounds are given by 
\begin{eqnarray}
c_i = x_{\rm min} + i \Delta c ~ ,~ {\text{for}} ~ i=0\ldots n ~. \label{eq:linBin}
\end{eqnarray}
During the binning procedure, a particular element belongs to 
bin $C_i$, where the integer identifier of the bin is given 
by $i= \lfloor x/\Delta c \rfloor$.

\subsection*{Logarithmic binning}
Considering \emph{logarithmic binning}, the whole range of data $[x_{\rm min},x_{\rm max})$ is collected
within $n$ bins that have equal width on a logarithmic scale, i.e.,  
$\log(c_{i+1})=\log(c_i)+\Delta c^\prime$ where $\Delta c^\prime = \log(x_{\rm max}/x_{\rm min})/n$.
In case of logarithmic binning,
a particular bin $C_i$ accumulates all elements in the interval $[c_i,c_{i+1})$,
where the interval bounds are consequently given by 
\begin{eqnarray}
c_i = c_0 \times \exp\{i \Delta c^\prime \} ~ ,~ {\text{for}} ~ i=0 \ldots n~.\label{eq:logBin}
\end{eqnarray}
During the histogram build-up, a particular element belongs to 
bin $C_i$, where $i= \lfloor \log(x/x_{\rm min})/\Delta c^\prime \rfloor$.
Note that on a linear scale, the width of the bins increases exponentially, i.e.\
$\Delta c_i = c_i \times(\exp\{ \Delta c^\prime\}-1)\propto c_i$.
Such a binning is especially well suited to represent power-law distributed data, 
see the example below.

A general drawback of any binning procedure is that many data in a 
given range $[c_{i},c_{i+1})$ are represented by only a single representative
$x_i^\prime$ of that interval. As a consequence, data binning always comes to the expense of information loss.
\medskip

\begin{figure}
\centering
\includegraphics[width=0.95\textwidth]{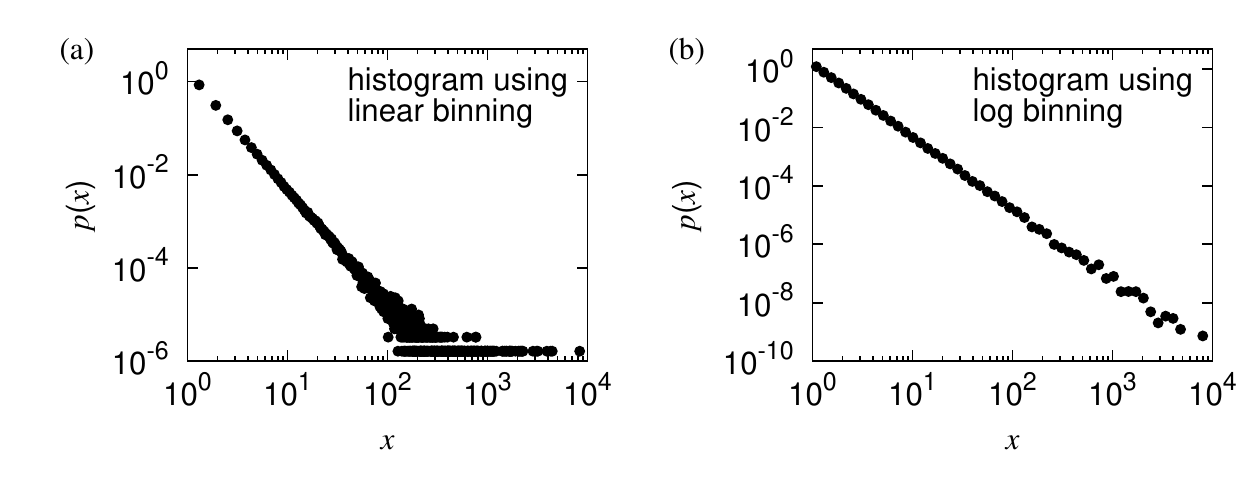}
\caption{
The plots illustrate a data set consisting of $10^6$ random numbers, 
drawn from the power-law distribution, see Eq.\ (\ref{eq:powerLaw}), with $\alpha=2.5$ and $x_0=1$. 
To give a visual account of the underlying probability distribution, 
the data was binned to yield
(a) a histogram using linear binning (displayed on log scale) and an overall number
of $n=2\times 10^4$ bins, and,
(b) a histogram using logarithmic binning and $n=55$ bins (for the same data as in (a)).
In either histogram, the center of a bin (on a linear scale) was chosen to represent
all values that fall within that bin.
\label{fig:hist_powerLaw}}
\end{figure}

\begin{example}{Data binning}

A small code-snippet that illustrates a {\tt python} implementation of a 
histogram using linear binning is listed below as \verb!hist_linBinning()!:
\SourceCodeLines{99}
\begin{Verbatim}[fontfamily=txtt,commandchars=\\\{\}]
 \PY{k}{def} \PY{n+nf}{hist\PYZus{}linBinning}\PY{p}{(}\PY{n}{rawData}\PY{p}{,}\PY{n}{xMin}\PY{p}{,}\PY{n}{xMax}\PY{p}{,}\PY{n}{nBins}\PY{o}{=}\PY{l+m+mi}{10}\PY{p}{)}\PY{p}{:}
	\PY{n}{h} \PY{o}{=} \PY{p}{[}\PY{l+m+mi}{0}\PY{p}{]}\PY{o}{*}\PY{n}{nBins}

	\PY{n}{dx} \PY{o}{=} \PY{p}{(}\PY{n}{xMax}\PY{o}{-}\PY{n}{xMin}\PY{p}{)}\PY{o}{/}\PY{n}{nBins}
	\PY{k}{def} \PY{n+nf}{binId}\PY{p}{(}\PY{n}{val}\PY{p}{)}\PY{p}{:}   \PY{k}{return} \PY{n+nb}{int}\PY{p}{(}\PY{n}{floor}\PY{p}{(}\PY{p}{(}\PY{n}{val}\PY{o}{-}\PY{n}{xMin}\PY{p}{)}\PY{o}{/}\PY{n}{dx}\PY{p}{)}\PY{p}{)}
	\PY{k}{def} \PY{n+nf}{bdry}\PY{p}{(}\PY{n}{i}\PY{p}{)}\PY{p}{:}      \PY{k}{return} \PY{n}{xMin}\PY{o}{+}\PY{n}{i}\PY{o}{*}\PY{n}{dx}\PY{p}{,} \PY{n}{xMin}\PY{o}{+}\PY{p}{(}\PY{n}{i}\PY{o}{+}\PY{l+m+mi}{1}\PY{p}{)}\PY{o}{*}\PY{n}{dx}
	\PY{k}{def} \PY{n+nf}{GErr}\PY{p}{(}\PY{n}{q}\PY{p}{,}\PY{n}{n}\PY{p}{,}\PY{n}{dx}\PY{p}{)}\PY{p}{:} \PY{k}{return} \PY{n}{sqrt}\PY{p}{(}\PY{n}{q}\PY{o}{*}\PY{p}{(}\PY{l+m+mi}{1}\PY{o}{-}\PY{n}{q}\PY{p}{)}\PY{o}{/}\PY{p}{(}\PY{n}{N}\PY{o}{-}\PY{l+m+mi}{1}\PY{p}{)}\PY{p}{)}\PY{o}{/}\PY{n}{dx} 

	\PY{k}{for} \PY{n}{value} \PY{o+ow}{in} \PY{n}{rawData}\PY{p}{:}
	   \PY{k}{if} \PY{l+m+mi}{0} \PY{o}{<}\PY{o}{=} \PY{n}{binId}\PY{p}{(}\PY{n}{value}\PY{p}{)} \PY{o}{<} \PY{n}{nBins}\PY{p}{:}
	      \PY{n}{h}\PY{p}{[}\PY{n}{binId}\PY{p}{(}\PY{n}{value}\PY{p}{)}\PY{p}{]} \PY{o}{+}\PY{o}{=} \PY{l+m+mi}{1}
	
	\PY{n}{N} \PY{o}{=} \PY{n+nb}{sum}\PY{p}{(}\PY{n}{h}\PY{p}{)}
	\PY{k}{for} \PY{n+nb}{bin} \PY{o+ow}{in} \PY{n+nb}{range}\PY{p}{(}\PY{n}{nBins}\PY{p}{)}\PY{p}{:}
	   \PY{n}{hRel}   \PY{o}{=} \PY{n+nb}{float}\PY{p}{(}\PY{n}{h}\PY{p}{[}\PY{n+nb}{bin}\PY{p}{]}\PY{p}{)}\PY{o}{/}\PY{n}{N}
	   \PY{n}{low}\PY{p}{,}\PY{n}{up} \PY{o}{=} \PY{n}{bdry}\PY{p}{(}\PY{n+nb}{bin}\PY{p}{)}
	   \PY{n}{width}  \PY{o}{=} \PY{n}{up}\PY{o}{-}\PY{n}{low}
	   \PY{k}{print} \PY{n}{low}\PY{p}{,} \PY{n}{up}\PY{p}{,} \PY{n}{hRel}\PY{o}{/}\PY{n}{width}\PY{p}{,} \PY{n}{GErr}\PY{p}{(}\PY{n}{hRel}\PY{p}{,}\PY{n}{N}\PY{p}{,}\PY{n}{width}\PY{p}{)} 
\end{Verbatim}

The first argument in the function call, see line 1 of the code listing,
indicates a list of the raw data, followed by the minimal and maximal 
variable value that should be considered during the binning procedure.
The last argument in the function call specifies the number of bins 
that shall be used therein (the default value is set to 10).
Based on the supplied data range and number of bins, the uniform bin 
width is computed in line 4.
Note that within the function \verb!hist_linBinning()!, 3 more functions are
defined. Those facilitate the calculation of the integer bin id that corresponds
to an element of the raw data (line 5), the lower and upper boundaries of 
a bin (line 6), and the Gaussian error bar for the respective data point (line 7).
In lines 9--11, the binning procedure is kicked off.
Since the upper bin boundary is exclusive, bear in mind that the numerical
value \verb!xMax! is identified with the bin index \verb!nBins+1!. As
a consequence it will not be considered during the histogram build-up.
Finally, in lines 13--18 the resulting normalized bin entries and their 
associated errors are sent to the standard out-stream.
The function is written in a quite general form, so that in order to 
implement a different kind of binning procedure only the definitions
in lines 4--7 have to be modified.
As regards this, for a more versatile variant that offers linear or 
logarithmic binning of the data, see the function \verb!hist! in the tiny
library \verb!MCS2012_lib.py!.

\paragraph{Binning of the data obtained for the $2D$ random walk:}
The approximate pdf of the average distance to the starting point for $10^5$ independent 
$100$-step $2D$ random walks, see Fig.\ \ref{fig:2DrandWalk_pdf}, 
was obtained by a linear binning procedure using the function 
\verb!hist_linBinning()! outlined above. For that task, the 
small script \verb!hist.py! listed below (see supplementary material) was used:
\SourceCodeLines{9}
\begin{Verbatim}[fontfamily=txtt,commandchars=\\\{\}]
 \PY{k+kn}{import} \PY{n+nn}{sys}
 \PY{k+kn}{from} \PY{n+nn}{MCS2012\PYZus{}lib} \PY{k+kn}{import} \PY{n}{fetchData}\PY{p}{,} \PY{n}{hist\PYZus{}linBinning}

 \PY{n}{fName} \PY{o}{=} \PY{n}{sys}\PY{o}{.}\PY{n}{argv}\PY{p}{[}\PY{l+m+mi}{1}\PY{p}{]}
 \PY{n}{col}   \PY{o}{=} \PY{n+nb}{int}\PY{p}{(}\PY{n}{sys}\PY{o}{.}\PY{n}{argv}\PY{p}{[}\PY{l+m+mi}{2}\PY{p}{]}\PY{p}{)}
 \PY{n}{nBins} \PY{o}{=} \PY{n+nb}{int}\PY{p}{(}\PY{n}{sys}\PY{o}{.}\PY{n}{argv}\PY{p}{[}\PY{l+m+mi}{3}\PY{p}{]}\PY{p}{)}
 \PY{n}{myData} \PY{o}{=} \PY{n}{fetchData}\PY{p}{(}\PY{n}{fName}\PY{p}{,}\PY{n}{col}\PY{p}{,}\PY{n+nb}{float}\PY{p}{)}
 \PY{n}{hist\PYZus{}linBinning}\PY{p}{(}\PY{n}{myData}\PY{p}{,}\PY{n+nb}{min}\PY{p}{(}\PY{n}{myData}\PY{p}{)}\PY{p}{,}\PY{n+nb}{max}\PY{p}{(}\PY{n}{myData}\PY{p}{)}\PY{p}{,}\PY{n}{nBins}\PY{p}{)} 
\end{Verbatim}

In principle, the Gaussian error bars are adequate for that data. However,
for a clearer presentation of the results, the error bars are not shown 
in Fig.\ \ref{fig:2DrandWalk_pdf}(b).

\paragraph{Binning of power-law distributed data:}
To illustrate the pros and cons of the binning types introduced above,
consider a data set consisting of $N=10^6$ random numbers drawn
from the power-law pdf, Eq.\ (\ref{eq:powerLaw}), with $\alpha=2.5$ and $x_0=1$.
Fig.\ \ref{fig:hist_powerLaw}(a) shows a histogram of the data using linear
binning.
In order to cover the whole range of data, 
the histogram uses $n=2\times 10^4$ bins with equal bin width $\Delta c\approx0.36$.
In principle, for a finite set of data and for power-law exponents $\alpha>1$, the number of samples per bin
decreases as the bin-index $i$, see Eq.\ (\ref{eq:linBin}), increases. For a
more elaborate discussion of the latter issue you might also want to consult Ref.\ \cite{newman2005}.
Consequently, the tail of the (binned) distribution is rather noisy. 
Thus, as evident from Fig.\ \ref{fig:hist_powerLaw}(a), a linear binning procedure
appears to be inadequate for power-law distributed data.
One can do better by means of logarithmic binning, see Fig.\ \ref{fig:hist_powerLaw}(b). 
Considering log-binning, the bins in the tail accumulate more samples, resulting 
in reduced statistical noise. 
In comparison to linear binning, data points in the tail are less scattered and the power law
decay of the data can be followed further to smaller probabilities. As a further 
benefit, on a logarithmic scale one has bins with equal width.
As a drawback note that any binning procedure involves a loss of information.
Also note that for the case of logarithmic binning, the Gaussian error bars
are not adequate.

\end{example}
\bigskip
\bigskip
\bigskip

\subsection[Bootstrap resampling]{Bootstrap resampling: the unbiased way to estimate errors}
\label{sect:resampling}

In a previous section, we discussed different parameters
that may serve to characterize the statistical properties of 
a finite set of data. Amongst others, we discussed 
estimators for the sample mean and standard deviation.
While there was a straight forward estimator for the 
standard error of the sample mean, there was no such 
measure for the standard deviation.
As a remedy, the current section illustrates a method that proves to be 
highly valuable when it comes to the issue of estimating errors
related to quite a lot of observables (for some more details you might 
want to consult chapter $15.6$ of Ref.\ \cite{nrc1992}), in an unbiased way.
To get an idea about the subtleties of the method, picture
the following situation: you perform a sequence of random experiments for 
a given model system and generate a sample $x$, consisting of $N$ 
statistically independent numbers.

Your aim is to measure some quantity $q$, e.g.\ some function 
$q\equiv f(x)$, that characterizes the simulated data.
At this point, bear in mind that this does not yield the \emph{true} quantity 
$q^\star$ that characterizes the model under consideration.
Instead, the numerical value of $q$ (very likely) differs from 
the latter value to some extent. 
To provide an error estimate that quantifies how good the observed 
value $q$ approximates the true value $q^\star$, the 
\emph{bootstrap method} utilizes a Monte Carlo simulation. 
This can be decomposed into the following two-step procedure:
Given a data set $x$ that consists of $N$ numerical values, where the 
corresponding random variables are assumed to be independent and identically 
distributed:
\begin{itemize}
\item[(i)] generate a number of $M$ auxiliary bootstrap data sets 
$\tilde{x}^{(k)}$, $k=0\ldots M-1$ by means of a resampling procedure. 
To obtain one such data set, draw $N$ data points (with 
replacement) from the original set $x$. During the construction 
procedure of a particular data set, some of the elements contained 
in $x$ will be chosen multiple times, while others won't appear
at all;

\item[(ii)] measure the observable of interest for each auxiliary data set,
to yield the set of estimates $\tilde{q}=\{\tilde{q}_k\}_{k=0}^{M-1}$.
Estimate the value of the desired observable using the original
data set and compute the corresponding error as the standard deviation 
\begin{eqnarray}
{\rm sDev}(\tilde{q}) = \Big( \frac{1}{M-1} \sum_{k=0}^{M-1} [\tilde{q}_k - {\rm av}(\tilde{q})]^2\Big)^{1/2} \qquad \text{(bootstrap error estimate)}
\end{eqnarray}
of the $M$ resampled (auxiliary) bootstrap data sets.
\end{itemize}

The finite sample $x$ only allows to get a coarse-grained glimpse on the 
probability distribution underlying the data. In principle, 
the latter one is not known.
Now, the basic assumption on which the bootstrap method relies is that 
the values $\tilde{q}_k$ (as obtained from the auxiliary data sets 
$\tilde{x}^{(k)}$) are distributed around the value $q$ (obtained from $x$) 
in a way similar to how further estimates of the observable obtained from 
further independent simulations are distributed around $q^\star$.
From a practical point of view, the procedure outlined above works quite well.

\begin{figure}
\includegraphics[width=0.94\textwidth]{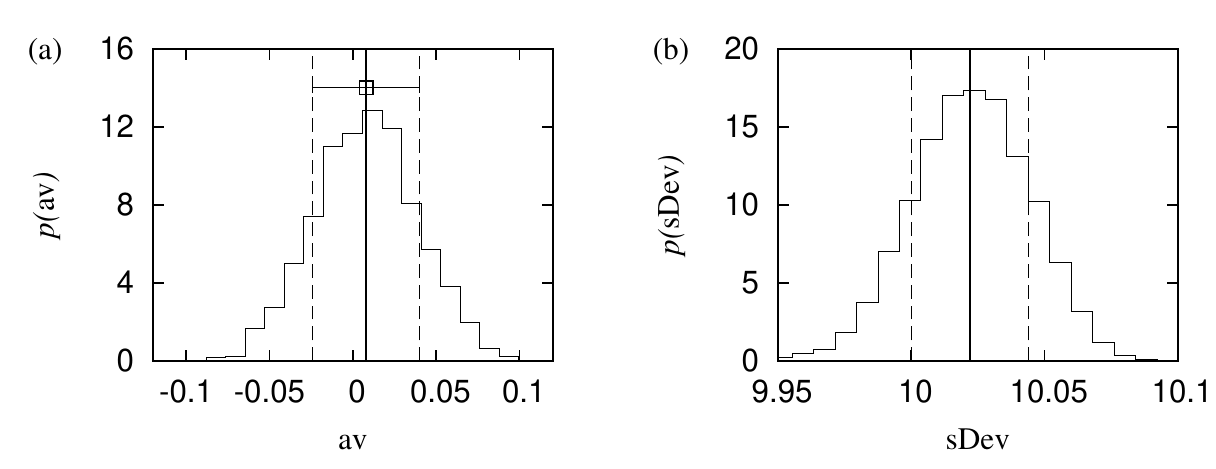}
\caption{
Result of the bootstrap resampling procedure ($M=1024$ auxiliary data sets) 
for the data characterizing the end point distribution for the symmetric $1D$ random walk.
(a) Approximate pdf (histogram using 18 bins) of the resampled
average value. The non-filled square marks the mean value and the 
associated error bar indicates the standard error as obtained in the 
example on ``basic statistics'' in Section \ref{sect:pf}.
(b) Approximate pdf (histogram using 18 bins) of the resampled
standard deviation.
The estimate of both quantities, as obtained from the original data 
set, is indicated by a solid vertical line, while the bootstrap error
bounds are shown by dashed vertical lines.
\label{fig:bootstrap}}
\end{figure}

\begin{example}{Error estimation via bootstrap resampling}
	
The code snippet below lists a {\tt python} implementation of the bootstrap
method outlined above. 
It is most convenient to amend the tiny library by the function \verb!bootstrap()!.
Note that it makes reference to \verb!basicStatistics()!, already defined in 
\verb!MCS2012_lib.py!.
\SourceCodeLines{99}
\begin{Verbatim}[fontfamily=txtt,commandchars=\\\{\}]
 \PY{k}{def} \PY{n+nf}{bootstrap}\PY{p}{(}\PY{n}{myData}\PY{p}{,}\PY{n}{estimFunc}\PY{p}{,}\PY{n}{M}\PY{o}{=}\PY{l+m+mi}{128}\PY{p}{)}\PY{p}{:}
        \PY{n}{N}        \PY{o}{=} \PY{n+nb}{len}\PY{p}{(}\PY{n}{myData}\PY{p}{)}		
        \PY{n}{h}        \PY{o}{=} \PY{p}{[}\PY{l+m+mf}{0.0}\PY{p}{]}\PY{o}{*}\PY{n}{M} 
        \PY{n}{bootSamp} \PY{o}{=} \PY{p}{[}\PY{l+m+mf}{0.0}\PY{p}{]}\PY{o}{*}\PY{n}{N}
        \PY{k}{for} \PY{n}{sample} \PY{o+ow}{in} \PY{n+nb}{range}\PY{p}{(}\PY{n}{M}\PY{p}{)}\PY{p}{:}
 	   \PY{k}{for} \PY{n}{val} \PY{o+ow}{in} \PY{n+nb}{range}\PY{p}{(}\PY{n}{N}\PY{p}{)}\PY{p}{:}
 	      \PY{n}{bootSamp}\PY{p}{[}\PY{n}{val}\PY{p}{]} \PY{o}{=} \PY{n}{myData}\PY{p}{[}\PY{n}{randint}\PY{p}{(}\PY{l+m+mi}{0}\PY{p}{,}\PY{n}{N}\PY{o}{-}\PY{l+m+mi}{1}\PY{p}{)}\PY{p}{]}
 	   \PY{n}{h}\PY{p}{[}\PY{n}{sample}\PY{p}{]} \PY{o}{=} \PY{n}{estimFunc}\PY{p}{(}\PY{n}{bootSamp}\PY{p}{)}
        \PY{n}{origEstim} \PY{o}{=} \PY{n}{estimFunc}\PY{p}{(}\PY{n}{myData}\PY{p}{)}
        \PY{n}{resError} \PY{o}{=} \PY{n}{basicStatistics}\PY{p}{(}\PY{n}{h}\PY{p}{)}\PY{p}{[}\PY{l+m+mi}{1}\PY{p}{]}
        \PY{k}{return} \PY{n}{origEstim}\PY{p}{,}\PY{n}{resError}
\end{Verbatim}

In lines 5--8, a number of $M$ auxiliary bootstrap data sets are 
obtained (the default number of auxiliary data sets is set to $M=128$, see 
function call in line 1). 
In line 9, the desired quantity, implemented by the function
\verb!estimFunc!, is computed for the original data set. Finally, the 
corresponding error is found as the standard deviation of the $M$ 
estimates of \verb!estimFunc! for the auxiliary data sets in line 10.
Note that the function \verb!bootstrap()! needs integer random numbers, uniformly
drawn in the interval $0\ldots N-1$, in order to generate the auxiliary data
sets. For this purpose the line \verb!from random import randint! must be 
included at the beginning of the file \verb!MCS2012_lib.py!.

As an example we may write the following small script that computes the mean 
and standard deviation, along with an error for those quantities computed using
the bootstrap method, for the data accumulated earlier for the symmetric $1D$
random walk (\verb!bootstrap.py!, see supplementary material).
\SourceCodeLines{99}
\begin{Verbatim}[fontfamily=txtt,commandchars=\\\{\}]
 \PY{k+kn}{import} \PY{n+nn}{sys}
 \PY{k+kn}{from} \PY{n+nn}{MCS2012\PYZus{}lib} \PY{k+kn}{import} \PY{o}{*}
 
 \PY{n}{fileName} \PY{o}{=} \PY{n}{sys}\PY{o}{.}\PY{n}{argv}\PY{p}{[}\PY{l+m+mi}{1}\PY{p}{]}
 \PY{n}{M}        \PY{o}{=} \PY{n+nb}{int}\PY{p}{(}\PY{n}{sys}\PY{o}{.}\PY{n}{argv}\PY{p}{[}\PY{l+m+mi}{2}\PY{p}{]}\PY{p}{)}
 \PY{n}{rawData}  \PY{o}{=} \PY{n}{fetchData}\PY{p}{(}\PY{n}{fileName}\PY{p}{,}\PY{l+m+mi}{1}\PY{p}{)}
 
 \PY{k}{def} \PY{n+nf}{mean}\PY{p}{(}\PY{n}{array}\PY{p}{)}\PY{p}{:} \PY{k}{return} \PY{n}{basicStatistics}\PY{p}{(}\PY{n}{array}\PY{p}{)}\PY{p}{[}\PY{l+m+mi}{0}\PY{p}{]}
 \PY{k}{def} \PY{n+nf}{sDev}\PY{p}{(}\PY{n}{array}\PY{p}{)}\PY{p}{:} \PY{k}{return} \PY{n}{basicStatistics}\PY{p}{(}\PY{n}{array}\PY{p}{)}\PY{p}{[}\PY{l+m+mi}{1}\PY{p}{]}
 
 \PY{k}{print} \PY{l+s}{"}\PY{l+s}{\PYZsh{} estimFunc: q +/- dq}\PY{l+s}{"}
 \PY{k}{print} \PY{l+s}{"}\PY{l+s}{mean: }\PY{l+s+si}{\PYZpc{}5.3lf}\PY{l+s}{ +/- }\PY{l+s+si}{\PYZpc{}4.3lf}\PY{l+s}{"}\PY{o}{\PYZpc{}}\PY{n}{bootstrap}\PY{p}{(}\PY{n}{rawData}\PY{p}{,}\PY{n}{mean}\PY{p}{,}\PY{n}{M}\PY{p}{)}
 \PY{k}{print} \PY{l+s}{"}\PY{l+s}{sDev: }\PY{l+s+si}{\PYZpc{}5.3lf}\PY{l+s}{ +/- }\PY{l+s+si}{\PYZpc{}4.3lf}\PY{l+s}{"}\PY{o}{\PYZpc{}}\PY{n}{bootstrap}\PY{p}{(}\PY{n}{rawData}\PY{p}{,}\PY{n}{sDev}\PY{p}{,}\PY{n}{M}\PY{p}{)}
\end{Verbatim}

Note that the statement \verb!from MCS2012_lib import *! imports
all functions from the indicated file and makes them available for data 
post-processing.
Invoking the script on the command line via 
\begin{Verbatim}
 > python bootstrap.py N100_n100000.dat 1024
\end{Verbatim}
where the latter number specifies the desired number of bootstrap samples,
the bootstrap method yields the results:
\begin{Verbatim}
 # estimFunc: q +/- dq
 mean:  0.008 +/- 0.032
 sDev: 10.022 +/- 0.022
\end{Verbatim}
For the sample mean, the bootstrap error is in good agreement with the 
standard error ${\rm sErr}(x)=0.032$ estimated in the ``basic statistics'' example 
(as it should be).
Regarding the bootstrap error for the standard deviation, the result is in 
agreement with the expectation $\sigma=10$ (for a number of $100$ steps
in an individual walk).
In Fig.\ \ref{fig:bootstrap}, the resulting distribution of the resampled estimates 
for the mean value (see Fig.\ \ref{fig:bootstrap}(a)) and standard error 
(see Fig.\ \ref{fig:bootstrap}(b)) are illustrated. 
For comparison, if we reduce the number of bootstrap samples to $M=24$ we 
obtain the two bootstrap errors $0.033$ and $0.024$ for the mean and standard error, respectively.

\end{example}


\subsection[The chi-square test]{The chi-square test: observed vs.\ expected}
\label{sect:chisquare}
The chi-square ($\chi^2$) goodness-of-fit test is commonly used to check whether the 
approximate pdf obtained from a set of sampled data appears to be 
equivalent to a theoretical density function. Here, ``equivalent'' is meant in 
the sense of ``the observed frequencies, as obtained from the data set, are 
consistent with the expected frequencies, as obtained from assuming the 
theoretical distribution function''.
More precisely,
assume that you obtained a set $\{(C_i,n_i)\}_{i=0}^{n-1}$ of 
binned data, summarizing an original data set of $N$ uncorrelated numerical values, where $C_i=[c_i,c_{i+1})$ 
signifies the $i$th bin with boundaries $c_i$ and $c_{i+1}$,
and $n_i$ is the number of \emph{observed events} that fall within that bin.
With reference to Section \ref{sect:hist}, the set of binned data is 
referred to as \emph{frequency histogram}.
Further, assume you already have a (more or less educated) guess about the 
expected limiting distribution function underlying the data, which
we call $f(x)$ in the following. 
Considering this expected limiting function, the number of \emph{expected
events} in the $i$th bin can be estimated as
\begin{eqnarray}
e_i = N\times \int_{c_i}^{c_{i+1}}f(x)~dx \qquad\text{(expected frequencies).}
\end{eqnarray}
To address the question whether the observed data might possibly 
be drawn from $f(x)$, the chi-square test compares the number of observed events in 
a given bin to the number of expected events in that bin by means of 
the expression
\begin{eqnarray}
\chi^2 = \sum_{i=0}^{n-1} \frac{(n_i - e_i)^2}{e_i} \qquad\text{(chi-square test).}
\end{eqnarray}

A further quantity that is important in relation to that test is the 
number of degrees of freedom, termed ${\rm dof}$. Usually it is equal 
to the number of bins less $1$, reflecting the fact that the sum of the
expected events has been re-normalized to match the sample size of the
original data set, i.e.\ $\sum_i e_i = N$.
However, if the function $f(x)$ involves additional free parameters that have
to be determined from the original data in order to properly represent
the expected limiting distribution function, each of these free parameters 
decreases the number of degrees of freedom by one.
Now, the wisdom behind the chi-square test is that for $\chi^2\approx {\rm dof}$,
one can consider the observed data as being consistent with the expected
limiting distribution function. If it holds that $\chi^2\gg {\rm dof}$, 
then the discrepancy between both is significant.
Besides listing the values of $\chi^2$ and ${\rm dof}$, a more convenient way 
to state the result of the chi-square test is to report the 
\emph{reduced chi-square}, i.e.\ chi-square per {\rm dof},
$\tilde{\chi}^2 = \chi^2 / {\rm dof}$. It is independent of the 
number of degrees of freedom and the observed data is consistent with the 
expected distribution function if $\tilde{\chi}^2\approx 1$.

For an account of more strict criteria that allow to assess the ``quality'' of the 
chi-square test and what to pay attention to if one attempts to compare
two sets of binned data that summarize data sets with possibly different 
sample sizes, see Refs.\ \cite{nrc1992,practicalGuide2009}. Let us leave it at that.
As a final note, notice that the chi-square test cannot be used to prove that the 
observed data is drawn from an expected distribution function, it merely reports whether the 
observed data is consistent with it.

\begin{example}{Chi-square test}
	
The code snippet below lists a {\tt python} implementation of the 
chi-square test outlined above. 
It is most convenient to amend \verb!MCS2012_lib.py! by 
the function definition
\SourceCodeLines{9}
\begin{Verbatim}[fontfamily=txtt,commandchars=\\\{\}]
 \PY{k}{def} \PY{n+nf}{chiSquare}\PY{p}{(}\PY{n}{obsFreq}\PY{p}{,}\PY{n}{expFreq}\PY{p}{,}\PY{n}{nConstr}\PY{p}{)}\PY{p}{:}
        \PY{n}{nBins} \PY{o}{=} \PY{n+nb}{len}\PY{p}{(}\PY{n}{obsFreq}\PY{p}{)}
        \PY{n}{chi2} \PY{o}{=} \PY{l+m+mf}{0.0}
        \PY{k}{for} \PY{n+nb}{bin} \PY{o+ow}{in} \PY{n+nb}{range}\PY{p}{(}\PY{n}{nBins}\PY{p}{)}\PY{p}{:}
           \PY{n}{dum}   \PY{o}{=} \PY{n}{obsFreq}\PY{p}{[}\PY{n+nb}{bin}\PY{p}{]}\PY{o}{-}\PY{n}{expFreq}\PY{p}{[}\PY{n+nb}{bin}\PY{p}{]}
           \PY{n}{chi2} \PY{o}{+}\PY{o}{=} \PY{n}{dum}\PY{o}{*}\PY{n}{dum}\PY{o}{/}\PY{n}{expFreq}\PY{p}{[}\PY{n+nb}{bin}\PY{p}{]}
        \PY{n}{dof} \PY{o}{=} \PY{n}{nBins}\PY{o}{-}\PY{n}{nConstr}
        \PY{k}{return} \PY{n}{dof}\PY{p}{,}\PY{n}{chi2}
\end{Verbatim}

A word of caution: it is a good advise to not trust a chi-square test for which some of the 
observed frequencies are less than, say, 4 (see Ref.\ \cite{bendat1971}). 
For such small frequencies, the statistical noise 
is simply too large and the respective terms might lead to an exploding value of $\chi^2$. 
As a remedy one can merge a couple of adjacent bins to form a ``super-bin'' containing 
more than just 4 events.

For an illustrational purpose we might perform a chi-square test for
the approximate pmf of the end positions of the symmetric $1D$ random 
walks, constructed earlier in Section \ref{sect:pf}.
In this regard, let $\{(x_i,n_i)\}_{i=0}^{n-1}$ be the 
binned set of data, summarizing an original data set 
(for a discrete random variable) with sample size $M$.
Therein, $n_i$ denotes the number of observed events that 
correspond to the value $x_i$. For the data at hand, the expected frequencies $e_i$
are simply proportional to the symmetric binomial distribution with variance $N/4$ 
(where $N$ is the number of steps in a single walk) at $x_i$. 
To be more precise, we have 
\begin{eqnarray}
e_i = M \times {\rm bin}(N,(x_i+N)/2) 2^{-N},
\end{eqnarray}
for $i=0\ldots n-1$, wherein $N$ specifies the number of steps in a single walk 
and ${\rm bin}(a,b)=a!/[(a-b)!b!]$ defines the binomial coefficients.
For the data on the symmetric $1D$ random walk, some of the 
observed frequencies in the tails of the distribution are smaller than $4$.
As mentioned above, this requires the pooling of several bins in the 
regions of low probability density. Practically speaking, we create 
two super-bins that lump together all frequencies 
that correspond to values $x_i\geq R$ or $x_i\leq -R$, respectively. 
This procedure works well, since the distribution is symmetric.
The following script implements this re-binning procedure and performs the 
chi-square test as defined above:

\SourceCodeLines{99}
\begin{Verbatim}[fontfamily=txtt,commandchars=\\\{\}]
 \PY{k+kn}{import} \PY{n+nn}{sys}
 \PY{k+kn}{from} \PY{n+nn}{math} \PY{k+kn}{import} \PY{n}{factorial} \PY{k}{as} \PY{n}{fac}
 \PY{k+kn}{from} \PY{n+nn}{MCS2012\PYZus{}lib} \PY{k+kn}{import} \PY{o}{*}
 \PY{k+kn}{import} \PY{n+nn}{scipy.special}
 
 \PY{n}{fileName} \PY{o}{=} \PY{n}{sys}\PY{o}{.}\PY{n}{argv}\PY{p}{[}\PY{l+m+mi}{1}\PY{p}{]}
 \PY{n}{col}      \PY{o}{=} \PY{n+nb}{int}\PY{p}{(}\PY{n}{sys}\PY{o}{.}\PY{n}{argv}\PY{p}{[}\PY{l+m+mi}{2}\PY{p}{]}\PY{p}{)}
 \PY{n}{R}        \PY{o}{=} \PY{n+nb}{int}\PY{p}{(}\PY{n}{sys}\PY{o}{.}\PY{n}{argv}\PY{p}{[}\PY{l+m+mi}{3}\PY{p}{]}\PY{p}{)}

 \PY{n}{rawData}  \PY{o}{=} \PY{n}{fetchData}\PY{p}{(}\PY{n}{fileName}\PY{p}{,}\PY{n}{col}\PY{p}{)}
 \PY{n}{pmf}      \PY{o}{=} \PY{n}{getPmf}\PY{p}{(}\PY{n}{rawData}\PY{p}{)}

 \PY{k}{def} \PY{n+nf}{bin}\PY{p}{(}\PY{n}{n}\PY{p}{,}\PY{n}{k}\PY{p}{)}\PY{p}{:} \PY{k}{return} \PY{n}{fac}\PY{p}{(}\PY{n}{n}\PY{p}{)}\PY{o}{/}\PY{p}{(}\PY{n}{fac}\PY{p}{(}\PY{n}{n}\PY{o}{-}\PY{n}{k}\PY{p}{)}\PY{o}{*}\PY{n}{fac}\PY{p}{(}\PY{n}{k}\PY{p}{)}\PY{p}{)}
 \PY{k}{def} \PY{n+nf}{f}\PY{p}{(}\PY{n}{x}\PY{p}{,}\PY{n}{N}\PY{p}{)}\PY{p}{:}   \PY{k}{return} \PY{n+nb}{bin}\PY{p}{(}\PY{n}{N}\PY{p}{,}\PY{p}{(}\PY{n}{x}\PY{o}{+}\PY{n}{N}\PY{p}{)}\PY{o}{*}\PY{l+m+mf}{0.5}\PY{p}{)}\PY{o}{*}\PY{l+m+mf}{0.5}\PY{o}{*}\PY{o}{*}\PY{n}{N}

 \PY{n}{N}\PY{o}{=}\PY{n+nb}{len}\PY{p}{(}\PY{n}{rawData}\PY{p}{)}
 \PY{n}{oFr}\PY{o}{=}\PY{p}{\PYZob{}}\PY{p}{\PYZcb{}}\PY{p}{;} \PY{n}{eFr}\PY{o}{=}\PY{p}{\PYZob{}}\PY{p}{\PYZcb{}}
 \PY{k}{for} \PY{n}{el} \PY{o+ow}{in} \PY{n}{pmf}\PY{p}{:}
  \PY{k}{if} \PY{n}{el} \PY{o}{>}\PY{o}{=} \PY{n}{R}\PY{p}{:}  
        \PY{k}{if} \PY{n}{R} \PY{o+ow}{not} \PY{o+ow}{in} \PY{n}{oFr}\PY{p}{:}
          \PY{n}{oFr}\PY{p}{[}\PY{n}{R}\PY{p}{]} \PY{o}{=} \PY{n}{eFr}\PY{p}{[}\PY{n}{R}\PY{p}{]} \PY{o}{=} \PY{l+m+mi}{0}
        \PY{n}{oFr}\PY{p}{[}\PY{n}{R}\PY{p}{]} \PY{o}{+}\PY{o}{=} \PY{n}{pmf}\PY{p}{[}\PY{n}{el}\PY{p}{]}\PY{o}{*}\PY{n}{N}
        \PY{n}{eFr}\PY{p}{[}\PY{n}{R}\PY{p}{]} \PY{o}{+}\PY{o}{=} \PY{n}{f}\PY{p}{(}\PY{n}{el}\PY{p}{,}\PY{l+m+mi}{100}\PY{p}{)}\PY{o}{*}\PY{n}{N}
  \PY{k}{elif} \PY{n}{el} \PY{o}{<}\PY{o}{=} \PY{o}{-}\PY{n}{R}\PY{p}{:}  
        \PY{k}{if} \PY{o}{-}\PY{n}{R} \PY{o+ow}{not} \PY{o+ow}{in} \PY{n}{oFr}\PY{p}{:}
          \PY{n}{oFr}\PY{p}{[}\PY{o}{-}\PY{n}{R}\PY{p}{]} \PY{o}{=} \PY{n}{eFr}\PY{p}{[}\PY{o}{-}\PY{n}{R}\PY{p}{]} \PY{o}{=} \PY{l+m+mi}{0}
        \PY{n}{oFr}\PY{p}{[}\PY{o}{-}\PY{n}{R}\PY{p}{]} \PY{o}{+}\PY{o}{=} \PY{n}{pmf}\PY{p}{[}\PY{n}{el}\PY{p}{]}\PY{o}{*}\PY{n}{N}
        \PY{n}{eFr}\PY{p}{[}\PY{o}{-}\PY{n}{R}\PY{p}{]} \PY{o}{+}\PY{o}{=} \PY{n}{f}\PY{p}{(}\PY{n}{el}\PY{p}{,}\PY{l+m+mi}{100}\PY{p}{)}\PY{o}{*}\PY{n}{N}
  \PY{k}{else}\PY{p}{:} 
        \PY{n}{oFr}\PY{p}{[}\PY{n}{el}\PY{p}{]} \PY{o}{=} \PY{n}{pmf}\PY{p}{[}\PY{n}{el}\PY{p}{]}\PY{o}{*}\PY{n}{N}
        \PY{n}{eFr}\PY{p}{[}\PY{n}{el}\PY{p}{]} \PY{o}{=} \PY{n}{f}\PY{p}{(}\PY{n}{el}\PY{p}{,}\PY{l+m+mi}{100}\PY{p}{)}\PY{o}{*}\PY{n}{N}

 \PY{n}{o} \PY{o}{=} \PY{n+nb}{map}\PY{p}{(}\PY{k}{lambda} \PY{n}{x}\PY{p}{:} \PY{n}{x}\PY{p}{[}\PY{l+m+mi}{1}\PY{p}{]}\PY{p}{,} \PY{n}{oFr}\PY{o}{.}\PY{n}{items}\PY{p}{(}\PY{p}{)}\PY{p}{)}
 \PY{n}{e} \PY{o}{=} \PY{n+nb}{map}\PY{p}{(}\PY{k}{lambda} \PY{n}{x}\PY{p}{:} \PY{n}{x}\PY{p}{[}\PY{l+m+mi}{1}\PY{p}{]}\PY{p}{,} \PY{n}{eFr}\PY{o}{.}\PY{n}{items}\PY{p}{(}\PY{p}{)}\PY{p}{)}
 
 \PY{n}{dof}\PY{p}{,}\PY{n}{chi2} \PY{o}{=} \PY{n}{chiSquare}\PY{p}{(}\PY{n}{o}\PY{p}{,}\PY{n}{e}\PY{p}{,}\PY{l+m+mi}{1}\PY{p}{)}
 \PY{k}{print} \PY{l+s}{"}\PY{l+s}{\PYZsh{} dof=}\PY{l+s+si}{\PYZpc{}d}\PY{l+s}{, chi2=}\PY{l+s+si}{\PYZpc{}5.3lf}\PY{l+s}{"}\PY{o}{\PYZpc{}}\PY{p}{(}\PY{n}{dof}\PY{p}{,}\PY{n}{chi2}\PY{p}{)}
 \PY{k}{print} \PY{l+s}{"}\PY{l+s}{\PYZsh{} reduced chi2=}\PY{l+s+si}{\PYZpc{}5.3lf}\PY{l+s}{"}\PY{o}{\PYZpc{}}\PY{p}{(}\PY{n}{chi2}\PY{o}{/}\PY{n}{dof}\PY{p}{)}

 \PY{n}{pVal} \PY{o}{=} \PY{n}{scipy}\PY{o}{.}\PY{n}{special}\PY{o}{.}\PY{n}{gammaincc}\PY{p}{(}\PY{n}{dof}\PY{o}{*}\PY{l+m+mf}{0.5}\PY{p}{,}\PY{n}{chi2}\PY{o}{*}\PY{l+m+mf}{0.5}\PY{p}{)}
 \PY{k}{print} \PY{l+s}{"}\PY{l+s}{\PYZsh{} p=}\PY{l+s}{"}\PY{p}{,}\PY{n}{pVal}
\end{Verbatim}

\smallskip

Therein, in lines 13--14, the symmetric binomial distribution 
is defined, and in lines 16--31 the re-binning of the data
is carried out. Finally, lines 33--34 prepare the re-binned data
for the chi-sqare test (line 36). 
Note that the script above extends the chi-square goodness-of-fit 
test by also computing the so-called $p$-value for the 
problem at hand (line 40).
The $p$-value is a standard way to assess the significance of the 
chi-square test, see Refs.\ \cite{nrc1992,practicalGuide2009}. 
In essence, the numerical value of $p$
gives the probability that the sum of the squares of a number of ${\rm dof}$ 
Gaussian random variables (zero mean and unit variance) will be
greater than $\chi^2$.
To compute the $p$-value an implementation of the
\emph{incomplete gamma function} is needed. Unfortunately, 
this is not contained in a standard {\tt python} package. 
However, the incomplete gamma function is available 
through the {\tt scipy}-package \cite{scipy}, which 
offers an extensive selection of special functions and
lots of scientific tools.
Considering the $p$-value, the observed frequencies are consistent with 
the expected frequencies, if the numerical value of $p$ is not smaller than,
say, $10^{-2}$.

If the script is called for $R=40$ it yields the result:
\begin{Verbatim}
 > python chiSquare.py N100_n100000.dat 1 40
 # dof=40, chi2=38.256
 # reduced chi2=0.956
 # p= 0.549
\end{Verbatim}
Hence, the pmf for the end point distribution of the symmetric 
$1D$ random walk appears to be consistent with the symmetric
binomial distribution with mean $0$ and variance $N/4$, 
wherein $N$ specifies the number of steps in a single walk.
\end{example}



\section{Object oriented programming in {\tt python}}

In addition to the built-in data types used earlier in these notes,
{\tt python} allows you to define your own custom data structures.
In doing so, {\tt python} makes it easy to follow an object oriented programming (OOP)
approach. The basic idea of the OOP approach is to use \emph{objects}
in order to design computer programs. Therein, the term 'object'
refers to a custom \emph{data structure} that has certain \emph{attributes}
and \emph{methods}, where the latter can be understood as functions 
that alter the attributes of the data structure.

By following an OOP approach, the first step consists in putting the
problem at hand under scrutiny and figuring out what the respective
objects might be.
Once this first step is accomplished one might go on and design custom 
data structures to represent the objects. In {\tt python} this is 
done using the concept of \emph{classes}.

In general, OOP techniques emphasize on \emph{data encapsulation},
\emph{inheritance}, and \emph{overloading}.
In less formal terms, data encapsulation means to 'hide the implementation'.
I.e., the access to the attributes of the data structures is typically restricted,
making them (somewhat) \emph{private}. 
Access to the attributes is only granted for certain methods that build
an \emph{interface} by means of which an user can alter the attributes.
The concept of data encapsulation slightly interferes with the 'everything is public' 
consensus of the {\tt python} community which follows the habit that even if one might be able to see the 
implementation one does not have to care about it.
Nevertheless, {\tt python} offers a kind of pseudo-encapsulation referred 
to as \emph{name-mangling} which will be illustrated in Subsection \ref{subsect:hist_oop}

In OOP terms, inheritance means to 'share code among classes'. 
It highly encourages code recycling and easily allows to \emph{extend} 
or \emph{specialize} existing classes, thereby generating a hierarchy
composed of classes and derived \emph{subclasses}.  

Finally, overloading means to 'redefine methods on subclass level'.
This makes it possible to adapt methods to their context. Also known
as \emph{polymorphism}, this offers the possibility to have several 
definitions of the same method on different levels of the class hierarchy.

\subsection{Implementing an undirected graph}

An example by means of which all three OOP techniques can be illustrated is
a \emph{graph data structure}. For the purpose of illustration consider 
an undirected graph $G=(V,E)$. Therein, the graph $G$
consists of an unordered set of $n=|V|$ nodes $i\in V$ and an unordered set 
of $m=|E|$ edges $\{i,j\}\in E$ (cf.\ the contribution by \textit{Hartmann}
in this volume).

\begin{figure}
\centering
\includegraphics[width=0.75\textwidth]{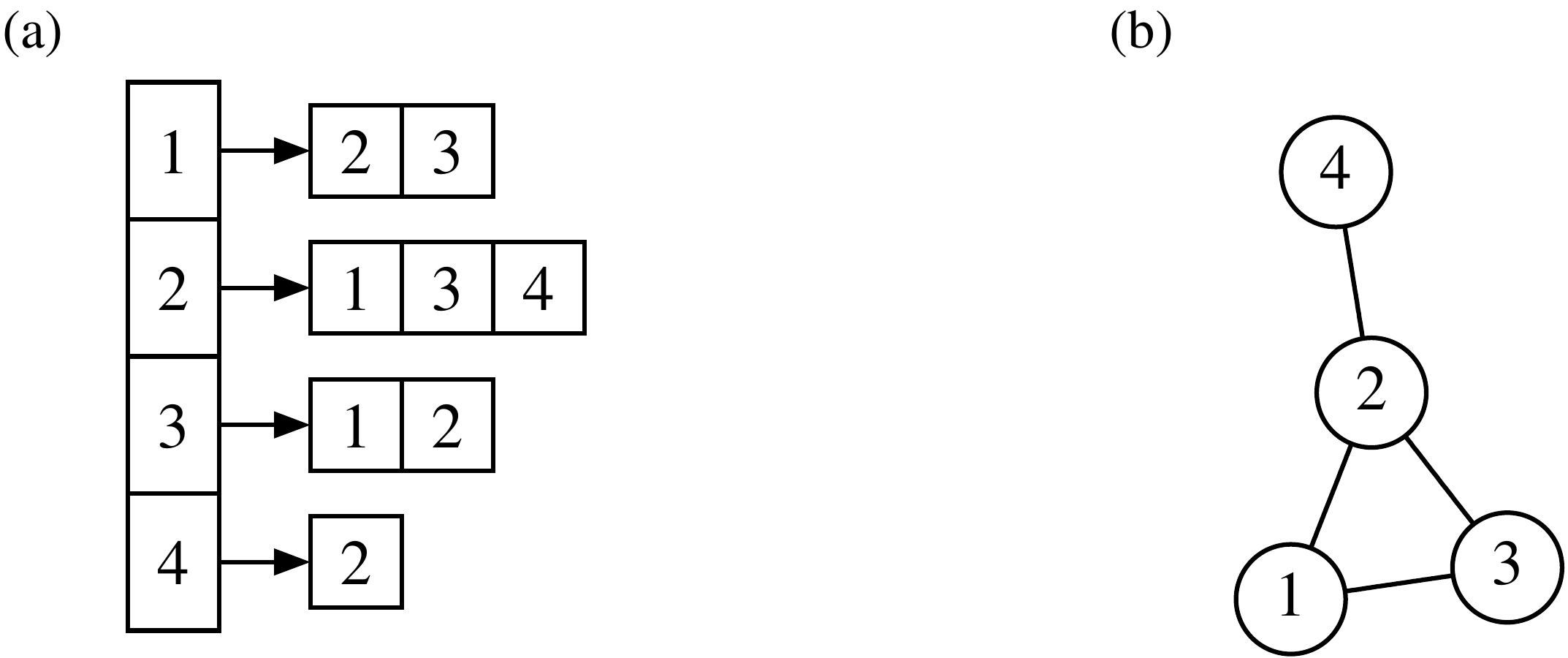}
\caption{
Undirected example graph $G=(V,E)$, consisting of four nodes and four edges.
(a) Adjacency list representation of $G$.
(b) Visual representation of $G$. 
\label{fig:adjList_graph}}
\end{figure}

According to the first step in the OOP plan one might now consider 
the entity \emph{graph} as an elementary object for which a data structure
should be designed.
Depending on the nature of the graph at hand there are different 'optimal'
ways to represent them. Here, let us consider sparse graphs, i.e.,
graphs with $O(m)\ll O(n^2)$. 
In terms of memory consumption it is most
beneficial to choose an \emph{adjacency list} representation for such graphs, see 
Fig.\ \ref{fig:adjList_graph}(a).
This only needs space $O(n+2m)$. An adjacency list representation for $G$ 
requires to maintain for each node $i\in V$ a list of its immediate neighbors.
I.e., the adjacency list for node $i\in V$ contains node $j\in V$ only if  
$\{i,j\}\in E$. 
Hence, as attributes we might consider the overall number of nodes and edges of $G$
as well as the adjacency lists for the nodes.
Further, we will implement six methods that serve as an interface to access and 
possibly alter the attributes. 

To get things going, we start with the class definition, the default 
constructor for an instance of the class and a few basic methods:
\SourceCodeLines{99}
\begin{Verbatim}[fontfamily=txtt,commandchars=\\\{\}]
 \PY{k}{class} \PY{n+nc}{myGraph}\PY{p}{(}\PY{n+nb}{object}\PY{p}{)}\PY{p}{:}


	\PY{k}{def} \PY{n+nf}{\PYZus{}\PYZus{}init\PYZus{}\PYZus{}}\PY{p}{(}\PY{n+nb+bp}{self}\PY{p}{)}\PY{p}{:}
		\PY{n+nb+bp}{self}\PY{o}{.}\PY{n}{\PYZus{}nNodes}  \PY{o}{=} \PY{l+m+mi}{0}
		\PY{n+nb+bp}{self}\PY{o}{.}\PY{n}{\PYZus{}nEdges}  \PY{o}{=} \PY{l+m+mi}{0}
		\PY{n+nb+bp}{self}\PY{o}{.}\PY{n}{\PYZus{}adjList} \PY{o}{=} \PY{p}{\PYZob{}}\PY{p}{\PYZcb{}}

	\PY{n+nd}{@property}
	\PY{k}{def} \PY{n+nf}{nNodes}\PY{p}{(}\PY{n+nb+bp}{self}\PY{p}{)}\PY{p}{:} 
		\PY{k}{return} \PY{n+nb+bp}{self}\PY{o}{.}\PY{n}{\PYZus{}nNodes}

	\PY{n+nd}{@nNodes.setter}
	\PY{k}{def} \PY{n+nf}{nNodes}\PY{p}{(}\PY{n+nb+bp}{self}\PY{p}{,}\PY{n}{val}\PY{p}{)}\PY{p}{:} 
		\PY{k}{print} \PY{l+s}{"}\PY{l+s}{will not change private attribute}\PY{l+s}{"}

	\PY{n+nd}{@property}
	\PY{k}{def} \PY{n+nf}{nEdges}\PY{p}{(}\PY{n+nb+bp}{self}\PY{p}{)}\PY{p}{:} 
		\PY{k}{return} \PY{n+nb+bp}{self}\PY{o}{.}\PY{n}{\PYZus{}nEdges}

	\PY{n+nd}{@nEdges.setter}
	\PY{k}{def} \PY{n+nf}{nEdges}\PY{p}{(}\PY{n+nb+bp}{self}\PY{p}{,}\PY{n}{val}\PY{p}{)}\PY{p}{:} 
		\PY{k}{print} \PY{l+s}{"}\PY{l+s}{will not change private attribute}\PY{l+s}{"}
	
	\PY{n+nd}{@property}
	\PY{k}{def} \PY{n+nf}{V}\PY{p}{(}\PY{n+nb+bp}{self}\PY{p}{)}\PY{p}{:} 
		\PY{k}{return} \PY{n+nb+bp}{self}\PY{o}{.}\PY{n}{\PYZus{}adjList}\PY{o}{.}\PY{n}{keys}\PY{p}{(}\PY{p}{)}

	\PY{k}{def} \PY{n+nf}{adjList}\PY{p}{(}\PY{n+nb+bp}{self}\PY{p}{,}\PY{n}{node}\PY{p}{)}\PY{p}{:}
		\PY{k}{return} \PY{n+nb+bp}{self}\PY{o}{.}\PY{n}{\PYZus{}adjList}\PY{p}{[}\PY{n}{node}\PY{p}{]}

	\PY{k}{def} \PY{n+nf}{deg}\PY{p}{(}\PY{n+nb+bp}{self}\PY{p}{,}\PY{n}{node}\PY{p}{)}\PY{p}{:}
		\PY{k}{return} \PY{n+nb}{len}\PY{p}{(}\PY{n+nb+bp}{self}\PY{o}{.}\PY{n}{\PYZus{}adjList}\PY{p}{[}\PY{n}{node}\PY{p}{]}\PY{p}{)}
\end{Verbatim}

In line 1, the class definition \verb!myGraph! inherits the propertis of 
\verb!object!, the latter being the most basic class type that allows to 
realize certain functionality for class methods (as, e.g., the \verb!@property!
decorators mentioned below).
The keyword 'self' is the first argument that appears in the argument list of any method
and makes a reference to the class itself (just like the 'this' pointer in {\tt C++}).
Note that by convention the leading underscore of the attributes 
signals that they are considered to be private. To access them,
we need to provide appropriate methods. E.g., in order to access the number
of nodes, the method \verb!nNodes(self)! is implemented and declared as a property (using
the \verb!@property! statement the precedes the method definition) of the class. 
As an effect it is now possible to receive the number of nodes by writing just
\verb![class reference].nNodes! instead of \verb![class reference].nNodes()!.
In the spirit of data encapsulation 
one has to provide a so-called \emph{setter} in order to alter the content of the private 
attribute \verb![class reference]._nNodes!. 
Here, for the number of nodes 
a setter similar to \verb!nNodes(self,val)!, indicated by the 
preceding \verb!@nNodes.setter! statement, might be implemented.
Similar methods can of course be defined for the number of edges.
These are examples of \emph{name-mangling}, the {\tt python} substitute
for data encapsulation.
 
Further, the above code snippet illustrates a method that returns a list 
that represents the node set of the graph (\verb!V(self)!), a method that
returns the adjacency list of a particular node (\verb!adjList(self,node)!),
and a method that returns the degree, i.e., the number of neighbors, 
of a node (\verb!deg(self)!).

Next, we might add some functionality to the graph. Therefore we might implement 
functions that add nodes and edges to an instance of the graph:
\SourceCodeLines{99}
\begin{Verbatim}[fontfamily=txtt,commandchars=\\\{\}]
	\PY{k}{def} \PY{n+nf}{addNode}\PY{p}{(}\PY{n+nb+bp}{self}\PY{p}{,}\PY{n}{node}\PY{p}{)}\PY{p}{:}
		\PY{k}{if} \PY{n}{node} \PY{o+ow}{not} \PY{o+ow}{in} \PY{n+nb+bp}{self}\PY{o}{.}\PY{n}{V}\PY{p}{:} 
		    \PY{n+nb+bp}{self}\PY{o}{.}\PY{n}{\PYZus{}adjList}\PY{p}{[}\PY{n}{node}\PY{p}{]}\PY{o}{=}\PY{p}{[}\PY{p}{]}
		    \PY{n+nb+bp}{self}\PY{o}{.}\PY{n}{\PYZus{}nNodes} \PY{o}{+}\PY{o}{=} \PY{l+m+mi}{1}

	\PY{k}{def} \PY{n+nf}{addEdge}\PY{p}{(}\PY{n+nb+bp}{self}\PY{p}{,}\PY{n}{fromNode}\PY{p}{,}\PY{n}{toNode}\PY{p}{)}\PY{p}{:}
		\PY{n}{flag}\PY{o}{=}\PY{l+m+mi}{0}
		\PY{n+nb+bp}{self}\PY{o}{.}\PY{n}{addNode}\PY{p}{(}\PY{n}{fromNode}\PY{p}{)}
		\PY{n+nb+bp}{self}\PY{o}{.}\PY{n}{addNode}\PY{p}{(}\PY{n}{toNode}\PY{p}{)}
		\PY{k}{if} \PY{p}{(}\PY{n}{fromNode} \PY{o}{!=} \PY{n}{toNode}\PY{p}{)} \PY{o+ow}{and}\PYZbs{}
		   \PY{p}{(}\PY{n}{toNode} \PY{o+ow}{not} \PY{o+ow}{in} \PY{n+nb+bp}{self}\PY{o}{.}\PY{n}{adjList}\PY{p}{(}\PY{n}{fromNode}\PY{p}{)}\PY{p}{)}\PY{p}{:}
		    \PY{n+nb+bp}{self}\PY{o}{.}\PY{n}{\PYZus{}adjList}\PY{p}{[}\PY{n}{fromNode}\PY{p}{]}\PY{o}{.}\PY{n}{append}\PY{p}{(}\PY{n}{toNode}\PY{p}{)}
		    \PY{n+nb+bp}{self}\PY{o}{.}\PY{n}{\PYZus{}adjList}\PY{p}{[}\PY{n}{toNode}\PY{p}{]}\PY{o}{.}\PY{n}{append}\PY{p}{(}\PY{n}{fromNode}\PY{p}{)}
		    \PY{n+nb+bp}{self}\PY{o}{.}\PY{n}{\PYZus{}nEdges} \PY{o}{+}\PY{o}{=} \PY{l+m+mi}{1}
		    \PY{n}{flag} \PY{o}{=} \PY{l+m+mi}{1}
		\PY{k}{return} \PY{n}{flag}

	\PY{n}{\PYZus{}\PYZus{}addEdge}\PY{o}{=}\PY{n}{addEdge}
\end{Verbatim}

The first method, \verb!addNode(self,node)!, attempts to add a node to the graph. 
If this node does not yet exist, it creates an empty adjacency list for the 
respective node and increases the number of nodes counter by one.
Since the adjacency list is designed by means of the built-in dictionary data structure,
the data type of \verb!node! can be one out of many types, e.g., an integer identifier 
that specifies the node or some string that represents the node.
The second method, \verb!addEdge(self,fromNode,toNode)!, attempts to add an edge 
to the graph and updates the adjacency lists of its terminal nodes as well as the
number of edges in the graph accordingly. 
If the nodes do not yet exist, it creates them first by calling the \verb!addNode()! method.
Note that the \verb!addEdge()! method returns a value of zero if the edge exists already
and returns a value of one if the edge did not yet exist. Further, the last line in the 
code snippet above constructs a local copy of the method. Neither the use of the 
edge existence flag, nor the use of the local copy is obvious at the moment. Both we will be clarified below.

To make the graph class even more flexible we might amend it by implementing methods 
that delete nodes and edges:
\SourceCodeLines{99}
\begin{Verbatim}[fontfamily=txtt,commandchars=\\\{\}]
	\PY{k}{def} \PY{n+nf}{delNode}\PY{p}{(}\PY{n+nb+bp}{self}\PY{p}{,}\PY{n}{node}\PY{p}{)}\PY{p}{:}
		\PY{k}{for} \PY{n}{nb} \PY{o+ow}{in} \PY{p}{[}\PY{n}{nbNodes} \PY{k}{for} \PY{n}{nbNodes} \PY{o+ow}{in}\PYZbs{}
				     \PY{n+nb+bp}{self}\PY{o}{.}\PY{n}{\PYZus{}adjList}\PY{p}{[}\PY{n}{node}\PY{p}{]}\PY{p}{]}\PY{p}{:}
		    \PY{n+nb+bp}{self}\PY{o}{.}\PY{n}{delEdge}\PY{p}{(}\PY{n}{nb}\PY{p}{,}\PY{n}{node}\PY{p}{)}
		\PY{k}{del} \PY{n+nb+bp}{self}\PY{o}{.}\PY{n}{\PYZus{}adjList}\PY{p}{[}\PY{n}{node}\PY{p}{]}
		\PY{n+nb+bp}{self}\PY{o}{.}\PY{n}{\PYZus{}nNodes} \PY{o}{-}\PY{o}{=} \PY{l+m+mi}{1}

	\PY{k}{def} \PY{n+nf}{delEdge}\PY{p}{(}\PY{n+nb+bp}{self}\PY{p}{,}\PY{n}{fromNode}\PY{p}{,}\PY{n}{toNode}\PY{p}{)}\PY{p}{:}
		\PY{n}{flag} \PY{o}{=} \PY{l+m+mi}{0}
		\PY{k}{if} \PY{n}{fromNode} \PY{o+ow}{in} \PY{n+nb+bp}{self}\PY{o}{.}\PY{n}{adjList}\PY{p}{(}\PY{n}{toNode}\PY{p}{)}\PY{p}{:}
		    \PY{n+nb+bp}{self}\PY{o}{.}\PY{n}{\PYZus{}adjList}\PY{p}{[}\PY{n}{fromNode}\PY{p}{]}\PY{o}{.}\PY{n}{remove}\PY{p}{(}\PY{n}{toNode}\PY{p}{)}
		    \PY{n+nb+bp}{self}\PY{o}{.}\PY{n}{\PYZus{}adjList}\PY{p}{[}\PY{n}{toNode}\PY{p}{]}\PY{o}{.}\PY{n}{remove}\PY{p}{(}\PY{n}{fromNode}\PY{p}{)}
		    \PY{n+nb+bp}{self}\PY{o}{.}\PY{n}{\PYZus{}nEdges} \PY{o}{-}\PY{o}{=} \PY{l+m+mi}{1}
		    \PY{n}{flag} \PY{o}{=} \PY{l+m+mi}{1}
		\PY{k}{return} \PY{n}{flag}

	\PY{n}{\PYZus{}\PYZus{}delEdge}\PY{o}{=}\PY{n}{delEdge}
\end{Verbatim}

In the above code snippet, the method \verb!delNode(self,node)! deletes a supplied node from 
the adjacency lists of all its neighbors, deletes the adjacency list of the node and finally 
decreases the number of nodes in the graph by one.
The second method, called \verb!delEdge(self,fromNode,toNode)!, deletes an edge from the graph.
Therefore, it modifies the adjacency lists of its terminal nodes and decreases the number of 
edges in the graph by one.

As a last point, we might want to print the graph. For that purpose we might implement the method 
\verb!__str__(self)! that is expected to compute a string representation of the graph object, to return 
the graph as string using the {\tt graphviz} {\tt dot}-language \cite{graphviz}:

\SourceCodeLines{99}
\begin{Verbatim}[fontfamily=txtt,commandchars=\\\{\}]
	\PY{k}{def} \PY{n+nf}{\PYZus{}\PYZus{}str\PYZus{}\PYZus{}}\PY{p}{(}\PY{n+nb+bp}{self}\PY{p}{)}\PY{p}{:}
		\PY{n}{string}  \PY{o}{=} \PY{l+s}{'}\PY{l+s}{graph G \PYZob{}}\PY{l+s+se}{\PYZbs{}}
\PY{l+s}{		  }\PY{l+s+se}{\PYZbs{}n}\PY{l+s}{  rankdir=LR;}\PY{l+s+se}{\PYZbs{}}
\PY{l+s}{		  }\PY{l+s+se}{\PYZbs{}n}\PY{l+s}{  node [shape = circle,size=0.5];}\PY{l+s+se}{\PYZbs{}}
\PY{l+s}{		  }\PY{l+s+se}{\PYZbs{}n}\PY{l+s}{  // graph attributes:}\PY{l+s+se}{\PYZbs{}}
\PY{l+s}{		  }\PY{l+s+se}{\PYZbs{}n}\PY{l+s}{  // nNodes=}\PY{l+s+si}{\PYZpc{}d}\PY{l+s+se}{\PYZbs{}}
\PY{l+s}{		  }\PY{l+s+se}{\PYZbs{}n}\PY{l+s}{  // nEdges=}\PY{l+s+si}{\PYZpc{}d}\PY{l+s+se}{\PYZbs{}}
\PY{l+s}{		  }\PY{l+s+se}{\PYZbs{}n}\PY{l+s}{'}\PY{o}{\PYZpc{}}\PY{p}{(}\PY{n+nb+bp}{self}\PY{o}{.}\PY{n}{nNodes}\PY{p}{,}\PY{n+nb+bp}{self}\PY{o}{.}\PY{n}{nEdges}\PY{p}{)}
		
		\PY{c}{\PYZsh{} for a more clear presentation: }
		\PY{c}{\PYZsh{} write node list fist}
		\PY{n}{string} \PY{o}{+}\PY{o}{=} \PY{l+s}{'}\PY{l+s+se}{\PYZbs{}n}\PY{l+s}{  // node-list:}\PY{l+s+se}{\PYZbs{}n}\PY{l+s}{'}
		\PY{k}{for} \PY{n}{n} \PY{o+ow}{in} \PY{n+nb+bp}{self}\PY{o}{.}\PY{n}{V}\PY{p}{:}
		  \PY{n}{string} \PY{o}{+}\PY{o}{=} \PY{l+s}{'}\PY{l+s}{  }\PY{l+s+si}{\PYZpc{}s}\PY{l+s}{; // deg=}\PY{l+s+si}{\PYZpc{}d}\PY{l+s+se}{\PYZbs{}n}\PY{l+s}{'}\PY{o}{\PYZpc{}}\PYZbs{}
					\PY{p}{(}\PY{n+nb}{str}\PY{p}{(}\PY{n}{n}\PY{p}{)}\PY{p}{,}\PY{n+nb+bp}{self}\PY{o}{.}\PY{n}{deg}\PY{p}{(}\PY{n}{n}\PY{p}{)}\PY{p}{)}

		\PY{c}{\PYZsh{} write edge list second }
		\PY{n}{string} \PY{o}{+}\PY{o}{=} \PY{l+s}{'}\PY{l+s+se}{\PYZbs{}n}\PY{l+s}{  // edge-list:}\PY{l+s+se}{\PYZbs{}n}\PY{l+s}{'}
		\PY{k}{for} \PY{n}{n1} \PY{o+ow}{in} \PY{n+nb+bp}{self}\PY{o}{.}\PY{n}{V}\PY{p}{:}
		  \PY{k}{for} \PY{n}{n2} \PY{o+ow}{in} \PY{n+nb+bp}{self}\PY{o}{.}\PY{n}{adjList}\PY{p}{(}\PY{n}{n1}\PY{p}{)}\PY{p}{:}
		    \PY{k}{if} \PY{n}{n1}\PY{o}{<}\PY{n}{n2}\PY{p}{:}
		      \PY{n}{string} \PY{o}{+}\PY{o}{=} \PY{l+s}{'}\PY{l+s}{  }\PY{l+s+si}{\PYZpc{}s}\PY{l+s}{ -- }\PY{l+s+si}{\PYZpc{}s}\PY{l+s}{ [len=1.5];}\PY{l+s+se}{\PYZbs{}n}\PY{l+s}{'}\PY{o}{\PYZpc{}}\PYZbs{}
					\PY{p}{(}\PY{n+nb}{str}\PY{p}{(}\PY{n}{n1}\PY{p}{)}\PY{p}{,}\PY{n+nb}{str}\PY{p}{(}\PY{n}{n2}\PY{p}{)}\PY{p}{)}
		\PY{n}{string} \PY{o}{+}\PY{o}{=} \PY{l+s}{'}\PY{l+s}{\PYZcb{}}\PY{l+s}{'}
		\PY{k}{return} \PY{n}{string}
\end{Verbatim}

A small script that illustrates some of the graph class functionality is easily written (\verb!graphExample.py!):
\SourceCodeLines{99}
\begin{Verbatim}[fontfamily=txtt,commandchars=\\\{\}]
 \PY{k+kn}{from} \PY{n+nn}{undirGraph\PYZus{}oop} \PY{k+kn}{import} \PY{o}{*}
	
 \PY{n}{g} \PY{o}{=} \PY{n}{myGraph}\PY{p}{(}\PY{p}{)}

 \PY{n}{g}\PY{o}{.}\PY{n}{addEdge}\PY{p}{(}\PY{l+m+mi}{1}\PY{p}{,}\PY{l+m+mi}{1}\PY{p}{)}
 \PY{n}{g}\PY{o}{.}\PY{n}{addEdge}\PY{p}{(}\PY{l+m+mi}{1}\PY{p}{,}\PY{l+m+mi}{2}\PY{p}{)}
 \PY{n}{g}\PY{o}{.}\PY{n}{addEdge}\PY{p}{(}\PY{l+m+mi}{1}\PY{p}{,}\PY{l+m+mi}{3}\PY{p}{)}
 \PY{n}{g}\PY{o}{.}\PY{n}{addEdge}\PY{p}{(}\PY{l+m+mi}{2}\PY{p}{,}\PY{l+m+mi}{3}\PY{p}{)}
 \PY{n}{g}\PY{o}{.}\PY{n}{addEdge}\PY{p}{(}\PY{l+m+mi}{2}\PY{p}{,}\PY{l+m+mi}{4}\PY{p}{)}

 \PY{k}{print} \PY{n}{g}
\end{Verbatim}

Invoking the script on the command-line yields the graph in terms of the {\tt dot}-language:

\SourceCodeLines{99}
\begin{Verbatim}[commandchars=\\\{\}]
 > python graphExample.py
 graph G \PY{o}{\PYZob{}}		           
  \PY{n+nv}{rankdir}\PY{o}{=}LR;			   
  node \PY{o}{[}\PY{n+nv}{shape} \PY{o}{=} circle,size\PY{o}{=}0.5\PY{o}{]};		           
  // graph attributes:		           
  // \PY{n+nv}{nNodes}\PY{o}{=}4		           
  // \PY{n+nv}{nEdges}\PY{o}{=}4			   

  // node-list:
  1; // \PY{n+nv}{deg}\PY{o}{=}2
  2; // \PY{n+nv}{deg}\PY{o}{=}3
  3; // \PY{n+nv}{deg}\PY{o}{=}2
  4; // \PY{n+nv}{deg}\PY{o}{=}1

  // edge-list:
  1 -- 2 \PY{o}{[}\PY{n+nv}{len}\PY{o}{=}1.5\PY{o}{]};
  1 -- 3 \PY{o}{[}\PY{n+nv}{len}\PY{o}{=}1.5\PY{o}{]};
  2 -- 3 \PY{o}{[}\PY{n+nv}{len}\PY{o}{=}1.5\PY{o}{]};
  2 -- 4 \PY{o}{[}\PY{n+nv}{len}\PY{o}{=}1.5\PY{o}{]};
 \PY{o}{\PYZcb{}}
\end{Verbatim}

Using the graph drawing command-line tools provided by the {\tt graphviz} library, the graph can 
be post-processed to yield Fig.\ \ref{fig:adjList_graph}(b). For that, simply pipe the {\tt dot}-language description
of the graph to a file, say \verb!graph.dot! and post-process the file via 
\verb!neato -Tpdf graph.dot > graph.pdf! to generate the pdf file that contains a visual representation of the graph. 

Further, we might use the class {\tt myGraph} as an base class to set up a subclass
{\tt myWeightedGraph}, allowing to handle graphs with edge weights. The respective 
class definition along with the default constructor for an instance of the class 
and some basic methods are listed below:
\SourceCodeLines{99}
\begin{Verbatim}[fontfamily=txtt,commandchars=\\\{\}]
 \PY{k}{class} \PY{n+nc}{myWeightedGraph}\PY{p}{(}\PY{n}{myGraph}\PY{p}{)}\PY{p}{:}
	\PY{k}{def} \PY{n+nf}{\PYZus{}\PYZus{}init\PYZus{}\PYZus{}}\PY{p}{(}\PY{n+nb+bp}{self}\PY{p}{)}\PY{p}{:}
		\PY{n}{myGraph}\PY{o}{.}\PY{n}{\PYZus{}\PYZus{}init\PYZus{}\PYZus{}}\PY{p}{(}\PY{n+nb+bp}{self}\PY{p}{)}
		\PY{n+nb+bp}{self}\PY{o}{.}\PY{n}{\PYZus{}wgt} \PY{o}{=} \PY{p}{\PYZob{}}\PY{p}{\PYZcb{}}

	\PY{n+nd}{@property}
	\PY{k}{def} \PY{n+nf}{E}\PY{p}{(}\PY{n+nb+bp}{self}\PY{p}{)}\PY{p}{:} 
		\PY{k}{return} \PY{n+nb+bp}{self}\PY{o}{.}\PY{n}{\PYZus{}wgt}\PY{o}{.}\PY{n}{keys}\PY{p}{(}\PY{p}{)}

	\PY{k}{def} \PY{n+nf}{wgt}\PY{p}{(}\PY{n+nb+bp}{self}\PY{p}{,}\PY{n}{fromNode}\PY{p}{,}\PY{n}{toNode}\PY{p}{)}\PY{p}{:}
		\PY{n}{sortedEdge} \PY{o}{=} \PY{p}{(}\PY{n+nb}{min}\PY{p}{(}\PY{n}{fromNode}\PY{p}{,}\PY{n}{toNode}\PY{p}{)}\PY{p}{,}\PYZbs{}
			      \PY{n+nb}{max}\PY{p}{(}\PY{n}{fromNode}\PY{p}{,}\PY{n}{toNode}\PY{p}{)}\PY{p}{)}
		\PY{k}{return} \PY{n+nb+bp}{self}\PY{o}{.}\PY{n}{\PYZus{}wgt}\PY{p}{[}\PY{n}{sortedEdge}\PY{p}{]}

	\PY{k}{def} \PY{n+nf}{setWgt}\PY{p}{(}\PY{n+nb+bp}{self}\PY{p}{,}\PY{n}{fromNode}\PY{p}{,}\PY{n}{toNode}\PY{p}{,}\PY{n}{wgt}\PY{p}{)}\PY{p}{:}
		\PY{k}{if} \PY{n}{toNode} \PY{o+ow}{in} \PY{n+nb+bp}{self}\PY{o}{.}\PY{n}{adjList}\PY{p}{(}\PY{n}{fromNode}\PY{p}{)}\PY{p}{:}
			\PY{n}{sortedEdge} \PY{o}{=} \PY{p}{(}\PY{n+nb}{min}\PY{p}{(}\PY{n}{fromNode}\PY{p}{,}\PY{n}{toNode}\PY{p}{)}\PY{p}{,}\PYZbs{}
				      \PY{n+nb}{max}\PY{p}{(}\PY{n}{fromNode}\PY{p}{,}\PY{n}{toNode}\PY{p}{)}\PY{p}{)}
			\PY{n+nb+bp}{self}\PY{o}{.}\PY{n}{\PYZus{}wgt}\PY{p}{[}\PY{n}{sortedEdge}\PY{p}{]}\PY{o}{=}\PY{n}{wgt}
\end{Verbatim}

Therein, the way in which the new class \verb!myWeightedGraph! incorporates the existing class \verb!myGraph!
is an example for \emph{inheritance}. In this manner, \verb!myWeightedGraph! extends the existing graph
class by edge weights that will be stored in the dictionary \verb!self._wgt!. In the default constructor, 
the latter is introduced as a private class attribute. Further, the code snippet shows three new methods
that are implemented to extend the base class. For the edge weight dictionary, it is necessary to 
explicitly store the edges, hence it comes at no extra 'cost' to provide a method (\verb!E(self)!) that 
serves to return a list of all edges in the graph. Further, the method \verb!wgt(self,fromNode,toNode)!
reports the weight of an edge if this exists already and the method \verb!setWgt(self,fromNode,toNode,wgt)!
sets the weight of an edge if the latter exists already.

Unfortunately, the existing base class methods for adding and deleting edges cannot be used in the 
'extended' context of weighted graphs. They have to be redefined in order to also make an entry
in the edge weight dictionary once an edge is constructed.
As an example for the OOP aspect of function \emph{overloading}, this is illustrated in the following code snippet:
\SourceCodeLines{99}
\begin{Verbatim}[fontfamily=txtt,commandchars=\\\{\}]
	\PY{k}{def} \PY{n+nf}{addEdge}\PY{p}{(}\PY{n+nb+bp}{self}\PY{p}{,}\PY{n}{fromNode}\PY{p}{,}\PY{n}{toNode}\PY{p}{,}\PY{n}{wgt}\PY{o}{=}\PY{l+m+mi}{1}\PY{p}{)}\PY{p}{:}
		\PY{k}{if} \PY{n+nb+bp}{self}\PY{o}{.}\PY{n}{\PYZus{}myGraph\PYZus{}\PYZus{}addEdge}\PY{p}{(}\PY{n}{fromNode}\PY{p}{,}\PY{n}{toNode}\PY{p}{)}\PY{p}{:}
			\PY{n+nb+bp}{self}\PY{o}{.}\PY{n}{setWgt}\PY{p}{(}\PY{n}{fromNode}\PY{p}{,}\PY{n}{toNode}\PY{p}{,}\PY{n}{wgt}\PY{p}{)}

	\PY{k}{def} \PY{n+nf}{delEdge}\PY{p}{(}\PY{n+nb+bp}{self}\PY{p}{,}\PY{n}{fromNode}\PY{p}{,}\PY{n}{toNode}\PY{p}{)}\PY{p}{:}
		\PY{k}{if} \PY{n+nb+bp}{self}\PY{o}{.}\PY{n}{\PYZus{}myGraph\PYZus{}\PYZus{}delEdge}\PY{p}{(}\PY{n}{fromNode}\PY{p}{,}\PY{n}{toNode}\PY{p}{)}\PY{p}{:}
			\PY{n}{sortedEdge} \PY{o}{=} \PY{p}{(}\PY{n+nb}{min}\PY{p}{(}\PY{n}{fromNode}\PY{p}{,}\PY{n}{toNode}\PY{p}{)}\PY{p}{,}\PYZbs{}
				      \PY{n+nb}{max}\PY{p}{(}\PY{n}{fromNode}\PY{p}{,}\PY{n}{toNode}\PY{p}{)}\PY{p}{)}
			\PY{k}{del} \PY{n+nb+bp}{self}\PY{o}{.}\PY{n}{\PYZus{}wgt}\PY{p}{[}\PY{n}{sortedEdge}\PY{p}{]}
\end{Verbatim}

There are two things to note. First, an odd-named method \verb!_myGraph__addEdge()! 
is called. This simply refers to the local copy \verb!__addEdge()! of the \verb!addEdge()! method on the level of 
the base class (hence the preceeing \verb!_myGraph!). This is still intact, even though the 
method \verb!addEdge()! is overloaded on the current level of the class hierarchy. Second, the
previously introduced edge-existence flag is utilized to make an entry in the weight dictionary only
if the edge is constructed during the current function call. In order to change an edge weight 
later on, the \verb!setWgt()! method must be used.
\smallskip

This completes the exemplary implementation of a graph data structure to illustrate the three OOP techniques
of data encapsulation, inheritance, and function overloading.
\smallskip

\subsection{A simple histogram data structure}
\label{subsect:hist_oop}

An example that is more useful for the purpose of data analysis is presented below (\verb!myHist_oop.py!). 
We will implement a histogram data structure that performs the same task as the 
\verb!hist_linBinning()! function implemented in Section \ref{sect:hist}, only in 
OOP style.

\newpage

\vspace*{-2ex}
\SourceCodeLines{99}
\begin{Verbatim}[fontfamily=txtt,commandchars=\\\{\}]
 \PY{k+kn}{from} \PY{n+nn}{math} \PY{k+kn}{import} \PY{n}{sqrt}\PY{p}{,}\PY{n}{floor}

 \PY{k}{class} \PY{n+nc}{simpleStats}\PY{p}{(}\PY{n+nb}{object}\PY{p}{)}\PY{p}{:}
	\PY{k}{def} \PY{n+nf}{\PYZus{}\PYZus{}init\PYZus{}\PYZus{}}\PY{p}{(}\PY{n+nb+bp}{self}\PY{p}{)}\PY{p}{:}
		\PY{n+nb+bp}{self}\PY{o}{.}\PY{n}{\PYZus{}N}    \PY{o}{=}\PY{l+m+mi}{0}	
		\PY{n+nb+bp}{self}\PY{o}{.}\PY{n}{\PYZus{}av}   \PY{o}{=}\PY{l+m+mf}{0.}   
		\PY{n+nb+bp}{self}\PY{o}{.}\PY{n}{\PYZus{}Q}    \PY{o}{=}\PY{l+m+mf}{0.}   
		
	\PY{k}{def} \PY{n+nf}{add}\PY{p}{(}\PY{n+nb+bp}{self}\PY{p}{,}\PY{n}{val}\PY{p}{)}\PY{p}{:}
		\PY{n+nb+bp}{self}\PY{o}{.}\PY{n}{\PYZus{}N}\PY{o}{+}\PY{o}{=}\PY{l+m+mi}{1}
		\PY{n}{dum} \PY{o}{=} \PY{n}{val}\PY{o}{-}\PY{n+nb+bp}{self}\PY{o}{.}\PY{n}{\PYZus{}av}
		\PY{n+nb+bp}{self}\PY{o}{.}\PY{n}{\PYZus{}av}\PY{o}{+}\PY{o}{=} \PY{n+nb}{float}\PY{p}{(}\PY{n}{dum}\PY{p}{)}\PY{o}{/}\PY{n+nb+bp}{self}\PY{o}{.}\PY{n}{\PYZus{}N}
		\PY{n+nb+bp}{self}\PY{o}{.}\PY{n}{\PYZus{}Q}\PY{o}{+}\PY{o}{=}\PY{n}{dum}\PY{o}{*}\PY{p}{(}\PY{n}{val}\PY{o}{-}\PY{n+nb+bp}{self}\PY{o}{.}\PY{n}{\PYZus{}av}\PY{p}{)}
	
        \PY{n+nd}{@property}
	\PY{k}{def} \PY{n+nf}{av}\PY{p}{(}\PY{n+nb+bp}{self}\PY{p}{)}\PY{p}{:} \PY{k}{return} \PY{n+nb+bp}{self}\PY{o}{.}\PY{n}{\PYZus{}av}
	
        \PY{n+nd}{@property}
	\PY{k}{def} \PY{n+nf}{sDev}\PY{p}{(}\PY{n+nb+bp}{self}\PY{p}{)}\PY{p}{:} \PY{k}{return} \PY{n}{sqrt}\PY{p}{(}\PY{n+nb+bp}{self}\PY{o}{.}\PY{n}{\PYZus{}Q}\PY{o}{/}\PY{p}{(}\PY{n+nb+bp}{self}\PY{o}{.}\PY{n}{\PYZus{}N}\PY{o}{-}\PY{l+m+mi}{1}\PY{p}{)}\PY{p}{)}
		
        \PY{n+nd}{@property}
	\PY{k}{def} \PY{n+nf}{sErr}\PY{p}{(}\PY{n+nb+bp}{self}\PY{p}{)}\PY{p}{:} \PY{k}{return} \PY{n+nb+bp}{self}\PY{o}{.}\PY{n}{sDev}\PY{o}{/}\PY{n}{sqrt}\PY{p}{(}\PY{n+nb+bp}{self}\PY{o}{.}\PY{n}{\PYZus{}N}\PY{p}{)}
	
	
 \PY{k}{class} \PY{n+nc}{myHist}\PY{p}{(}\PY{n+nb}{object}\PY{p}{)}\PY{p}{:}

	\PY{k}{def} \PY{n+nf}{\PYZus{}\PYZus{}init\PYZus{}\PYZus{}}\PY{p}{(}\PY{n+nb+bp}{self}\PY{p}{,}\PY{n}{xMin}\PY{p}{,}\PY{n}{xMax}\PY{p}{,}\PY{n}{nBins}\PY{p}{)}\PY{p}{:}
		\PY{n+nb+bp}{self}\PY{o}{.}\PY{n}{\PYZus{}min}   \PY{o}{=} \PY{n+nb}{min}\PY{p}{(}\PY{n}{xMin}\PY{p}{,}\PY{n}{xMax}\PY{p}{)}
		\PY{n+nb+bp}{self}\PY{o}{.}\PY{n}{\PYZus{}max}   \PY{o}{=} \PY{n+nb}{max}\PY{p}{(}\PY{n}{xMin}\PY{p}{,}\PY{n}{xMax}\PY{p}{)}
		\PY{n+nb+bp}{self}\PY{o}{.}\PY{n}{\PYZus{}cts}   \PY{o}{=} \PY{l+m+mi}{0}
		\PY{n+nb+bp}{self}\PY{o}{.}\PY{n}{\PYZus{}nBins} \PY{o}{=} \PY{n}{nBins}
		
		\PY{n+nb+bp}{self}\PY{o}{.}\PY{n}{\PYZus{}bin}   \PY{o}{=} \PY{p}{[}\PY{l+m+mi}{0}\PY{p}{]}\PY{o}{*}\PY{p}{(}\PY{n}{nBins}\PY{o}{+}\PY{l+m+mi}{2}\PY{p}{)}
		\PY{n+nb+bp}{self}\PY{o}{.}\PY{n}{\PYZus{}norm}  \PY{o}{=} \PY{n+nb+bp}{None}
		\PY{n+nb+bp}{self}\PY{o}{.}\PY{n}{\PYZus{}dx}    \PY{o}{=} \PYZbs{}
                   \PY{n+nb}{float}\PY{p}{(}\PY{n+nb+bp}{self}\PY{o}{.}\PY{n}{\PYZus{}max}\PY{o}{-}\PY{n+nb+bp}{self}\PY{o}{.}\PY{n}{\PYZus{}min}\PY{p}{)}\PY{o}{/}\PY{n+nb+bp}{self}\PY{o}{.}\PY{n}{\PYZus{}nBins}

		\PY{n+nb+bp}{self}\PY{o}{.}\PY{n}{\PYZus{}stats} \PY{o}{=} \PY{n}{simpleStats}\PY{p}{(}\PY{p}{)}
	
	\PY{k}{def} \PY{n+nf}{\PYZus{}\PYZus{}binId}\PY{p}{(}\PY{n+nb+bp}{self}\PY{p}{,}\PY{n}{val}\PY{p}{)}\PY{p}{:}   
		\PY{k}{return} \PY{n+nb}{int}\PY{p}{(}\PY{n}{floor}\PY{p}{(}\PY{p}{(}\PY{n}{val}\PY{o}{-}\PY{n+nb+bp}{self}\PY{o}{.}\PY{n}{\PYZus{}min}\PY{p}{)}\PY{o}{/}\PY{n+nb+bp}{self}\PY{o}{.}\PY{n}{\PYZus{}dx}\PY{p}{)}\PY{p}{)}

	\PY{k}{def} \PY{n+nf}{\PYZus{}\PYZus{}bdry}\PY{p}{(}\PY{n+nb+bp}{self}\PY{p}{,}\PY{n+nb}{bin}\PY{p}{)}\PY{p}{:}	  
		\PY{k}{return} \PY{n+nb+bp}{self}\PY{o}{.}\PY{n}{\PYZus{}min}\PY{o}{+}\PY{n+nb}{bin}\PY{o}{*}\PY{n+nb+bp}{self}\PY{o}{.}\PY{n}{\PYZus{}dx}\PY{p}{,}\PYZbs{}
                       \PY{n+nb+bp}{self}\PY{o}{.}\PY{n}{\PYZus{}min}\PY{o}{+}\PY{p}{(}\PY{n+nb}{bin}\PY{o}{+}\PY{l+m+mi}{1}\PY{p}{)}\PY{o}{*}\PY{n+nb+bp}{self}\PY{o}{.}\PY{n}{\PYZus{}dx}

	\PY{k}{def} \PY{n+nf}{\PYZus{}\PYZus{}GErr}\PY{p}{(}\PY{n+nb+bp}{self}\PY{p}{,}\PY{n+nb}{bin}\PY{p}{)}\PY{p}{:} 
		\PY{n}{q}\PY{o}{=}\PY{n+nb+bp}{self}\PY{o}{.}\PY{n}{\PYZus{}norm}\PY{p}{[}\PY{n+nb}{bin}\PY{p}{]}\PY{o}{*}\PY{n+nb+bp}{self}\PY{o}{.}\PY{n}{\PYZus{}dx}
		\PY{k}{return} \PY{n}{sqrt}\PY{p}{(}\PY{n}{q}\PY{o}{*}\PY{p}{(}\PY{l+m+mi}{1}\PY{o}{-}\PY{n}{q}\PY{p}{)}\PY{o}{/}\PY{p}{(}\PY{n+nb+bp}{self}\PY{o}{.}\PY{n}{\PYZus{}cts}\PY{o}{-}\PY{l+m+mi}{1}\PY{p}{)}\PY{p}{)}\PY{o}{/}\PY{n+nb+bp}{self}\PY{o}{.}\PY{n}{\PYZus{}dx} 

	\PY{k}{def} \PY{n+nf}{\PYZus{}\PYZus{}normalize}\PY{p}{(}\PY{n+nb+bp}{self}\PY{p}{)}\PY{p}{:}
		\PY{n+nb+bp}{self}\PY{o}{.}\PY{n}{\PYZus{}norm} \PY{o}{=} \PY{n+nb}{map}\PY{p}{(}\PY{k}{lambda} \PY{n}{x}\PY{p}{:} \PY{n+nb}{float}\PY{p}{(}\PY{n}{x}\PY{p}{)}\PY{o}{/}\PYZbs{}
                            \PY{p}{(}\PY{n+nb+bp}{self}\PY{o}{.}\PY{n}{\PYZus{}cts}\PY{o}{*}\PY{n+nb+bp}{self}\PY{o}{.}\PY{n}{\PYZus{}dx}\PY{p}{)}\PY{p}{,}\PY{n+nb+bp}{self}\PY{o}{.}\PY{n}{\PYZus{}bin}\PY{p}{)}

        \PY{n+nd}{@property}
	\PY{k}{def} \PY{n+nf}{av}\PY{p}{(}\PY{n+nb+bp}{self}\PY{p}{)}\PY{p}{:} \PY{k}{return} \PY{n+nb+bp}{self}\PY{o}{.}\PY{n}{\PYZus{}stats}\PY{o}{.}\PY{n}{av}
	
        \PY{n+nd}{@property}
	\PY{k}{def} \PY{n+nf}{sDev}\PY{p}{(}\PY{n+nb+bp}{self}\PY{p}{)}\PY{p}{:} \PY{k}{return} \PY{n+nb+bp}{self}\PY{o}{.}\PY{n}{\PYZus{}stats}\PY{o}{.}\PY{n}{sDev}
	
        \PY{n+nd}{@property}
	\PY{k}{def} \PY{n+nf}{sErr}\PY{p}{(}\PY{n+nb+bp}{self}\PY{p}{)}\PY{p}{:} \PY{k}{return} \PY{n+nb+bp}{self}\PY{o}{.}\PY{n}{\PYZus{}stats}\PY{o}{.}\PY{n}{sErr}

	\PY{k}{def} \PY{n+nf}{addValue}\PY{p}{(}\PY{n+nb+bp}{self}\PY{p}{,} \PY{n}{val}\PY{p}{)}\PY{p}{:}
		\PY{n+nb+bp}{self}\PY{o}{.}\PY{n}{\PYZus{}stats}\PY{o}{.}\PY{n}{add}\PY{p}{(}\PY{n}{val}\PY{p}{)}
		\PY{k}{if} \PY{n}{val}\PY{o}{<}\PY{n+nb+bp}{self}\PY{o}{.}\PY{n}{\PYZus{}min}\PY{p}{:}
			\PY{n+nb+bp}{self}\PY{o}{.}\PY{n}{\PYZus{}bin}\PY{p}{[}\PY{n+nb+bp}{self}\PY{o}{.}\PY{n}{\PYZus{}nBins}\PY{p}{]}\PY{o}{+}\PY{o}{=}\PY{l+m+mi}{1}
		\PY{k}{elif} \PY{n}{val}\PY{o}{>}\PY{o}{=}\PY{n+nb+bp}{self}\PY{o}{.}\PY{n}{\PYZus{}max}\PY{p}{:}
			\PY{n+nb+bp}{self}\PY{o}{.}\PY{n}{\PYZus{}bin}\PY{p}{[}\PY{n+nb+bp}{self}\PY{o}{.}\PY{n}{\PYZus{}nBins}\PY{o}{+}\PY{l+m+mi}{1}\PY{p}{]}\PY{o}{+}\PY{o}{=}\PY{l+m+mi}{1}
		\PY{k}{else}\PY{p}{:}
			\PY{n+nb+bp}{self}\PY{o}{.}\PY{n}{\PYZus{}bin}\PY{p}{[}\PY{n+nb+bp}{self}\PY{o}{.}\PY{n}{\PYZus{}\PYZus{}binId}\PY{p}{(}\PY{n}{val}\PY{p}{)}\PY{p}{]}\PY{o}{+}\PY{o}{=}\PY{l+m+mi}{1}
		\PY{n+nb+bp}{self}\PY{o}{.}\PY{n}{\PYZus{}cts}\PY{o}{+}\PY{o}{=}\PY{l+m+mi}{1}

	\PY{k}{def} \PY{n+nf}{\PYZus{}\PYZus{}str\PYZus{}\PYZus{}}\PY{p}{(}\PY{n+nb+bp}{self}\PY{p}{)}\PY{p}{:}
		\PY{l+s+sd}{"""represent histogram as string"""}
		\PY{n+nb+bp}{self}\PY{o}{.}\PY{n}{\PYZus{}\PYZus{}normalize}\PY{p}{(}\PY{p}{)}
		\PY{n}{myStr} \PY{o}{=}\PY{l+s}{'}\PY{l+s}{\PYZsh{} min  = }\PY{l+s+si}{\PYZpc{}lf}\PY{l+s+se}{\PYZbs{}n}\PY{l+s}{'}\PY{o}{\PYZpc{}}\PY{p}{(}\PY{n+nb}{float}\PY{p}{(}\PY{n+nb+bp}{self}\PY{o}{.}\PY{n}{\PYZus{}min}\PY{p}{)}\PY{p}{)}
		\PY{n}{myStr}\PY{o}{+}\PY{o}{=}\PY{l+s}{'}\PY{l+s}{\PYZsh{} max  = }\PY{l+s+si}{\PYZpc{}lf}\PY{l+s+se}{\PYZbs{}n}\PY{l+s}{'}\PY{o}{\PYZpc{}}\PY{p}{(}\PY{n+nb}{float}\PY{p}{(}\PY{n+nb+bp}{self}\PY{o}{.}\PY{n}{\PYZus{}max}\PY{p}{)}\PY{p}{)}
		\PY{n}{myStr}\PY{o}{+}\PY{o}{=}\PY{l+s}{'}\PY{l+s}{\PYZsh{} dx   = }\PY{l+s+si}{\PYZpc{}lf}\PY{l+s+se}{\PYZbs{}n}\PY{l+s}{'}\PY{o}{\PYZpc{}}\PY{p}{(}\PY{n+nb}{float}\PY{p}{(}\PY{n+nb+bp}{self}\PY{o}{.}\PY{n}{\PYZus{}dx}\PY{p}{)}\PY{p}{)}
		\PY{n}{myStr}\PY{o}{+}\PY{o}{=}\PY{l+s}{'}\PY{l+s}{\PYZsh{} av   = }\PY{l+s+si}{\PYZpc{}lf}\PY{l+s}{ (sErr = }\PY{l+s+si}{\PYZpc{}lf}\PY{l+s}{)}\PY{l+s+se}{\PYZbs{}n}\PY{l+s}{'}\PY{o}{\PYZpc{}}\PYZbs{}
                                       \PY{p}{(}\PY{n+nb+bp}{self}\PY{o}{.}\PY{n}{av}\PY{p}{,}\PY{n+nb+bp}{self}\PY{o}{.}\PY{n}{sErr}\PY{p}{)}
		\PY{n}{myStr}\PY{o}{+}\PY{o}{=}\PY{l+s}{'}\PY{l+s}{\PYZsh{} sDev = }\PY{l+s+si}{\PYZpc{}lf}\PY{l+s+se}{\PYZbs{}n}\PY{l+s}{'}\PY{o}{\PYZpc{}}\PY{p}{(}\PY{n+nb+bp}{self}\PY{o}{.}\PY{n}{sDev}\PY{p}{)}
		\PY{n}{myStr}\PY{o}{+}\PY{o}{=}\PY{l+s}{'}\PY{l+s}{\PYZsh{} xL xH p(xL<=x<xH) Gauss\PYZus{}err}\PY{l+s+se}{\PYZbs{}n}\PY{l+s}{'}
		\PY{k}{for} \PY{n+nb}{bin} \PY{o+ow}{in} \PY{n+nb}{range}\PY{p}{(}\PY{n+nb+bp}{self}\PY{o}{.}\PY{n}{\PYZus{}nBins}\PY{p}{)}\PY{p}{:} 
		  \PY{n}{low}\PY{p}{,}\PY{n}{up}\PY{o}{=}\PY{n+nb+bp}{self}\PY{o}{.}\PY{n}{\PYZus{}\PYZus{}bdry}\PY{p}{(}\PY{n+nb}{bin}\PY{p}{)}
	          \PY{n}{myStr}\PY{o}{+}\PY{o}{=}\PY{l+s}{'}\PY{l+s+si}{\PYZpc{}lf}\PY{l+s}{ }\PY{l+s+si}{\PYZpc{}lf}\PY{l+s}{ }\PY{l+s+si}{\PYZpc{}lf}\PY{l+s}{ }\PY{l+s+si}{\PYZpc{}lf}\PY{l+s+se}{\PYZbs{}n}\PY{l+s}{'}\PY{o}{\PYZpc{}}\PYZbs{}
                    \PY{p}{(}\PY{n}{low}\PY{p}{,}\PY{n}{up}\PY{p}{,}\PY{n+nb+bp}{self}\PY{o}{.}\PY{n}{\PYZus{}norm}\PY{p}{[}\PY{n+nb}{bin}\PY{p}{]}\PY{p}{,} \PY{n+nb+bp}{self}\PY{o}{.}\PY{n}{\PYZus{}\PYZus{}GErr}\PY{p}{(}\PY{n+nb}{bin}\PY{p}{)}\PY{p}{)}
		\PY{k}{return} \PY{n}{myStr}
\end{Verbatim}

\newpage

In the above code snippet, the first class definition (line 3) defines the data structure
\verb!simple statistics!. It implements methods to compute the average, 
standard deviation and standard error of the supplied values in single-pass fashion. 
Numerical values are added by calling the class method \verb!add()!, defined in lines 9--13.

The second class definition (line 25) belongs to the histogram data structure.
The class implements a histogram using linear binning of the
supplied data. It computes the probability density function that 
approximates the supplied data and further provides some 
simple statistics such as average, standard deviation and 
standard error.

Upon initialization an instance of the class requires
three numerical values, as evident from the default constructor for an instance 
of the class (lines 27--38). These are 
\verb!xMin!, a lower bound (inclusive) on values that are considered in order to
accumulate frequency statistics for the histogram,
\verb!xMax!, an upper bound (exclusive) on values that are considered to
accumulate frequency statistics for the histogram, and 
\verb!nBins!, the number of bins used to set up the histogram.
All values $x$ that are either smaller than \verb!xMin! or greater or equal to \verb!xMax! are not considered
for the calculation of frequency statistics, but are considered for 
the normalization of the histogram to yield a probability 
density function (pdf). Further, all values are used to compute
the average, standard deviation and standard error.

In the scope of the histogram class, several ``private'' methods are 
defined (signified by preceding double underscores), that, e.g.,
compute the integer identifier of the bin that corresponds to a 
supplied numerical value (lines 40--41), 
compute upper and lower boundaries of a bin (lines 43--45),
compute a Gaussian error bar for each bin (lines 47--49), and normalize the 
frequency statistics to yield a probability density (lines 51--53).

Further, the user interface consists of four methods, the 
first three of which are \verb!av()!, \verb!sDev()!, and \verb!sErr()!, providing access to 
the mean, standard deviation, and standard error of the supplied 
values, respectively.
Most important, lines 64--72 implement the fourth method \verb!addValue()!,
that adds a value to the histogram and updates the frequencies and overall 
statistics accordingly.
Finally, lines 74--88 implement the string representation of the 
histogram class.

As an example, we might generate a histogram from the data of the $2D$
random walks using the \verb!myHist_oop! class (\verb!rw2D_hist_oop.py!):

\SourceCodeLines{99}
\begin{Verbatim}[fontfamily=txtt,commandchars=\\\{\}]
 \PY{k+kn}{import} \PY{n+nn}{sys}
 \PY{k+kn}{from} \PY{n+nn}{myHist\PYZus{}oop} \PY{k+kn}{import} \PY{n}{myHist}
 \PY{k+kn}{from} \PY{n+nn}{MCS2012\PYZus{}lib} \PY{k+kn}{import} \PY{n}{fetchData}

 \PY{k}{def} \PY{n+nf}{main}\PY{p}{(}\PY{p}{)}\PY{p}{:}
        \PY{c}{\PYZsh{}\PYZsh{} parse command line args}
        \PY{n}{fName} \PY{o}{=} \PY{n}{sys}\PY{o}{.}\PY{n}{argv}\PY{p}{[}\PY{l+m+mi}{1}\PY{p}{]}
        \PY{n}{col}   \PY{o}{=} \PY{n+nb}{int}\PY{p}{(}\PY{n}{sys}\PY{o}{.}\PY{n}{argv}\PY{p}{[}\PY{l+m+mi}{2}\PY{p}{]}\PY{p}{)}
        \PY{n}{nBins} \PY{o}{=} \PY{n+nb}{int}\PY{p}{(}\PY{n}{sys}\PY{o}{.}\PY{n}{argv}\PY{p}{[}\PY{l+m+mi}{3}\PY{p}{]}\PY{p}{)}

        \PY{c}{\PYZsh{}\PYZsh{} get data from file}
        \PY{n}{myData} \PY{o}{=} \PY{n}{fetchData}\PY{p}{(}\PY{n}{fName}\PY{p}{,}\PY{n}{col}\PY{p}{,}\PY{n+nb}{float}\PY{p}{)}

        \PY{c}{\PYZsh{}\PYZsh{} assemble histogram}
        \PY{n}{h} \PY{o}{=} \PY{n}{myHist}\PY{p}{(}\PY{n+nb}{min}\PY{p}{(}\PY{n}{myData}\PY{p}{)}\PY{p}{,}\PY{n+nb}{max}\PY{p}{(}\PY{n}{myData}\PY{p}{)}\PY{p}{,}\PY{n}{nBins}\PY{p}{)}
        \PY{k}{for} \PY{n}{val} \PY{o+ow}{in} \PY{n}{myData}\PY{p}{:}
                \PY{n}{h}\PY{o}{.}\PY{n}{addValue}\PY{p}{(}\PY{n}{val}\PY{p}{)}
        \PY{k}{print} \PY{n}{h}

 \PY{n}{main}\PY{p}{(}\PY{p}{)}
\end{Verbatim}

\noindent
Invoking the script on the commandline yields:
\SourceCodeLines{9}
\begin{Verbatim}
 > python rw2D_hist_oop.py ./2dRW_N100_n100000.dat 1 22
 # min  = 0.039408
 # max  = 35.154325
 # dx   = 1.596133
 # av   = 8.892287 (sErr = 0.014664)
 # sDev = 4.637152
 # xLow xHigh p(xLow <= x < xHigh) Gaussian_error
 0.039408 1.635541 0.016371 0.000316
 ...
 31.962060 33.558193 0.000019 0.000011
 33.558193 35.154325 0.000006 0.000006
\end{Verbatim}



\section[Increase your efficiency using Scientific Python ({\tt scipy})]{Scientific Python ({\tt SciPy}): increase your efficiency in a blink!}

Basically all analyses illustrated above (apart from bootstrap resampling) can be  
done without the need to implement the core routines on your own.
The {\tt python} community is very active and hence there is a vast number of 
well tested and community approved modules, one of which might come in handy 
to accomplish your task.
A particular open source library that I use a lot in order to get certain 
things done quick and clean is {\tt scipy}, a {\tt python} module tailored towards scientific 
computing \cite{scipy}.
It is easy to use and offers lots of scientific tools, organized in sub-modules
that address, e.g., statistics, optimization, and linear algebra.

However, it is nevertheless beneficial to implement things on your own from time to 
time. I do not think of this as ``reinventing the wheel'', but more as 
an exercise that assures you get the coding routine needed as a programmer.
The more routine you have, the more efficient your planning, implementing, and 
testing cycles will become.

\subsection{Basic data analysis using {\tt scipy}}

As an example to illustrate the capabilities of the {\tt scipy} module, we will
re-perform and extend the analysis of the data for the $2D$ random walks.
To be more precise, we will compute basic parameters related to the data, 
generate a histogram to yield a pdf of the geometric distance $R_N$ the walkers
traveled after $N=100$ steps, and perform a $\chi^2$ goodness-of-fit test to
assess whether the approximate pdf appears to be equivalent to a Rayleigh 
distribution. A full script to perform the above analyses using 
the {\tt scipy} module reads (\verb!rw2d_analysis_scipy.py!):
\SourceCodeLines{99}
\begin{Verbatim}[fontfamily=txtt,commandchars=\\\{\}]
 \PY{k+kn}{import} \PY{n+nn}{sys}
 \PY{k+kn}{import} \PY{n+nn}{scipy}
 \PY{k+kn}{import} \PY{n+nn}{scipy.stats} \PY{k+kn}{as} \PY{n+nn}{sciStat}
 \PY{k+kn}{from} \PY{n+nn}{MCS2012\PYZus{}lib} \PY{k+kn}{import} \PY{n}{fetchData} 

 \PY{k}{def} \PY{n+nf}{myPdf}\PY{p}{(}\PY{n}{x}\PY{p}{)}\PY{p}{:} \PY{k}{return} \PY{n}{sciStat}\PY{o}{.}\PY{n}{rayleigh}\PY{o}{.}\PY{n}{pdf}\PY{p}{(}\PY{n}{x}\PY{p}{,}\PY{n}{scale}\PY{o}{=}\PY{n}{sigma}\PY{p}{)}
 \PY{k}{def} \PY{n+nf}{myCdf}\PY{p}{(}\PY{n}{x}\PY{p}{)}\PY{p}{:} \PY{k}{return} \PY{n}{sciStat}\PY{o}{.}\PY{n}{rayleigh}\PY{o}{.}\PY{n}{cdf}\PY{p}{(}\PY{n}{x}\PY{p}{,}\PY{n}{scale}\PY{o}{=}\PY{n}{sigma}\PY{p}{)}

 \PY{n}{fileName} \PY{o}{=} \PY{n}{sys}\PY{o}{.}\PY{n}{argv}\PY{p}{[}\PY{l+m+mi}{1}\PY{p}{]}
 \PY{n}{nBins}  \PY{o}{=} \PY{n+nb}{int}\PY{p}{(}\PY{n}{sys}\PY{o}{.}\PY{n}{argv}\PY{p}{[}\PY{l+m+mi}{2}\PY{p}{]}\PY{p}{)} 

 \PY{n}{rawData}  \PY{o}{=} \PY{n}{fetchData}\PY{p}{(}\PY{n}{fileName}\PY{p}{,}\PY{l+m+mi}{1}\PY{p}{,}\PY{n+nb}{float}\PY{p}{)}

 \PY{k}{print} \PY{l+s}{'}\PY{l+s}{\PYZsh{}\PYZsh{} Basic statistics using scipy}\PY{l+s}{'}
 \PY{k}{print} \PY{l+s}{'}\PY{l+s}{\PYZsh{} data file:}\PY{l+s}{'}\PY{p}{,} \PY{n}{fileName}
 \PY{n}{N}     \PY{o}{=} \PY{n+nb}{len}\PY{p}{(}\PY{n}{rawData}\PY{p}{)}
 \PY{n}{sigma} \PY{o}{=} \PY{n}{scipy}\PY{o}{.}\PY{n}{sqrt}\PY{p}{(}\PY{n+nb}{sum}\PY{p}{(}\PY{n+nb}{map}\PY{p}{(}\PY{k}{lambda} \PY{n}{x}\PY{p}{:}\PY{n}{x}\PY{o}{*}\PY{n}{x}\PY{p}{,}\PY{n}{rawData}\PY{p}{)}\PY{p}{)}\PY{o}{*}\PY{l+m+mf}{0.5}\PY{o}{/}\PY{n}{N}\PY{p}{)}
 \PY{n}{av}    \PY{o}{=} \PY{n}{scipy}\PY{o}{.}\PY{n}{mean}\PY{p}{(}\PY{n}{rawData}\PY{p}{)}
 \PY{n}{sDev}  \PY{o}{=} \PY{n}{scipy}\PY{o}{.}\PY{n}{std}\PY{p}{(}\PY{n}{rawData}\PY{p}{)}
 \PY{k}{print} \PY{l+s}{'}\PY{l+s}{\PYZsh{} N     =}\PY{l+s}{'}\PY{p}{,} \PY{n}{N}
 \PY{k}{print} \PY{l+s}{'}\PY{l+s}{\PYZsh{} av    =}\PY{l+s}{'}\PY{p}{,} \PY{n}{av} 
 \PY{k}{print} \PY{l+s}{'}\PY{l+s}{\PYZsh{} sDev  =}\PY{l+s}{'}\PY{p}{,} \PY{n}{sDev} 
 \PY{k}{print} \PY{l+s}{'}\PY{l+s}{\PYZsh{} sigma =}\PY{l+s}{'}\PY{p}{,} \PY{n}{sigma}

 \PY{k}{print} \PY{l+s}{'}\PY{l+s}{\PYZsh{}\PYZsh{} Histogram using scipy}\PY{l+s}{'}
 \PY{n}{limits} \PY{o}{=} \PY{p}{(}\PY{n+nb}{min}\PY{p}{(}\PY{n}{rawData}\PY{p}{)}\PY{p}{,}\PY{n+nb}{max}\PY{p}{(}\PY{n}{rawData}\PY{p}{)}\PY{p}{)}
 \PY{n}{freqObs}\PY{p}{,}\PY{n}{xMin}\PY{p}{,}\PY{n}{dx}\PY{p}{,}\PY{n}{nOut} \PY{o}{=}\PYZbs{}
        \PY{n}{sciStat}\PY{o}{.}\PY{n}{histogram}\PY{p}{(}\PY{n}{rawData}\PY{p}{,}\PY{n}{nBins}\PY{p}{,}\PY{n}{limits}\PY{p}{)}
 \PY{n}{bdry} \PY{o}{=} \PY{p}{[}\PY{p}{(}\PY{n}{xMin}\PY{o}{+}\PY{n}{i}\PY{o}{*}\PY{n}{dx} \PY{p}{,}\PY{n}{xMin}\PY{o}{+}\PY{p}{(}\PY{n}{i}\PY{o}{+}\PY{l+m+mi}{1}\PY{p}{)}\PY{o}{*}\PY{n}{dx}\PY{p}{)} \PY{k}{for} \PY{n}{i} \PY{o+ow}{in} \PY{n+nb}{range}\PY{p}{(}\PY{n}{nBins}\PY{p}{)}\PY{p}{]}

 \PY{k}{print} \PY{l+s}{'}\PY{l+s}{\PYZsh{} nBins = }\PY{l+s}{'}\PY{p}{,} \PY{n}{nBins}
 \PY{k}{print} \PY{l+s}{'}\PY{l+s}{\PYZsh{} xMin  = }\PY{l+s}{'}\PY{p}{,} \PY{n}{xMin} 
 \PY{k}{print} \PY{l+s}{'}\PY{l+s}{\PYZsh{} dx    = }\PY{l+s}{'}\PY{p}{,} \PY{n}{dx}   
 \PY{k}{print} \PY{l+s}{'}\PY{l+s}{\PYZsh{} (binCenter) (pdf-observed) (pdf-rayleigh distrib.)}\PY{l+s}{'}
 \PY{k}{for} \PY{n}{i} \PY{o+ow}{in} \PY{n+nb}{range}\PY{p}{(}\PY{n}{nBins}\PY{p}{)}\PY{p}{:}
        \PY{n}{x} \PY{o}{=} \PY{l+m+mf}{0.5}\PY{o}{*}\PY{p}{(}\PY{n}{bdry}\PY{p}{[}\PY{n}{i}\PY{p}{]}\PY{p}{[}\PY{l+m+mi}{0}\PY{p}{]}\PY{o}{+}\PY{n}{bdry}\PY{p}{[}\PY{n}{i}\PY{p}{]}\PY{p}{[}\PY{l+m+mi}{1}\PY{p}{]}\PY{p}{)} 
        \PY{k}{print} \PY{n}{x}\PY{p}{,} \PY{n}{freqObs}\PY{p}{[}\PY{n}{i}\PY{p}{]}\PY{o}{/}\PY{p}{(}\PY{n}{N}\PY{o}{*}\PY{n}{dx}\PY{p}{)}\PY{p}{,} \PY{n}{myPdf}\PY{p}{(}\PY{n}{x}\PY{p}{)} 

 \PY{k}{print} \PY{l+s}{'}\PY{l+s}{\PYZsh{}\PYZsh{} Chi2 - test using scipy}\PY{l+s}{'}
 \PY{n}{freqExp} \PY{o}{=} \PY{n}{scipy}\PY{o}{.}\PY{n}{array}\PY{p}{(}
           \PY{p}{[}\PY{n}{N}\PY{o}{*}\PY{p}{(}\PY{n}{myCdf}\PY{p}{(}\PY{n}{x}\PY{p}{[}\PY{l+m+mi}{1}\PY{p}{]}\PY{p}{)}\PY{o}{-}\PY{n}{myCdf}\PY{p}{(}\PY{n}{x}\PY{p}{[}\PY{l+m+mi}{0}\PY{p}{]}\PY{p}{)}\PY{p}{)} \PY{k}{for} \PY{n}{x} \PY{o+ow}{in} \PY{n}{bdry}\PY{p}{]}\PY{p}{)}

 \PY{n}{chi2}\PY{p}{,}\PY{n}{p} \PY{o}{=} \PY{n}{sciStat}\PY{o}{.}\PY{n}{chisquare}\PY{p}{(}\PY{n}{freqObs}\PY{p}{,}\PY{n}{freqExp}\PY{p}{,}\PY{n}{ddof}\PY{o}{=}\PY{l+m+mi}{1}\PY{p}{)}
 \PY{k}{print} \PY{l+s}{'}\PY{l+s}{\PYZsh{} chi2     = }\PY{l+s}{'}\PY{p}{,} \PY{n}{chi2}
 \PY{k}{print} \PY{l+s}{'}\PY{l+s}{\PYZsh{} chi2/dof = }\PY{l+s}{'}\PY{p}{,} \PY{n}{chi2}\PY{o}{/}\PY{p}{(}\PY{n}{nBins}\PY{o}{-}\PY{l+m+mi}{1}\PY{p}{)}
 \PY{k}{print} \PY{l+s}{'}\PY{l+s}{\PYZsh{} pVal     = }\PY{l+s}{'}\PY{p}{,} \PY{n}{p}
\end{Verbatim}

In lines 2 and 3, the {\tt scipy} module and its submodule {\tt scipy.stats}
are imported. The latter contains a large number of discrete and continuous probability distributions 
(that allow to draw random variates, or to evaluate the respective pdf and cdf)
and a huge library of statistical functions.
E.g., in lines 18 and 19 we use the {\tt scipy} functions \verb!scipy.mean(rawData)!
and \verb!scipy.std(rawData)!, respectively, to compute the mean and standard deviation of the 
individual values of $R_N$ stored in the array \verb!rawData!. 

However, be aware that the function \verb!scipy.std(x)! computes the 
square-root of the uncorrected variance ${\rm uVar}(x) = {\rm av}([x -{\rm av}(x)]^2)$.
Hence, the {\tt scipy} implementation of the above function yields no unbiased
estimate of the standard deviation. An unbiased estimate of the
standard deviation can be obtained using \verb!scipy.sqrt(N/(N-1))*scipy.std(x)!, where
\verb!N=len(x)!.

Next, in lines 25--37 a pdf of the geometric distance $R_N$ is computed. 
Therefore, in lines 27--28, a frequency histogram for the raw data 
is calculated using the function \verb!scipy.stats.histogram()!.
As arguments, the function takes the array \verb!rawData! with events that 
are to be binned, the integer \verb!nBins! specifying the number of histogram 
bins, and the tuple limits, holding the upper and lower limits of the histogram range.
The histogram function returns the four-tuple \verb!freqObs,xMin,dx,nOut!, where
\verb!freqObs! is an array containing the number of events for the individual bins, 
\verb!xMin! is the lower boundary value of the smallest bin, \verb!dx! is the bin width,
and \verb!nOut! counts the number of events that fall not within the range of the histogram.
In line 29 the boundary values of all individual bins are reconstructed from the output
of the histogram function by means of a list comprehension (i.e.\ an inline looping construct).
In lines 35--37, the frequencies are transformed to probability densities and listed as 
function of the bin centers.
As discussed earlier (see the example on the continuous $2D$ random walk in 
Section \ref{subsect:continuousVars}), for $N$ large enough, $R_N$ is properly described 
by the Rayleigh distribution with shape parameter $\sigma$.
In order to facilitate a comparison of the observed pdf to the pdf thus expected, the 
{\tt scipy} probability distribution \verb!scipy.stats.rayleigh.pdf(x,shape=sigma)!, 
abbreviated as \verb!myPdf(x)! in line 6, is listed, too.

From the corresponding cumulative distribution function (abbreviated \verb!myCdf(x)! in line 7),
\verb!scipy.stats.rayleigh.cdf(x,shape=sigma)!, the expected number of events within each bin
is computed in lines 40--41. Finally, in line 43, and by using the function \verb!scipy.stats.chisquare()!,
a chi-square goodness-of-fit test is 
performed to assess whether the observed distribution is in agreement with the expected 
limiting distribution.
In the above function, the first two arguments refer to the observed and expected frequencies 
of events, respectively, and the parameter \verb!ddof! serves to adjust the number 
of degrees of freedom for the $p$-value test. Here, the assignment \verb!ddof=1! ensures
that the number of degrees of freedom is set to \verb!nBins-1!, to account for the 
fact that the sum of the expected events has been re-normalized to $N$ 
(to further improve on the reliability of the chi-square test, one might further 
take care of those bins that have very small observed frequencies, see Section \ref{sect:chisquare}).

Invoking the above script for the data on $n=10^5$ individual $2D$ random walks of $N=100$ steps yields:
\SourceCodeLines{99}
\begin{Verbatim}
 > python rw2d_analysis_scipy.py 2dRW_N100_n100000.dat 30
 ## Basic statistics using scipy
 # data file: ../EX_2DrandWalk/2dRW_N100_n100000.dat
 # N     = 100000
 # av    = 8.89228749223
 # sDev  = 4.63712834129
 # sigma = 7.09139394259
 ## Histogram using scipy
 # nBins =  30
 # xMin  =  0.0394084293183
 # dx    =  1.17049722938
 # (binCenter) (pdf-observed) (pdf-rayleigh distrib.)
 0.624657044008 0.0122084868219 0.0123735271797
 1.79515427339 0.0346604835806 0.0345718951134
 ...
 34.569076696 1.70867555241e-05 4.75360576683e-06
 ## Chi2 - test using scipy
 # chi2     =  36.6268754006
 # chi2/dof =  1.26299570347
 # pVal     =  0.127319735292
\end{Verbatim}
Hence, in view of a reduced chi-square value of $\approx 1.26$ and a $p$-value of 
$O(10^{-1})$, the hypotheses that the pdf is consistent with a Rayleigh distribution
might be accepted.

\subsection{Least-squares parameter estimation using {\tt scipy}}
Up to now, we only have an estimate of the scale parameter $\sigma$ without error bars.
In the following, we will compute an error estimate $\Delta \sigma$ using two different methods:
(i)  via resampling using the formula $\sigma^2=(2n)^{-1}\sum_{i=0}^{n-1}R_{N,i}^2$, and, 
(ii) via resampling of least-squares parameter estimates obtained by fitting a model function to the data. 
In the latter case, independent 
estimates of $\sigma$ are obtained by fitting a function of the form $f(x)=(x/\sigma)^2\, \exp[-x^2/(2\sigma^2)]$,
see Eq.\ (\ref{eq:Rayleigh}), to resampled data sets.
This will be accomplished by minimizing the \emph{sum-of-squares error}, referring to the difference
between observations (based on the data) and expectations (based on a model).  
Here, method (ii) is used to illustrate the {\tt scipy} submodule
\verb!scipy.optimize!. A small script that performs tasks (i) and (ii) is illustrated below (\verb!scipy_fitSigma.py!):
\SourceCodeLines{99}
\begin{Verbatim}[fontfamily=txtt,commandchars=\\\{\}]
 \PY{k+kn}{import} \PY{n+nn}{sys}
 \PY{k+kn}{import} \PY{n+nn}{scipy}
 \PY{k+kn}{import} \PY{n+nn}{scipy.optimize} \PY{k+kn}{as} \PY{n+nn}{sciOpt}
 \PY{k+kn}{import} \PY{n+nn}{scipy.stats} \PY{k+kn}{as} \PY{n+nn}{sciStat} 
 \PY{k+kn}{from} \PY{n+nn}{MCS2012\PYZus{}lib} \PY{k+kn}{import} \PY{n}{fetchData}\PY{p}{,} \PY{n}{bootstrap}

 \PY{k}{def} \PY{n+nf}{sigma}\PY{p}{(}\PY{n}{rawData}\PY{p}{)}\PY{p}{:}
        \PY{n}{N} \PY{o}{=} \PY{n+nb}{len}\PY{p}{(}\PY{n}{rawData}\PY{p}{)}
        \PY{n}{sum2} \PY{o}{=} \PY{n+nb}{sum}\PY{p}{(}\PY{n+nb}{map}\PY{p}{(}\PY{k}{lambda} \PY{n}{x}\PY{p}{:}\PY{n}{x}\PY{o}{*}\PY{n}{x}\PY{p}{,}\PY{n}{rawData}\PY{p}{)}\PY{p}{)}
        \PY{k}{return} \PY{n}{scipy}\PY{o}{.}\PY{n}{sqrt}\PY{p}{(}\PY{n}{sum2}\PY{o}{*}\PY{l+m+mf}{0.5}\PY{o}{/}\PY{n}{N}\PY{p}{)}

 \PY{k}{def} \PY{n+nf}{myFit}\PY{p}{(}\PY{n}{data}\PY{p}{,}\PY{n}{nBins}\PY{o}{=}\PY{l+m+mi}{30}\PY{p}{)}\PY{p}{:}

        \PY{n}{freqObs}\PY{p}{,}\PY{n}{xMin}\PY{p}{,}\PY{n}{dx}\PY{p}{,}\PY{n}{nOut} \PY{o}{=} \PY{n}{sciStat}\PY{o}{.}\PY{n}{histogram}\PY{p}{(}\PY{n}{data}\PY{p}{,}\PY{n}{nBins}\PY{p}{)}

        \PY{n}{N} \PY{o}{=} \PY{n+nb}{len}\PY{p}{(}\PY{n}{data}\PY{p}{)}
        \PY{n}{xVals} \PY{o}{=} \PY{p}{[}\PY{n}{xMin} \PY{o}{+} \PY{p}{(}\PY{n}{i}\PY{o}{+}\PY{l+m+mf}{0.5}\PY{p}{)}\PY{o}{*}\PY{n}{dx} \PY{k}{for} \PY{n}{i} \PY{o+ow}{in} \PY{n+nb}{range}\PY{p}{(}\PY{n}{nBins}\PY{p}{)}\PY{p}{]}
        \PY{n}{yVals} \PY{o}{=} \PY{p}{[}\PY{n}{freqObs}\PY{p}{[}\PY{n}{i}\PY{p}{]}\PY{o}{/}\PY{p}{(}\PY{n}{N}\PY{o}{*}\PY{n}{dx}\PY{p}{)} \PY{k}{for} \PY{n}{i} \PY{o+ow}{in} \PY{n+nb}{range}\PY{p}{(}\PY{n}{nBins}\PY{p}{)}\PY{p}{]}
        
        \PY{n}{fitFunc} \PY{o}{=} \PY{k}{lambda} \PY{n}{s}\PY{p}{,}\PY{n}{x}\PY{p}{:} \PY{n}{sciStat}\PY{o}{.}\PY{n}{rayleigh}\PY{o}{.}\PY{n}{pdf}\PY{p}{(}\PY{n}{x}\PY{p}{,}\PY{n}{scale}\PY{o}{=}\PY{n}{s}\PY{p}{)}
        \PY{n}{objFunc} \PY{o}{=} \PY{k}{lambda} \PY{n}{s}\PY{p}{,}\PY{n}{x}\PY{p}{,}\PY{n}{y}\PY{p}{:} \PY{p}{(}\PY{n}{y} \PY{o}{-} \PY{n}{fitFunc}\PY{p}{(}\PY{n}{s}\PY{p}{,}\PY{n}{x}\PY{p}{)}\PY{p}{)}

        \PY{n}{s0}\PY{o}{=}\PY{l+m+mf}{10.}
        \PY{n}{s}\PY{p}{,}\PY{n}{flag} \PY{o}{=} \PY{n}{sciOpt}\PY{o}{.}\PY{n}{leastsq}\PY{p}{(}\PY{n}{objFunc}\PY{p}{,}\PY{n}{s0}\PY{p}{,}\PY{n}{args}\PY{o}{=}\PY{p}{(}\PY{n}{xVals}\PY{p}{,}\PY{n}{yVals}\PY{p}{)}\PY{p}{)}

        \PY{k}{return} \PY{n}{s}\PY{p}{[}\PY{l+m+mi}{0}\PY{p}{]}

 \PY{n}{fileName}  \PY{o}{=} \PY{n}{sys}\PY{o}{.}\PY{n}{argv}\PY{p}{[}\PY{l+m+mi}{1}\PY{p}{]}
 \PY{n}{nBootSamp} \PY{o}{=} \PY{n+nb}{int}\PY{p}{(}\PY{n}{sys}\PY{o}{.}\PY{n}{argv}\PY{p}{[}\PY{l+m+mi}{2}\PY{p}{]}\PY{p}{)} 

 \PY{n}{rawData}  \PY{o}{=} \PY{n}{fetchData}\PY{p}{(}\PY{n}{fileName}\PY{p}{,}\PY{l+m+mi}{1}\PY{p}{,}\PY{n+nb}{float}\PY{p}{)}
 \PY{n}{s1}\PY{p}{,}\PY{n}{s1\PYZus{}Err}  \PY{o}{=} \PY{n}{bootstrap}\PY{p}{(}\PY{n}{rawData}\PY{p}{,}\PY{n}{sigma}\PY{p}{,}\PY{n}{nBootSamp}\PY{p}{)}
 \PY{n}{s2}\PY{p}{,}\PY{n}{s2\PYZus{}Err}  \PY{o}{=} \PY{n}{bootstrap}\PY{p}{(}\PY{n}{rawData}\PY{p}{,}\PY{n}{myFit}\PY{p}{,}\PY{n}{nBootSamp}\PY{p}{)}

 \PY{k}{print} \PY{l+s}{"}\PY{l+s}{sigma = }\PY{l+s+si}{\PYZpc{}lf}\PY{l+s}{ +/- }\PY{l+s+si}{\PYZpc{}lf}\PY{l+s}{"}\PY{o}{\PYZpc{}}\PY{p}{(}\PY{n}{s1}\PY{p}{,}\PY{n}{s1\PYZus{}Err}\PY{p}{)}
 \PY{k}{print} \PY{l+s}{"}\PY{l+s}{sigma = }\PY{l+s+si}{\PYZpc{}lf}\PY{l+s}{ +/- }\PY{l+s+si}{\PYZpc{}lf}\PY{l+s}{ (least-squares fit)}\PY{l+s}{"}\PY{o}{\PYZpc{}}\PY{p}{(}\PY{n}{s2}\PY{p}{,}\PY{n}{s2\PYZus{}Err}\PY{p}{)}
\end{Verbatim}

In lines 3 and 4, the optimization and statistics submodules are imported under the 
names \verb!sciOpt! and \verb!sciStat!, respectively.
Lines 7--10 implement the simple formula that characterizes method (i), while 
lines 12--26 define a routine that uses the {\tt scipy} function \verb!scipy.optimize.leastsq()!
in order to fit the function $f(x)$ to the pdf that approximates the raw data.
More precisely, in line 14, the raw data is binned to obtain a frequency histogram using
\verb!nBins! histogram bins. Note that if no limits for the histogram range are specified, the 
full range of data is considered. As evident from the argument list in line 14, the value 
of \verb!nBins! defaults to 30. In lines 17 and 18, the observed $x$-values (i.e.\ bin centers) 
and $y$-values (probability densitites for the respective bins) are 
computed using list comprehensions. Line 20 defines the fit function, and line 21 defines 
the objective funtion as the vertical difference between the observed data and the fit function.
In line 24, a least-squares fit is performed using the function \verb!scipy.optimize.leastsq()!.
As shown here, the least-squares function takes three arguments (in principle, it takes up to 13 arguments!):
\verb!objFunc!, referring to the objective funtion, \verb!s0!, specifying an initial guess
for the fit parameter (in principle, this might also be a tuple), and the tuple \verb!args!, 
comprising all additional parameters to the objective function.
If the least-squares function terminates succesfully, indicated by a return value of \verb!flag! $>0$, 
the tuple \verb!s! contains the found solution.

Invoking the script on the commandline using $20$ bootstrap samples yields the output:
\SourceCodeLines{9}
\begin{Verbatim}
 > python scipy_fitSigma.py 2dRW_N100_n100000.dat 20
 sigma = 7.091394 +/- 0.010746
 sigma = 7.131461 +/- 0.013460 (least-squares fit)
\end{Verbatim}
Albeit not within error bars, the results are in reasonable agreement (considering the comparatively small sample size;
we further made no attempt to study the influence of the number of histogram bins).
The results of the least-squares fit is shown in Figs.\ \ref{fig:scipy_LSE_MLE_abcd}(a),(b).

A similar analysis for a synthetic data set consisting of a number of $10^5$ standard normal 
variates, where also the bootstrap resampling routine was implemented using the {\tt scipy} module,
is available at {\tt Scipy Central} \cite{scipy_central}, a code-sharing and software list site 
focused on Scientific Python. In that example, entitled ``Error estimates for fit parameters resulting from 
least-squares fits using bootstrap resampling'', the distribution of the resampled observable as 
well as the influence of the number of bootstrap data sets was studied.
Apparently, a number of $O(10)$ bootstrap data sets usually suffices in order to obtain a reasonable error estimate.

\subsection{Maximum-likelihood parameter estimation using {\tt scipy}}

Maximum-likelihood estimation (MLE) represents a standard approach to the problem
of \emph{parameter estimation}. Generally speaking, it is concerned with the
following \emph{inverse problem} \cite{myung2003}:
given the observed sample $x=\{x_i\}_{i=1}^N$ and a model, 
comprising a parametrized family of probability distributions, find the model
parameters that yield the pdf that is \emph{most likely} to have generated 
the data.

In this regard, let $p(x| \underline{\omega})$ specify the 
total probability density function (pdf) of observing the sample, given the 
$k$-component parameter vector $\underline{\omega}=(\omega_1,\ldots,\omega_k)$.
If the individual observations $x_i\in x$ are statistically independent, the 
total pdf is simply the product of all the single-observation pdfs:
\begin{align}
p(x| \underline{\omega})= \prod_{i=1}^N p_i(x_i| \underline{\omega}) \label{eq:fullPdf}
\end{align}
To solve the above inverse problem one makes use of the \emph{likelihood function}. 
From a conceptual point of view, the latter is obtained by interchanging the role of 
the data set and the parameters to yield the function  
\begin{align}
\mathcal{L}(\underline{\omega}| x) = p(x| \underline{\omega}).\label{eq:likelihood}
\end{align}
In the above equation, the rhs 
represents the pdf of the observed data
$x$ given the parameter vector $\underline{\omega}$, defined on the \emph{data scale}.
In contrast to this, the lhs 
represents the likelihood (i.e.\ an unnormalized probability measure)
of the parameter vector $\underline{\omega}$ given the observed data $x$, consequently defined on 
the \emph{parameter scale}. 
In general, $\mathcal{L}(\underline{\omega}| x)$ is a $k$-dimensional surface over a 
$k$-dimensional hyperplane spanned by the probability vector $\underline{\omega}$.

Now, the basic principle of maximum-likelihood estimation is that the target pdf is the one that
makes the observed sample $x$ ``most likely''.
Accordingly, the idea of MLE is to maximizes the likelihood function 
$\mathcal{L}(\underline{\omega}| x)$ for a given sample $x$ with respect to the parameter vector $\underline{\omega}$. 
The result of this optimization procedure is referred to as $\underline{\omega}_{\rm MLE}$.
Instead of maximizing the likelihood function itself, it is usually more convenient to 
maximize its logarithm, the log-likelihood function $\ln(\mathcal{L}(\underline{\omega}| x))$.
The respective transformation is monotonous, thus preserving the location of the maxima.

If $\underline{\omega}_{\rm MLE}$ exists, it must necessarily satisfy the so-called \emph{likelihood equation}:
\begin{align}
\left. \frac{\partial }{\partial \omega_i} \ln(\mathcal{L}(\underline{\omega}| x)) \right|_{\omega_{i,{\rm MLE}}} = 0, ~ i=1\ldots k. \label{eq:likelihoodEq}
\end{align}
To further ensure that $\ln(\mathcal{L}(\underline{\omega}| x))$ is a maximum, it needs to satisfy
\begin{align}
\left. \frac{\partial^2 }{\partial \omega_i^2} \ln(\mathcal{L}(\underline{\omega}| x)) \right|_{\omega_{i,{\rm MLE}}} < 0, ~  i=1\ldots k. \label{eq:likelihoodEqIsMax}
\end{align}
In some cases, i.e., when the model is sufficiently simple, the two equations above might yield an analytic solution
for the MLE parameter estimates \cite{clauset2007}.
However, from a more general point of view the MLE parameter estimate has to be obtained by numerical methods using 
non-linear optimization algorithms. The subsequent script (\verb!scipy_MLE_sigma.py!) shows how to 
cope with that task by using the scipy submodule 
\verb!scipy.optimize!:
\SourceCodeLines{99}
\begin{Verbatim}[fontfamily=txtt,commandchars=\\\{\}]
 \PY{k+kn}{import} \PY{n+nn}{sys}
 \PY{k+kn}{import} \PY{n+nn}{scipy}
 \PY{k+kn}{import} \PY{n+nn}{scipy.stats} \PY{k+kn}{as} \PY{n+nn}{sciStat}
 \PY{k+kn}{import} \PY{n+nn}{scipy.optimize} \PY{k+kn}{as} \PY{n+nn}{sciOpt}
 \PY{k+kn}{from} \PY{n+nn}{MCS2012\PYZus{}lib} \PY{k+kn}{import} \PY{n}{fetchData}\PY{p}{,} \PY{n}{bootstrap}

 \PY{k}{def} \PY{n+nf}{myMleEstimate}\PY{p}{(}\PY{n}{myFunc}\PY{p}{,}\PY{n}{par}\PY{p}{,}\PY{n}{data}\PY{p}{)}\PY{p}{:}

        \PY{k}{def} \PY{n+nf}{lnL\PYZus{}av}\PY{p}{(}\PY{n}{x}\PY{p}{,}\PY{n}{par}\PY{p}{)}\PY{p}{:}
           \PY{n}{N} \PY{o}{=} \PY{n+nb}{len}\PY{p}{(}\PY{n}{x}\PY{p}{)}
           \PY{n}{lnL} \PY{o}{=} \PY{l+m+mf}{0.}
           \PY{k}{for} \PY{n}{i} \PY{o+ow}{in} \PY{n+nb}{range}\PY{p}{(}\PY{n}{N}\PY{p}{)}\PY{p}{:}
              \PY{n}{lnL} \PY{o}{+}\PY{o}{=}  \PY{n}{scipy}\PY{o}{.}\PY{n}{log}\PY{p}{(}\PY{n}{myFunc}\PY{p}{(}\PY{n}{par}\PY{p}{,}\PY{n}{x}\PY{p}{[}\PY{n}{i}\PY{p}{]}\PY{p}{)}\PY{p}{)}
           \PY{k}{return} \PY{n}{lnL}\PY{o}{/}\PY{n}{N}

        \PY{n}{objFunc} \PY{o}{=} \PY{k}{lambda} \PY{n}{s}\PY{p}{:} \PY{o}{-}\PY{n}{lnL\PYZus{}av}\PY{p}{(}\PY{n}{data}\PY{p}{,}\PY{n}{s}\PY{p}{)}
        \PY{n}{par\PYZus{}mle} \PY{o}{=} \PY{n}{sciOpt}\PY{o}{.}\PY{n}{fmin}\PY{p}{(}\PY{n}{objFunc}\PY{p}{,}\PY{n}{par}\PY{p}{,}\PY{n}{disp}\PY{o}{=}\PY{l+m+mi}{0}\PY{p}{)}
        \PY{k}{return} \PY{n}{par\PYZus{}mle}

 \PY{n}{fileName}   \PY{o}{=} \PY{n}{sys}\PY{o}{.}\PY{n}{argv}\PY{p}{[}\PY{l+m+mi}{1}\PY{p}{]}
 \PY{n}{nBoot}      \PY{o}{=} \PY{n+nb}{int}\PY{p}{(}\PY{n}{sys}\PY{o}{.}\PY{n}{argv}\PY{p}{[}\PY{l+m+mi}{2}\PY{p}{]}\PY{p}{)} 

 \PY{n}{rawData}    \PY{o}{=} \PY{n}{fetchData}\PY{p}{(}\PY{n}{fileName}\PY{p}{,}\PY{l+m+mi}{1}\PY{p}{,}\PY{n+nb}{float}\PY{p}{)}

 \PY{n}{Rayleigh}   \PY{o}{=} \PY{k}{lambda} \PY{n}{s}\PY{p}{,}\PY{n}{x}\PY{p}{:} \PY{n}{sciStat}\PY{o}{.}\PY{n}{rayleigh}\PY{o}{.}\PY{n}{pdf}\PY{p}{(}\PY{n}{x}\PY{p}{,}\PY{n}{scale}\PY{o}{=}\PY{n}{s}\PY{p}{)}
 \PY{n}{objFunc}    \PY{o}{=} \PY{k}{lambda} \PY{n}{x}\PY{p}{:} \PY{n}{myMleEstimate}\PY{p}{(}\PY{n}{Rayleigh}\PY{p}{,}\PY{l+m+mf}{7.1}\PY{p}{,}\PY{n}{x}\PY{p}{)}

 \PY{n}{s\PYZus{}av}\PY{p}{,}\PY{n}{s\PYZus{}Err} \PY{o}{=} \PY{n}{bootstrap}\PY{p}{(}\PY{n}{rawData}\PY{p}{,}\PY{n}{objFunc}\PY{p}{,}\PY{n}{nBoot}\PY{p}{)}
 \PY{k}{print} \PY{l+s}{"}\PY{l+s}{\PYZsh{} sigma = }\PY{l+s+si}{\PYZpc{}lf}\PY{l+s}{ +/- }\PY{l+s+si}{\PYZpc{}lf}\PY{l+s}{ (MLE)}\PY{l+s}{"}\PY{o}{\PYZpc{}}\PY{p}{(}\PY{n}{s\PYZus{}av}\PY{p}{,}\PY{n}{s\PYZus{}Err}\PY{p}{)}
\end{Verbatim}

\begin{figure}
\centering
\includegraphics[width=\textwidth]{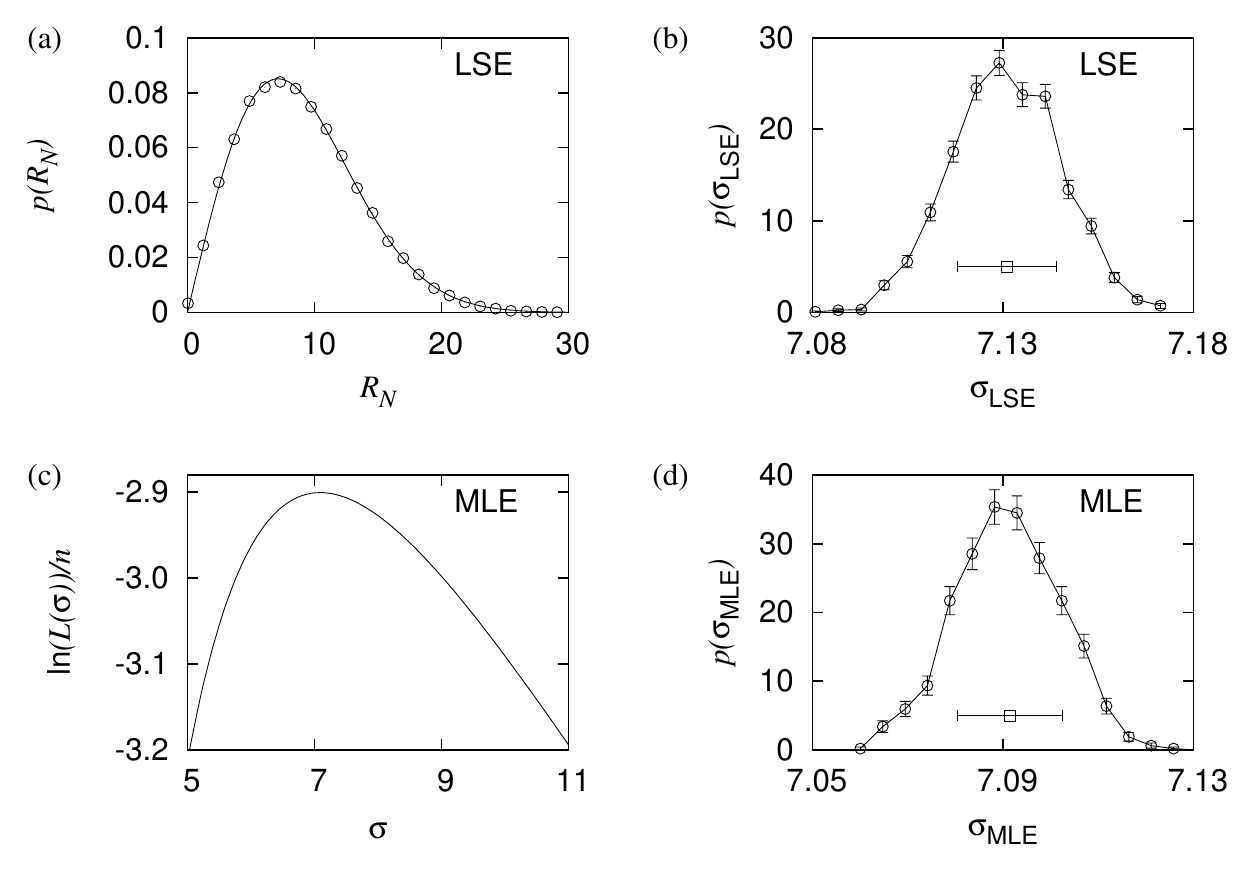}
\caption{
Results of the least-squares and maximum-likelihood parameter estimation approaches.
(a) Approximate pdf for the distance to the starting point (using $30$ 
bins), estimated from $n=10^5$ $2D$ random walks after $N=100$ steps (data points)
together with the most accurate fit to a Rayleigh distribution function 
obtained by a least-squares fit (solid line).
(b) Approximate pdf (histogram using 16 bins) of the resampled
parameter $\sigma$ as result of the bootstrap resampling procedure ($1000$ auxiliary data sets). 
The non-filled square indicates the result $\sigma=7.131(13)$.
(c) Log-likelihood function as function of the single parameter $\sigma$. As evident 
from the figure, the maximum is approximately located at $\sigma=7.09$.
(d) Approximate pdf (histogram using 16 bins) of the resampled
parameter $\sigma_{\rm MLE}$ as result of the bootstrap resampling procedure ($1000$ auxiliary data sets). 
The non-filled square indicates the result $\sigma=7.091(11)$.
\label{fig:scipy_LSE_MLE_abcd}}
\end{figure}

In lines 3 and 4, the optimization and statistics submodules are imported under the 
names \verb!sciStat! and \verb!sciOpt!, respectively.
Lines 7--18 implement the function \verb!myMleEstim!
which estimates the parameter \verb!par! for the model function \verb!myFunc! 
that yields the pdf that is most likely to have generated the sample \verb!data!.
For that purpose it maximizes the average log-likelihood per observation (defined in lines 9 through 14) 
in line 17 (more precisely, it minimizes the negative thereof).

Say, we aim to obtain MLE estimates for the parameter $\sigma$ that enters the Rayleigh distribution,
presumably describing the pdf of the distances $R_N$ traveled by the $2D$ random walk after $N$ 
steps. For that purpose, in line 25 the \verb!scipy! implementation of the Rayleigh distribution is 
used and declared as the (permanent) first argument of the \verb!myMleEstim! function (line 26).
Similarly, the initial guess for $\sigma$ is permanently set to the value $7.1$.
Finally, in order to obtain an error estimate on the model parameter $\sigma$, empirical bootstrap resampling
is used (line 28).

Invoking the script on the command-line using $20$ bootstrap samples yields the output:
\SourceCodeLines{9}
\begin{Verbatim}
 > python scipy_MLE_sigma.py 2dRW_N100_n100000.dat 20
 sigma = 7.091420 +/- 0.012335 (MLE)
\end{Verbatim}
In effect, this MLE parameter estimate appears to yield a slightly smaller numerical value for $\sigma$
as compared to the least-squares fit above.
The results of the MLE procedure is shown in Figs.\ \ref{fig:scipy_LSE_MLE_abcd}(c),(d).

To compare the least-squares parameter estimation (LSE) approach to the maxi\-mum-likelihood one, consider the objective
of both procedures:
In LSE one aims to obtain a \emph{most accurate} description of the data by means of the model, considering a minimization 
of the sum-of-squares error between observations (based on the data) and expectations (based on the model).
In MLE, one aims at those parameters that maximize the log-likelihood and consequently yield those parameters 
that have \emph{most likely} generated the data. Generally, MLE and LSE estimates differ from each other.
In any case, a MLE estimate should be preferred over an LSE estimate. 
Finally, note that for statistically independent observations that are normally distributed, the maximization of the 
log-likelihood is equivalent to a minimization of the sum-of-squares error function.

\subsection{Hierarchical cluster analysis using {\tt scipy}}

A cluster analysis might be used to classify a given set of ``objects'' 
with regard to their ``distance''. Therein, the measure of distance between either two 
objects can be quite general (not necessarily metric). 
E.g., below we will run a cluster analysis on a set of 
$6$ points located in a $2D$ plane,
wherein the euclidean distance between two points is used 
to measure their distance
(in this particular example the objects are vector-valued observations and
the distance measure is metric).

In order to perform a cluster analysis, many different
algorithmic schemes are available. Here we will illustrate one particular 
algorithm, falling into the realm of agglomerative clustering methods, that
generates a hierarchical clustering for a supplied distance matrix \cite{gan2007}. 
The respective algorithm assembles a hierarchical cluster starting from 
single-object clusters and can be found in the {\tt cluster} submodule
of {\tt scipy}. The particular function we will use here is called
\verb!scipy.cluster.hierarchy.linkage!.
The subsequent script (\verb!scipy_clusterAnalysis_cophDist.py!) comprises a basic example 
that illustrates the usage of the above function:
\SourceCodeLines{99}
\begin{Verbatim}[fontfamily=txtt,commandchars=\\\{\}]
 \PY{k+kn}{import} \PY{n+nn}{sys}
 \PY{k+kn}{import} \PY{n+nn}{scipy}
 \PY{k+kn}{import} \PY{n+nn}{scipy.cluster.hierarchy} \PY{k+kn}{as} \PY{n+nn}{sch}
 \PY{k+kn}{import} \PY{n+nn}{scipy.spatial.distance} \PY{k+kn}{as} \PY{n+nn}{ssd}
 \PY{k+kn}{from} \PY{n+nn}{dumpDendrogram} \PY{k+kn}{import} \PY{n}{droPyt\PYZus{}distMat\PYZus{}dendrogram\PYZus{}sciPy}

 \PY{k}{def} \PY{n+nf}{myDistMatrix\PYZus{}2D}\PY{p}{(}\PY{n}{N}\PY{p}{,}\PY{n}{s}\PY{p}{)}\PY{p}{:}
        \PY{n}{scipy}\PY{o}{.}\PY{n}{random}\PY{o}{.}\PY{n}{seed}\PY{p}{(}\PY{n}{s}\PY{p}{)}
        \PY{n}{xVec} \PY{o}{=} \PY{n}{scipy}\PY{o}{.}\PY{n}{rand}\PY{p}{(}\PY{n}{N}\PY{p}{,}\PY{l+m+mi}{2}\PY{p}{)}
        \PY{n}{Dij}  \PY{o}{=} \PY{n}{scipy}\PY{o}{.}\PY{n}{zeros}\PY{p}{(}\PY{p}{(}\PY{n}{N}\PY{p}{,}\PY{n}{N}\PY{p}{)}\PY{p}{,}\PY{n}{dtype}\PY{o}{=}\PY{l+s}{'}\PY{l+s}{d}\PY{l+s}{'}\PY{p}{)}
        \PY{k}{for} \PY{n}{i} \PY{o+ow}{in} \PY{n+nb}{range}\PY{p}{(}\PY{n}{N}\PY{p}{)}\PY{p}{:}
          \PY{k}{for} \PY{n}{j} \PY{o+ow}{in} \PY{n+nb}{range}\PY{p}{(}\PY{n}{N}\PY{p}{)}\PY{p}{:}
            \PY{n}{Dij}\PY{p}{[}\PY{n}{i}\PY{p}{,}\PY{n}{j}\PY{p}{]} \PY{o}{=} \PY{n}{ssd}\PY{o}{.}\PY{n}{euclidean}\PY{p}{(}\PY{n}{xVec}\PY{p}{[}\PY{n}{i}\PY{p}{]}\PY{p}{,}\PY{n}{xVec}\PY{p}{[}\PY{n}{j}\PY{p}{]}\PY{p}{)} 
        \PY{k}{return} \PY{n}{Dij}

 \PY{k}{def} \PY{n+nf}{main}\PY{p}{(}\PY{p}{)}\PY{p}{:}
        \PY{c}{\PYZsh{} construct distance matrix}
        \PY{n}{N} \PY{o}{=} \PY{l+m+mi}{6}
        \PY{n}{Dij\PYZus{}sq} \PY{o}{=} \PY{n}{myDistMatrix\PYZus{}2D}\PY{p}{(}\PY{n}{N}\PY{p}{,}\PY{l+m+mi}{0}\PY{p}{)}

        \PY{c}{\PYZsh{} obtain hierarchical clustering via scipy}
        \PY{n}{Dij\PYZus{}cd} \PY{o}{=} \PY{n}{ssd}\PY{o}{.}\PY{n}{squareform}\PY{p}{(}\PY{n}{Dij\PYZus{}sq}\PY{p}{)}
        \PY{n}{clusterResult} \PY{o}{=} \PY{n}{sch}\PY{o}{.}\PY{n}{linkage}\PY{p}{(}\PY{n}{Dij\PYZus{}cd}\PY{p}{,} \PY{n}{method}\PY{o}{=}\PY{l+s}{'}\PY{l+s}{average}\PY{l+s}{'}\PY{p}{)}
        \PY{n}{corr}\PY{p}{,}\PY{n}{Cij\PYZus{}cd}   \PY{o}{=} \PY{n}{sch}\PY{o}{.}\PY{n}{cophenet}\PY{p}{(}\PY{n}{clusterResult}\PY{p}{,}\PY{n}{Dij\PYZus{}cd}\PY{p}{)}
        \PY{n}{Cij\PYZus{}sq} \PY{o}{=} \PY{n}{ssd}\PY{o}{.}\PY{n}{squareform}\PY{p}{(}\PY{n}{Cij\PYZus{}cd}\PY{p}{)}

        \PY{c}{\PYZsh{} print dendrogram on top of cophenetic distance }
        \PY{c}{\PYZsh{} matrix to standard outstream}
        \PY{n}{droPyt\PYZus{}distMat\PYZus{}dendrogram\PYZus{}sciPy}\PY{p}{(}\PY{n}{Cij\PYZus{}sq}\PY{p}{,}\PY{n}{clusterResult}\PY{p}{,}\PY{n}{N}\PY{p}{)}

 \PY{n}{main}\PY{p}{(}\PY{p}{)}
\end{Verbatim}

In lines 3 and 4, two {\tt scipy} submodules are imported that 
are useful in the context of cluster analysis.
The first one, {\tt scipy.cluster}, contains
a further submodule, implementing a collection of functions for hierarchical 
clustering, referred to as {\tt scipy.cluster.hierarchy}.
It allows to compute distance matrices from vector-valued observations,
to generate hierarchical clusters from distance matrices, and to post-process
the results obtained therein.
The module {\tt scipy.spatial} implements various data structures that 
facilitate ``spatial'' computations. In particular, its submodule 
{\tt scipy.spatial.distance} contains functions that allow, e.g., to 
compute distance matrices from observations, and it implements
a bunch of different distance functions, measuring the distance between 
two vector-valued observations.
However, first and foremost, it allows to
convert full (i.e.\ square) distance matrices to condensed (i.e.\ upper triangular) 
ones and vice versa.
This is important since most functions in {\tt scipy.cluster} operate 
on condensed distance matrices.
In line 5, the {\tt python} module {\tt dumpDendrogram} is imported. It 
implements several functions that allow to display the results of 
a hierarchical cluster analysis as \emph{scalable vector graphics} (SVG).
(Note that {\tt scipy} also offers means to display clusters via 
Dendrograms. However, sometimes it is useful to implement things on 
your own to dive into the nitty-gritty details and to arrive at a 
more deep understanding of the matter.)

In the main routine, comprising lines 16 through 29, a full distance 
matrix, stemming from $N=6$ vector-valued observations in $2D$, is 
computed (line 19). For this purpose, the function \verb!myDistMatrix_2D!, 
defined in lines 7 through 14, is called. In line 13, this 
function makes use of the euclidean distance function, contained in 
the {\tt scipy.spatial.distance} submodule (of course, one could have
implemented this oneself). However, please note the way the multidimensional
{\tt scipy} array \verb!Dij! is defined (line 10) and indexed (line 13).
This is special to {\tt scipy} arrays and is supposed to provide
a slightly faster access to the array elements.
As mentioned above, most {\tt scipy.cluster} functions assume condensed
distance matrices. For this reason, in line 22, the function 
{\tt scipy.spatial.squareform} is used to convert the full to a condensed 
distance matrix.

\begin{figure}
\centering
\includegraphics[width=0.95\textwidth]{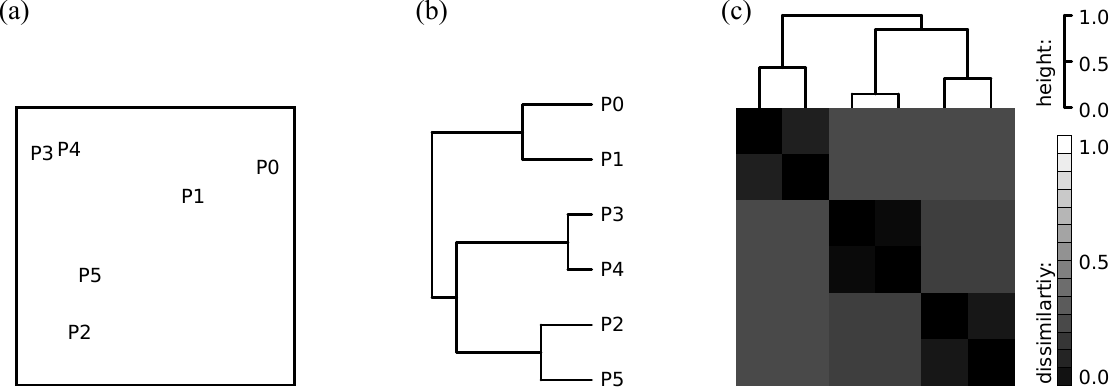}
\caption{
Exemplary analysis of 6 data points in a plane using a hierarchical cluster analysis.
(a) Location of points P0 through P5 in the plane.
(b) Dendrogram illustrating the ``distance'' between the data points as obtained
by hierarchical clustering (the distance between the data points is given by the
euclidean distance between the points).
(c) Dendrogram on top of the cophenetic distance matrix of the data points (in the distance
matrix, a bright (dark) color indicate a significant dissimilarity (similarity) of 
the data points in terms of the distance function). The cophenetic distance between 
two data points is simply the distance between the respective clusters (see text).
\label{fig:clusterEX_N6}}
\end{figure}

In line 23, the scipy function \verb!linkage! is used to generate 
a hierarchical clustering from the condensed distance matrix.
In principal, it takes three arguments,
the first of which is the condensed distance matrix. The 
second argument specifies the method by which the 
clustering will be computed, affecting the way distances
between clusters are measured. It implements various linkage methods, e.g.,
\begin{itemize}
\item ``single'' assigns the distance
$d(u,v) = \min({\rm dist}(u[i],v[j]))$
for all points $i$ in cluster $u$ and
$j$ in cluster $v$ (also referred to as the 
\emph{nearest point algorithm}),
\item ``complete'' assigns the distance
$d(u, v) = \max({\rm dist}(u[i],v[j]))$
for all points $i$ in cluster $u$ and $j$ in
cluster $v$ (also referred to as the \emph{farthest point
algorithm}), and 
\item ``average'' assigns the distance
$d(u,v) = \sum_{ij} \frac{{\rm dist}(u[i], v[j])}
                                   {(|u|*|v|)}$
for all points $i$ and $j$ where $|u|$
and $|v|$ are the number of points in the respective clusters
(also referred to as the \emph{unweighted pair group with
arithmetic mean algorithm} (UPGMA)).
\end{itemize}
The argument ``method'' defaults to ``single''.
The third argument specifies the metric to be used for 
measuring the distance between individual points, and 
it defaults to the euclidean distance.
Finally, the function returns the result of the cluster analysis 
as linkage matrix. To be more precise, the linkage matrix is a 
4 by $N-1$ matrix $Z$ that encodes the progress of the 
analysis (see example below). I.e., at the $i$-th iteration of the 
hierarchical cluster algorithm, the entries $Z[i,0]$ and $Z[i,1]$
specify the clusters which are merged to form a cluster with 
integer identifier $N+i$ (note that a cluster with identifier
smaller than $N$ corresponds to one of the original observations).
The distance between the two merged clusters is stored in $Z[i,2]$, 
and $Z[i,3]$ contains the number of original observations in the 
new cluster. 

In line 24, the cophenetic distances between the individual observations
are computed and returned as condensed cophenetic distance matrix.
Say, $i$ and $j$ label original observations in
disjoint clusters $u$ and $v$, respectively, and
$u$ and $v$ are joined by a direct parent cluster
$w$. Then, the cophenetic distance between $i$ and $j$ 
is simply the distance between the clusters $u$ and $v$.
In line 25, the condensed cophenetic distance matrix is transformed to a
full one and converted to a SVG image in line 29.
The resulting SVG image shows the Dendrogram that corresponds to the
linkage matrix on top of the cophenetic distance matrix (see example below).

Invoking the script on the command line might yield the following output.
The full distance matrix \verb!Dij_sq! (computed in line 19; largest distance is set to 
1 and all other distances are scaled accordingly) might read
\SourceCodeLines{9}
\begin{Verbatim}
 [[ 0.     0.319  1.     0.903  0.796  0.829]
  [ 0.319  0.     0.709  0.627  0.528  0.518]
  [ 1.     0.709  0.     0.731  0.733  0.234]
  [ 0.903  0.627  0.731  0.     0.109  0.522]
  [ 0.796  0.528  0.733  0.109  0.     0.508]
  [ 0.829  0.518  0.234  0.522  0.508  0.   ]],
\end{Verbatim}
for which the result of the cluster analysis (obtained in line 23) yields
the linkage matrix \verb!clusterResult!
\SourceCodeLines{9}
\begin{Verbatim}
 [[ 3.          4.          0.109  2.   ]
  [ 2.          5.          0.234  2.   ]
  [ 0.          1.          0.319  2.   ]
  [ 6.          7.          0.624  4.   ]
  [ 8.          9.          0.739  6.   ]].
\end{Verbatim}
Finally, the full cophenetic distance matrix \verb!Cij_sq! (obtained in line 25) reads
\SourceCodeLines{9}
\begin{Verbatim}
 [[ 0.     0.319  0.739  0.739  0.739  0.739]
  [ 0.319  0.     0.739  0.739  0.739  0.739]
  [ 0.739  0.739  0.     0.624  0.624  0.234]
  [ 0.739  0.739  0.624  0.     0.109  0.624]
  [ 0.739  0.739  0.624  0.109  0.     0.624]
  [ 0.739  0.739  0.234  0.624  0.624  0.   ]],
\end{Verbatim}
and the resulting SVG image looks as shown in Fig.\ \ref{fig:clusterEX_N6}(c)
while Figs.\ \ref{fig:clusterEX_N6}(a),(b) show the distribution of the 6 points 
and the Dendrogram obtained by the hierarchical cluster analysis, respectively.

There are several things to note that might facilitate the interpretation 
of Dendrograms. A Dendrogram displays the hierarchical structure from single-object
clusters up to the full final cluster composed of all considered objects.
Therein, the joining 'height' of two clusters indicates the distance between the 
two clusters. The earlier two clusters merge, the more similar they are, while long
branches indicate a large difference between the joined clusters. Usually, 
the order of the Dendrogram leaves has no particular relevance. Finally, a
classification of the initial objects can be obtained by cutting the tree 
at a \emph{meaningful} height. But that is an entirely different topic, which will not be covered here.



\section{Struggling for speed}

\subsection[Combining {\tt C} and {\tt python} ({\tt Cython})]{Combining {\tt C} and {\tt python} ({\tt Cython}): sciencing as fast as you can!}

To reimplement the example in the following section, you will need to
have the {\tt cython} extension module installed, see \cite{cython,behnel2010}.
\medskip

As should be clear by now, code development in {\tt python} is usually 
rather fast and the result is very concise and almost as readable as 
pseudocode. However, when it comes to scientific computing and 
computationally rather involved tasks in data analysis (as e.g.\ 
bootstrap resampling), {\tt python} scripts tend to be comparatively slow.
One possible reason for this might be that the program spends lots of time
in looping constructs which are costly in terms of computing time (using
pure {\tt python}). As a remedy one might use {\tt cython} \cite{cython}, an extension to the 
{\tt python} language. It offers the possibility to directly compile slightly modified 
{\tt python} code to C, and to link the resulting C code against
{\tt python}. This yields compiled code that can be imported and executed by pure
{\tt python} scripts.
In doing so, {\tt cython} makes it possible to speed up {\tt python} code and 
to directly interact with external C code.
 
In order to benefit from {\tt cython}, there is no need to fully change 
your {\tt python} programming habits. This means that in a first step you can 
assemble your software project as you would usually do in {\tt python}, and 
in a second step you can identify those parts that might be optimized 
further by means of {\tt cython}. 
In this way you can selectively speed up your code where it is most 
beneficial.

To prepare pure {\tt python} scripts for use with {\tt cython},
you might amend the existing code by statically declaring C data types
for some variables using {\tt cython} specific syntax. 
E.g., the \verb!cdef! statement and the possibility to prepend data 
types in argument lists are {\tt cython} extensions to the original 
{\tt python} language.

In effect, {\tt cython} combines the speed of {\tt C} with the flexibility and
conciseness of {\tt python} \cite{behnel2010}.

As an example to illustrate the extent to which an existing script needs to be altered in order
to achieve a valuable speed-up using {\tt cython}, we will perform a small 
benchmark test of {\tt python} vs.\ {\tt cython}. 
For this purpose, consider the following pure {\tt python} code (\verb!boxMueller_python.py!) 
for the generation of Gaussian random numbers using the \emph{Box-M\"uller} method:
\SourceCodeLines{99}
\begin{Verbatim}[fontfamily=txtt,commandchars=\\\{\}]
 \PY{k+kn}{from} \PY{n+nn}{math} \PY{k+kn}{import} \PY{n}{pi}\PY{p}{,}\PY{n}{sqrt}\PY{p}{,}\PY{n}{log}\PY{p}{,}\PY{n}{sin}
 \PY{k+kn}{from} \PY{n+nn}{random} \PY{k+kn}{import} \PY{n}{seed}\PY{p}{,}\PY{n}{random}

 \PY{k}{def} \PY{n+nf}{gaussRand}\PY{p}{(}\PY{n}{mu}\PY{p}{,}\PY{n}{sigma}\PY{p}{)}\PY{p}{:}
        \PY{n}{u1} \PY{o}{=} \PY{n}{random}\PY{p}{(}\PY{p}{)}
        \PY{n}{u2} \PY{o}{=} \PY{n}{random}\PY{p}{(}\PY{p}{)}
        \PY{n}{r}   \PY{o}{=} \PY{n}{sqrt}\PY{p}{(}\PY{o}{-}\PY{l+m+mf}{2.}\PY{o}{*}\PY{n}{log}\PY{p}{(}\PY{n}{u1}\PY{p}{)}\PY{p}{)}
        \PY{n}{phi} \PY{o}{=} \PY{l+m+mf}{2.}\PY{o}{*}\PY{n}{pi}\PY{o}{*}\PY{n}{u2}
        \PY{k}{return} \PY{n}{mu}\PY{o}{+}\PY{n}{sigma}\PY{o}{*}\PY{n}{r}\PY{o}{*}\PY{n}{sin}\PY{p}{(}\PY{n}{phi}\PY{p}{)}

 \PY{k}{def} \PY{n+nf}{main\PYZus{}python}\PY{p}{(}\PY{n}{N}\PY{p}{,}\PY{n}{mu}\PY{p}{,}\PY{n}{sigma}\PY{p}{)}\PY{p}{:}
        \PY{n}{seed}\PY{p}{(}\PY{l+m+mi}{0}\PY{p}{)}
        \PY{k}{for} \PY{n}{i} \PY{o+ow}{in} \PY{n+nb}{range}\PY{p}{(}\PY{n}{N}\PY{p}{)}\PY{p}{:}
                \PY{n}{z1} \PY{o}{=} \PY{n}{gaussRand}\PY{p}{(}\PY{n}{mu}\PY{p}{,}\PY{n}{sigma}\PY{p}{)}
\end{Verbatim}

In lines 1 and 2, several {\tt python} functions from the \verb!math! and \verb!random! libraries 
are imported. Lines 4 through 9 define the function \verb!gaussRand()! that uses two 
random numbers uniformly distributed in $[0,1)$ in order to generate one 
random deviate drawn from a Gaussian distribution with mean \verb!mu! and width \verb!sigma!.
(Note: in principle, one can generated two independent Gaussian random numbers by means
of two independent uniformly distributed ones, but for the purpose of illustration let us just generate one of them.)
Finally, lines 11 through 14 seed the random number generator and call the above function
\verb!N! times using a looping construct. 

A modified script for use with {\tt cython} might look as the following 
code snippet (\verb!boxMueller_cython.pyx!; {\tt cython} scripts usually have the suffix \verb!.pyx!):
\SourceCodeLines{99}
\begin{Verbatim}[fontfamily=txtt,commandchars=\\\{\}]
 \PY{k}{cdef} \PY{k+kr}{extern} \PY{k}{from} \PY{l+s}{"}\PY{l+s}{math.h}\PY{l+s}{"}\PY{p}{:}
        \PY{n}{double} \PY{n}{sin}\PY{p}{(}\PY{n}{double}\PY{p}{)}
        \PY{n}{double} \PY{n}{log}\PY{p}{(}\PY{n}{double}\PY{p}{)}
        \PY{n}{double} \PY{n}{sqrt}\PY{p}{(}\PY{n}{double}\PY{p}{)}
        \PY{n}{double} \PY{n}{M\PYZus{}PI}

 \PY{k}{cdef} \PY{k+kr}{extern} \PY{k}{from} \PY{l+s}{"}\PY{l+s}{stdlib.h}\PY{l+s}{"}\PY{p}{:}
         \PY{n}{double} \PY{n}{drand48}\PY{p}{(}\PY{p}{)}
         \PY{n}{void} \PY{n}{srand48}\PY{p}{(}\PY{n}{unsigned} \PY{n+nb}{int} \PY{n}{SEED}\PY{p}{)}

 \PY{k}{cdef} \PY{k+kt}{double} \PY{n+nf}{gaussRand}\PY{p}{(}\PY{n}{double} \PY{n}{mu}\PY{p}{,}\PY{n}{double} \PY{n}{sigma}\PY{p}{)}\PY{p}{:}
        \PY{k}{cdef} \PY{k+kt}{double} \PY{n+nf}{u1}\PY{p}{,}\PY{n+nf}{u2}\PY{p}{,}\PY{n+nf}{r}\PY{p}{,}\PY{n+nf}{phi}
        \PY{n}{u1} \PY{o}{=} \PY{n}{drand48}\PY{p}{(}\PY{p}{)}
        \PY{n}{u2} \PY{o}{=} \PY{n}{drand48}\PY{p}{(}\PY{p}{)}
        \PY{n}{r}   \PY{o}{=} \PY{n}{sqrt}\PY{p}{(}\PY{o}{-}\PY{l+m+mf}{2.}\PY{o}{*}\PY{n}{log}\PY{p}{(}\PY{n}{u1}\PY{p}{)}\PY{p}{)}
        \PY{n}{phi} \PY{o}{=} \PY{l+m+mf}{2.}\PY{o}{*}\PY{n}{M\PYZus{}PI}\PY{o}{*}\PY{n}{u2}
        \PY{k}{return} \PY{n}{mu}\PY{o}{+}\PY{n}{sigma}\PY{o}{*}\PY{n}{r}\PY{o}{*}\PY{n}{sin}\PY{p}{(}\PY{n}{phi}\PY{p}{)}

 \PY{k}{def} \PY{n+nf}{main\PYZus{}cython}\PY{p}{(}\PY{n}{N}\PY{p}{,}\PY{n}{mu}\PY{p}{,}\PY{n}{sigma}\PY{p}{)}\PY{p}{:}
        \PY{k}{cdef} \PY{k+kt}{int} \PY{n+nf}{i}
        \PY{n}{srand48}\PY{p}{(}\PY{l+m+mf}{0}\PY{p}{)}
        \PY{k}{for} \PY{n}{i} \PY{k}{from} \PY{l+m+mf}{0}\PY{o}{<}\PY{o}{=}\PY{n}{i}\PY{o}{<}\PY{n}{N}\PY{p}{:}
                \PY{n}{z} \PY{o}{=} \PY{n}{gaussRand}\PY{p}{(}\PY{n}{mu}\PY{p}{,}\PY{n}{sigma}\PY{p}{)}
\end{Verbatim}

Therein, in lines 1 through 9, the same functions as in the pure {\tt python} script are 
imported. However, they are not imported from the ``standard'' {\tt python} libraries but 
from the C header files \verb!math.h! and \verb!stdlib.h!. This illustrates the 
ability of {\tt cython} to directly interact with C code. In comparison to the 
{\tt python} implementation, the C implementation of these functions are quite fast (since they
lack the typical {\tt python} overheads spent by calling {\tt python} functions).
Regarding the function definition on line 11, the return value of the \verb!gaussRand()! function is 
declared as \verb!double!, and it is defined via the \verb!cdef! statement. The latter 
results in faster function calling via {\tt cython}. 
Further, in the argument list of the function the data types of the variables 
are prepended. More precisely, both arguments are declared to be of type \verb!double!. 
In the block of statements that follows the function definition, the data types
of all the variables are declared which also results in a significant speed-up of the 
resulting C code. 
Note that in principle one can also mix declaration and value assignment, i.e., statements
such as \verb!cdef double phi = 2.*M_PI*u2! work very well.
Finally, in line 20 the data type of the index \verb!idx! is declared to be of type \verb!int!
and in line 22 a faster version of the \verb!for ... in range(...):! statement,
namely the \verb!for ... from ... :! is implemented.

Post-processing of the resulting \verb!.pyx!-{\tt cython} script uses the setup script \verb!setup.py!:

\SourceCodeLines{9}
\begin{Verbatim}[fontfamily=txtt,commandchars=\\\{\}]
 \PY{k+kn}{from} \PY{n+nn}{distutils.core} \PY{k+kn}{import} \PY{n}{setup}
 \PY{k+kn}{from} \PY{n+nn}{distutils.extension} \PY{k+kn}{import} \PY{n}{Extension}
 \PY{k+kn}{from} \PY{n+nn}{Cython.Distutils} \PY{k+kn}{import} \PY{n}{build\PYZus{}ext}

 \PY{n}{setup}\PY{p}{(}
	\PY{n}{cmdclass}    \PY{o}{=} \PY{p}{\PYZob{}}\PY{l+s}{'}\PY{l+s}{build\PYZus{}ext}\PY{l+s}{'}\PY{p}{:} \PY{n}{build\PYZus{}ext}\PY{p}{\PYZcb{}}\PY{p}{,}
	\PY{n}{ext\PYZus{}modules} \PY{o}{=} \PY{p}{[}\PY{n}{Extension}\PY{p}{(}\PY{l+s}{"}\PY{l+s}{boxMueller\PYZus{}cython}\PY{l+s}{"}\PY{p}{,}\PYZbs{}
                           \PY{p}{[}\PY{l+s}{"}\PY{l+s}{boxMueller\PYZus{}cython.pyx}\PY{l+s}{"}\PY{p}{]}\PY{p}{)}\PY{p}{]}\PY{p}{,}  
 \PY{p}{)}
\end{Verbatim}

Invoking the script by \verb!python setup.py build_ext --inplace! yields the shared library \verb!boxMueller_cython.so!, 
which you can import and use within any {\tt python} script. 
The above \verb!setup.py! script uses the \verb!distutils! module which provides
support for building extension modules for {\tt python}. Furthermore, it uses the \verb!Cython! 
module to extend the capabilities of {\tt python}.

Now, the benchmark test might be performed as follows: 
create a minimal {\tt python} script that executes the 
functions \verb!main_python! and \verb!main_cython! for the parameters $N=10^7$, $\mu=0.$, 
$\sigma=1.$, and determines the respective running times
using the \verb!timeit! module \cite{py_timeit} (\verb!timeit! offers the possibility to time small 
pieces of {\tt python} code; as an alternative you might use the common unix tool \verb!time!). 
This might look as follows (\verb!benchmark_boxMueller.py!):
\SourceCodeLines{9}
\begin{Verbatim}[fontfamily=txtt,commandchars=\\\{\}]
 \PY{k+kn}{from} \PY{n+nn}{timeit} \PY{k+kn}{import} \PY{n}{Timer}

 \PY{n}{N}\PY{o}{=}\PY{l+m+mi}{10000000}

 \PY{c}{\PYZsh{} time PYTHON code}
 \PY{n}{t\PYZus{}py} \PY{o}{=} \PY{n}{Timer}\PY{p}{(}\PY{l+s}{"}\PY{l+s}{main\PYZus{}python(}\PY{l+s+si}{\PYZpc{}d}\PY{l+s}{,0.,1.)}\PY{l+s}{"}\PY{o}{\PYZpc{}}\PY{p}{(}\PY{n}{N}\PY{p}{)}\PY{p}{,}
             \PY{l+s}{"}\PY{l+s}{from boxMueller\PYZus{}python import main\PYZus{}python}\PY{l+s}{"}
             \PY{p}{)}\PY{o}{.}\PY{n}{timeit}\PY{p}{(}\PY{n}{number}\PY{o}{=}\PY{l+m+mi}{1}\PY{p}{)}

 \PY{c}{\PYZsh{} time CYTHON code}
 \PY{n}{t\PYZus{}cy} \PY{o}{=} \PY{n}{Timer}\PY{p}{(}\PY{l+s}{"}\PY{l+s}{main\PYZus{}cython(}\PY{l+s+si}{\PYZpc{}d}\PY{l+s}{,0.,1.)}\PY{l+s}{"}\PY{o}{\PYZpc{}}\PY{p}{(}\PY{n}{N}\PY{p}{)}\PY{p}{,}
             \PY{l+s}{"}\PY{l+s}{from boxMueller\PYZus{}cython import main\PYZus{}cython}\PY{l+s}{"}
             \PY{p}{)}\PY{o}{.}\PY{n}{timeit}\PY{p}{(}\PY{n}{number}\PY{o}{=}\PY{l+m+mi}{1}\PY{p}{)}

 \PY{k}{print} \PY{l+s}{"}\PY{l+s}{\PYZsh{} PYTHON vs CYTHON: Box-Mueller Benchmark test}\PY{l+s}{"}
 \PY{k}{print} \PY{l+s}{"}\PY{l+s}{\PYZsh{} number of random deviates N=}\PY{l+s}{"}\PY{p}{,}\PY{n}{N} 
 \PY{k}{print} \PY{l+s}{"}\PY{l+s}{\PYZsh{} PYTHON: t\PYZus{}PY =}\PY{l+s}{"}\PY{p}{,}\PY{n}{t\PYZus{}py}
 \PY{k}{print} \PY{l+s}{"}\PY{l+s}{\PYZsh{} CYTHON: t\PYZus{}CY =}\PY{l+s}{"}\PY{p}{,}\PY{n}{t\PYZus{}cy}
 \PY{k}{print} \PY{l+s}{"}\PY{l+s}{\PYZsh{} speed-up factor t\PYZus{}PY/t\PYZus{}CY =}\PY{l+s}{"}\PY{p}{,}\PY{n}{t\PYZus{}py}\PY{o}{/}\PY{n}{t\PYZus{}cy}
\end{Verbatim}

Calling it on the command line yields:
\SourceCodeLines{9}
\begin{Verbatim}
 > python benchmark_boxMueller.py
 # PYTHON vs CYTHON: Box-Mueller Benchmark test
 # number of random deviates N= 10000000
 # PYTHON: t_PY = 10.7791149616
 # CYTHON: t_CY = 0.316421031952
 # speed-up factor t_PY/t_CY = 34.0657348063
\end{Verbatim}
Hence, the {\tt cython} code outperforms the pure {\tt python} code by a factor of approximately $34$.
Given the few changes that were necessary in order to generate the shared library \verb!boxMueller_cython.so!, this is an impressive speed-up.

\subsection[Parallel computing and {\tt python}]{Parallel computing and {\tt python}: do both!}
Another way to speed up computations is to perform them in parallel, if possible.
For this purpose, {\tt python} offers several modules such as, e.g., the {\tt threading} and
{\tt multiprocessing} modules.
However, if you want to benefit from the speed-up that can be achieved by using 
multiple cores simultaneously, you might want to go for the latter one.
The problem with the {\tt threading} module is that it suffers from the ``Global 
Interpreter Lock'' (GIT), which effectively limits the performance of the code since it 
allows only one thread at a time to execute {\tt python} code.
The {\tt multiprocessing} module does not suffer from a GIT and yields a speed-up
that depends on the number of cores that are available.

As an example to illustrate the {\tt multiprocessing} module we 
here perform a parallel simulation of random walks. This is a 
rather simple application where subprocesses do not need 
to communicate 
(\verb!mp_2DrandWalk_oop.py!): 
\SourceCodeLines{99}
\begin{Verbatim}[fontfamily=txtt,commandchars=\\\{\}]
 \PY{k+kn}{from} \PY{n+nn}{math} \PY{k+kn}{import} \PY{n}{sqrt}\PY{p}{,}\PY{n}{cos}\PY{p}{,}\PY{n}{sin}\PY{p}{,}\PY{n}{pi}
 \PY{k+kn}{import} \PY{n+nn}{random}
 \PY{k+kn}{import} \PY{n+nn}{multiprocessing} 

 \PY{k}{class} \PY{n+nc}{myProcess}\PY{p}{(}\PY{n}{multiprocessing}\PY{o}{.}\PY{n}{Process}\PY{p}{)}\PY{p}{:}
        \PY{k}{def} \PY{n+nf}{\PYZus{}\PYZus{}init\PYZus{}\PYZus{}}\PY{p}{(}\PY{n+nb+bp}{self}\PY{p}{,}\PY{n}{myLock}\PY{p}{,}\PY{n}{myRng}\PY{p}{,}\PY{n}{mySeed}\PY{p}{,}\PY{n}{Nsteps}\PY{p}{)}\PY{p}{:}
                \PY{n+nb+bp}{self}\PY{o}{.}\PY{n}{myLock} \PY{o}{=} \PY{n}{myLock}
                \PY{n+nb+bp}{self}\PY{o}{.}\PY{n}{myRng}  \PY{o}{=} \PY{n}{myRng}
                \PY{n+nb+bp}{self}\PY{o}{.}\PY{n}{mySeed} \PY{o}{=} \PY{n}{mySeed}
                \PY{n+nb+bp}{self}\PY{o}{.}\PY{n}{Nsteps} \PY{o}{=} \PY{n}{Nsteps}
                \PY{n}{multiprocessing}\PY{o}{.}\PY{n}{Process}\PY{o}{.}\PY{n}{\PYZus{}\PYZus{}init\PYZus{}\PYZus{}}\PY{p}{(}\PY{n+nb+bp}{self}\PY{p}{)}

        \PY{k}{def} \PY{n+nf}{run}\PY{p}{(}\PY{n+nb+bp}{self}\PY{p}{)}\PY{p}{:}
                \PY{n+nb+bp}{self}\PY{o}{.}\PY{n}{myRng}\PY{o}{.}\PY{n}{seed}\PY{p}{(}\PY{n+nb+bp}{self}\PY{o}{.}\PY{n}{mySeed}\PY{p}{)}
                \PY{n}{x}\PY{o}{=}\PY{n}{y}\PY{o}{=}\PY{l+m+mf}{0.}
                \PY{k}{for} \PY{n}{i} \PY{o+ow}{in} \PY{n+nb}{range}\PY{p}{(}\PY{n+nb+bp}{self}\PY{o}{.}\PY{n}{Nsteps}\PY{p}{)}\PY{p}{:}
                   \PY{n}{phi}  \PY{o}{=} \PY{n+nb+bp}{self}\PY{o}{.}\PY{n}{myRng}\PY{o}{.}\PY{n}{random}\PY{p}{(}\PY{p}{)}\PY{o}{*}\PY{l+m+mf}{2.}\PY{o}{*}\PY{n}{pi}
                   \PY{n}{x}   \PY{o}{+}\PY{o}{=} \PY{n}{cos}\PY{p}{(}\PY{n}{phi}\PY{p}{)}
                   \PY{n}{y}   \PY{o}{+}\PY{o}{=} \PY{n}{sin}\PY{p}{(}\PY{n}{phi}\PY{p}{)}
                \PY{n+nb+bp}{self}\PY{o}{.}\PY{n}{myLock}\PY{o}{.}\PY{n}{acquire}\PY{p}{(}\PY{p}{)}
                \PY{k}{print} \PY{n+nb+bp}{self}\PY{o}{.}\PY{n}{mySeed}\PY{p}{,}\PY{n}{sqrt}\PY{p}{(}\PY{n}{x}\PY{o}{*}\PY{n}{x}\PY{o}{+}\PY{n}{y}\PY{o}{*}\PY{n}{y}\PY{p}{)}
                \PY{n+nb+bp}{self}\PY{o}{.}\PY{n}{myLock}\PY{o}{.}\PY{n}{release}\PY{p}{(}\PY{p}{)}

 \PY{k}{def} \PY{n+nf}{main}\PY{p}{(}\PY{p}{)}\PY{p}{:}
        \PY{n}{N}       \PY{o}{=} \PY{l+m+mi}{10000}
        \PY{n}{nSamp}   \PY{o}{=} \PY{l+m+mi}{5}
        \PY{n}{myLock}  \PY{o}{=} \PY{n}{multiprocessing}\PY{o}{.}\PY{n}{Lock}\PY{p}{(}\PY{p}{)}
        \PY{k}{for} \PY{n}{s} \PY{o+ow}{in} \PY{n+nb}{range}\PY{p}{(}\PY{n}{nSamp}\PY{p}{)}\PY{p}{:}
                \PY{n}{myRng}   \PY{o}{=} \PY{n}{random}\PY{o}{.}\PY{n}{Random}\PY{p}{(}\PY{p}{)}
                \PY{n}{process} \PY{o}{=} \PY{n}{myProcess}\PY{p}{(}\PY{n}{myLock}\PY{p}{,}\PY{n}{myRng}\PY{p}{,}\PY{n}{s}\PY{p}{,}\PY{n}{N}\PY{p}{)}
                \PY{n}{process}\PY{o}{.}\PY{n}{start}\PY{p}{(}\PY{p}{)}

        \PY{k}{for} \PY{n}{p} \PY{o+ow}{in} \PY{n}{multiprocessing}\PY{o}{.}\PY{n}{active\PYZus{}children}\PY{p}{(}\PY{p}{)}\PY{p}{:}
                \PY{k}{print} \PY{l+s}{"}\PY{l+s}{random walk ID: }\PY{l+s}{"}\PY{p}{,} \PY{n}{p}\PY{o}{.}\PY{n}{mySeed}

 \PY{n}{main}\PY{p}{(}\PY{p}{)}
\end{Verbatim}

The pivotal object, contained in the {\tt multiprocessing}
module, is the {\tt Process} class.
In lines 5--22 of the above example,
we derived the subclass \verb!myProcess! that overwrites the default 
constructor of the {\tt Process} class (lines 6--11) in order
to generate an instance of the class that has several additional 
properties: 
\verb!myLock!, a lock that prevents multiple processes to write
to the standard output simultaneously,
\verb!myRng! and \verb!mySeed!, 
specifying its own instance of a random number generator and 
the respective seed, and \verb!Nstep!, referring to the number of steps in the walk.
The method \verb!run!, defined in lines 13--22, finally implements
the $2D$ random walk and prints the seed and the geometric distance
to the starting point to the standard output (thereby using the lock 
in order to make sure that the output is not messed up by other processes).
In line 31 of the function \verb!main!, the \verb!start! method of the 
{\tt Process} class is called. In effect, the start method invokes the
\verb!run! method of the class and starts the activity of the process.
All active processes can be addressed by a call to \verb!active_children!,
which returns a list of all active children of the current process.
This is illustrated in lines 33 and 34. Finally, invoking the 
script on the command line yields the output:
\smallskip

\SourceCodeLines{99}
\begin{Verbatim}
 > python mp_2DrandWalk_oop.py 
 random walk ID:  2
 random walk ID:  4
 random walk ID:  0
 random walk ID:  3
 random walk ID:  1
 0 105.760414243
 2 61.8775421529
 1 80.2195530965
 3 144.975510642
 4 66.9035741041
\end{Verbatim}

\begin{figure}
\centering
\includegraphics[width=\textwidth]{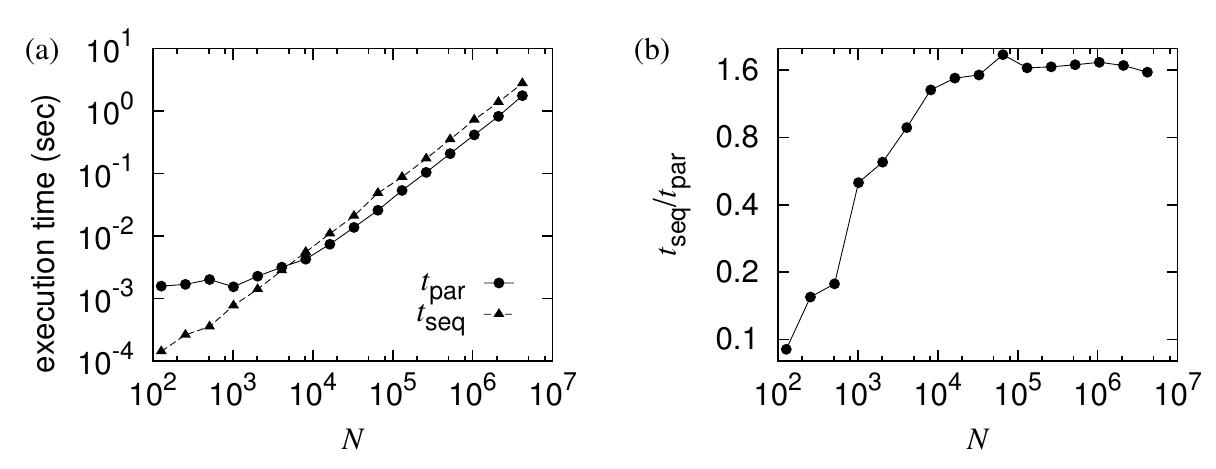}
\caption{
Speed-up obtained by parallelization of the random walk code
using the {\tt multiprocessing} module.
(a) Execution times $t_{\rm seq}$ and $t_{\rm par}$ for random walks of length $N$
for a sequential and parallel implementation, respectively. For short execution
times, the running time of the parallel code is dominated by a 
time overhead that is due to the parallel implementation.
(b) Speed-up obtained on a 1.7 GHz Intel Core i5 with 2 cores.
\label{fig:mp_randWalk_oop}}
\end{figure}

The speed-up obtained by parallelizing your code will depend on several 
factors. First and foremost, it depends on the number of cores available and on the 
number of additional processes that are running on the machine. Further, 
the running time of the started processes also plays a role, since 
the parallelization comes to the expense of a small time overhead during 
the initialization phase. This means that if the running time of the
individual processes is rather small, the overall running time is dominated
by the time overhead. In Fig.\ \ref{fig:mp_randWalk_oop}(a) this is 
illustrated for random walks with an increasing number of steps $N$
(data obtained using the script \verb!mp_randWalk_timer.py!, again using
the {\tt timeit} module).
As evident from Fig.\ \ref{fig:mp_randWalk_oop}(b), the speed-up at 
large values of $N$ is approximately $1.8$ 
(obtained on a 1,7 GHz Intel Core i5 with 2 cores). 


\bigskip

\section{Further reading}

In the following I will 
list a selection of references wherein some more background 
(or simply further useful information) on the topics covered in 
Section \ref{sect:basicDataAnalysis} can be found.

In order to prepare Section \ref{sect:pf}, I used Refs.\ \cite{clrs2001} 
(see appendix {\rm C}) and \cite{practicalGuide2009} (see chapter $7.1$).
In particular, the part on ``Basic parameters related to random variables'' was prepared
using Ref.\ \cite{nrc1992} (see chapter $14.1$).
The part on ``Estimators with(out) bias'' picked up a little bit of 
stem from Ref.\ \cite{bendat1971}.
Section \ref{sect:hist} was prepared using Refs.\ \cite{practicalGuide2009} 
(see chapter $7.3.3$) and \cite{newman2005}. The latter reference is a 
research article with focus on analyzing power law distributed data.
Section \ref{sect:resampling} on unbiased error estimation
via bootstrap resampling could benefit from Refs.\ \cite{nrc1992}
(see chapter $15.6$) and \cite{practicalGuide2009} (see chapter $7.3.4$).
Also, Ref.\ \cite{gould2007} considers the bootstrap method in terms
of an example problem (see problem $11.12$), dealing with error estimation.
Finally, Section \ref{sect:chisquare} on the chi-square test 
was partly prepared using Refs.\ \cite{nrc1992} (see chapter $14.3$), 
\cite{practicalGuide2009} (see chapter $7.5.1$), and \cite{bendat1971} (see chapter $4.6$).

\paragraph{Acknowledgements.} I would like to thank A.\ K.\ Hartmann, 
C.\ Norrenbrock, and P.\ Fieth for critically reading the manuscript and for
providing comments and suggestions that helped to amend the document.
%

\bibliographystyle{plain}      
\bibliography{MCS_LNMelchertRefs}      

\end{document}